\documentclass[12pt,journal,draftclsnofoot,onecolumn]{IEEEtran}
\IEEEoverridecommandlockouts

\makeatletter
\def\markboth#1#2{\def\leftmark{\@IEEEcompsoconly{\sffamily}\MakeUppercase{\protect#1}}%
\def\rightmark{\@IEEEcompsoconly{\sffamily}\MakeUppercase{\protect#2}}}
\makeatother

\usepackage[english]{babel}
\usepackage{color}
\usepackage{lipsum}
\usepackage{caption}
\usepackage{cite}
\usepackage[pdftex]{graphicx}
\usepackage[font=footnotesize]{subcaption}
\usepackage{amsmath}
\usepackage{amsfonts}
\usepackage{array}
\newcolumntype{P}[1]{>{\centering\arraybackslash}p{#1}}
\usepackage{verbatim}
\usepackage{listings}
\usepackage{algorithm}
\usepackage{algpseudocode}
\usepackage{hyperref}
\usepackage{url}
\usepackage{enumerate}
\usepackage{multirow}
\usepackage{mathtools}
\usepackage{makecell}

\usepackage{epsfig}
\usepackage{epstopdf}
\usepackage{multicol}
\usepackage[font=footnotesize]{caption}

\usepackage{ntheorem}
\usepackage{indentfirst}

\DeclareMathOperator*{\argmax}{arg\,max}
\renewcommand{\arraystretch}{2}

\usepackage{amssymb}
\usepackage{mleftright}

\newcommand{\bi}{\begin{itemize}}
\newcommand{\ei}{\end{itemize}}
\newcommand{\be}{\begin{equation}}
\newcommand{\ee}{\end{equation}}

\addto\captionsenglish{}

\addto\captionsenglish{}

\theoremheaderfont{\hspace*{\parindent}\itshape }
\theorembodyfont{\normalfont}
\theoremseparator{:}

\theoremheaderfont{\hspace*{\parindent}\bfseries}
\theorembodyfont{\normalfont}
\theoremseparator{:}
\theoremstyle{remark}

\theoremheaderfont{\hspace*{\parindent} \normalfont \bfseries}
\theorembodyfont{\itshape}
\theoremseparator{:}
\newtheorem{theorem}{Theorem}
\newtheorem{lemma}{Lemma}

\def\beq{\begin{equation}}
\def\eeq{\end{equation}}
\def\beqa{\begin{eqnarray}}
\def\eeqa{\end{eqnarray}}
\def\beqan{\begin{eqnarray*}}
\def\eeqan{\end{eqnarray*}}

\usepackage{adjustbox}

\newcommand{\captionaboveof}[3][]{%
    \vskip-\abovecaptionskip
    \vskip+\belowcaptionskip
    \def\@captype{#2}%
    \ifx\@nnil#1\@nnil
        \caption{#3}%
    \else
        \caption[#1]{#3}%
    \fi
    \vskip+\abovecaptionskip
    \vskip-\belowcaptionskip
}
\makeatother

\newcommand{\ts}{$T_{\rm S }$ }

\usepackage{threeparttable}
\usepackage{xfrac}
\usepackage[table,xcdraw]{xcolor}
\usepackage{tabularx}
   \usepackage{multirow}
   \usepackage{booktabs}

   \usepackage{graphicx}
   \usepackage{array, blindtext}
   \usepackage{wrapfig}
\usepackage{pdfpages}

\usepackage{tikz}
\usepackage{pgfplots}
\pgfplotsset{compat=newest} 
\pgfplotsset{plot coordinates/math parser=false} 
\newlength\fheight
\newlength\fwidth
\usetikzlibrary{patterns,decorations.pathreplacing,backgrounds,calc,spy,arrows.meta}
\definecolor{SchoolColor}{RGB}{0.71, 0, 0.106}
\definecolor{chaptergrey}{rgb}{0.61, 0, 0.09} 
\definecolor{midgrey}{rgb}{0.4, 0.4, 0.4}
\definecolor{chaptergreen}{rgb}{0.09, 0.612, 0}
\definecolor{chapterpurple}{rgb}{0.522, 0, 0.612}
\definecolor{chapterlightgreen}{rgb}{0, 0.612, 0.522}

            \usepgfplotslibrary{colorbrewer}
            
            \def \squeeze_eq{
            \beq
            \mathcal{L}^{(s)}_{I_i^j}(t){=} \exp\left( -2\lambda_b \int_{ \left( \frac{C_{j}}{C_i} r^{\alpha_{i}} \right)^{\frac{1}{\alpha_{j}}}}^\infty \left[ 1-\left( \frac{\sfrac{1}{\mu}(\theta_b/\pi)}{\sfrac{1}{\mu}+tv^{-\alpha_j}C_j G_b G_{\rm VN}} + \frac{\sfrac{1}{\mu}[1-(\theta_b/\pi)]}{\sfrac{1}{\mu}+tv^{-\alpha_j}C_j g_b g_{\rm VN}} \right) \right] p^{(s)}_j(v) \mathrm{d}v  \right).
            \label{eq:appendix_P_cov_2}
            \eeq
            }

            \def \Laplace{
             \begin{align}
            \mathbb{P}&\left[|h_1|^2 > \frac{ \Big[(I_L + I_N) + \sigma^2\Big]\Gamma r^{\alpha_i}}{\Delta_1 C_i}  \mathrel{\bigg|} r \right] = \mathbb{E}_{r,I_L, I_N}\left\lbrace \mathbb{P}\left[|h_1|^2 > \frac{ \Big[(I_L + I_N) + \sigma^2\Big]\Gamma r^{\alpha_i}}{\Delta_1 C_i}  \mathrel{\bigg|} r,I_L, I_N \right]  \right\rbrace \notag \\
            &= \mathbb{E}_{I_L, I_N}\left[  \exp\left( \frac{-\mu  \sigma^2\Gamma r^{\alpha_i}}{\Delta_1 C_i}   \right) \exp\left( \frac{-\mu  I_L\Gamma r^{\alpha_i}}{\Delta_1 C_i}   \right) \exp\left( \frac{-\mu  I_N\Gamma r^{\alpha_i}}{\Delta_1 C_i}   \right) \mathrel{\bigg|} I_L, I_N  \right] \notag \\
            &\stackrel{(b)}{=} \exp\left( \frac{-\mu\sigma^2 \Gamma r^{\alpha_i}}{\Delta_1 C_i} \right)
             \mathcal{L}^{(s)}_{I_i^L}\left(  \frac{\mu \Gamma r^{\alpha_i}}{\Delta_1 C_i} \right) 
             \mathcal{L}^{(s)}_{I_i^N}\left(  \frac{\mu \Gamma r^{\alpha_i}}{\Delta_1 C_i} \right),
             \label{eq:appendix_P_cov_laplace}
            \end{align}
            }
            
             \def \eq_text{
            \beq
            \mathcal{L}^{(s)}_{I_i^j}(t){=} \exp\left( -2\lambda_b \int_{ \left( \frac{C_{j}}{C_i} r^{\alpha_{i}} \right)^{\frac{1}{\alpha_{j}}}}^\infty \left[ 1-\left( \frac{\sfrac{1}{\mu}(\theta_b/\pi)}{\sfrac{1}{\mu}+tv^{-\alpha_j}C_j G_b G_{\rm VN}} + \frac{\sfrac{1}{\mu}[1-(\theta_b/\pi)]}{\sfrac{1}{\mu}+tv^{-\alpha_j}C_j g_b g_{\rm VN}} \right) \right] p^{(s)}_j(v) \mathrm{d}v  \right).           
            \label{eq:L_I}
                        \eeq
            }

            \def \sinrfinal{
                \begin{equation}
                \mathbb{P}\Big[ \text{SINR} > \Gamma,\, n^* \in \Phi_{i} \Big] =  \int_W^\infty \exp\left( \frac{-\mu\sigma^2 \Gamma r^{\alpha_i}}{\Delta_1 C_i} \right)
               \mathcal{L}^{(s)}_{I_i^L}\left(  \frac{\mu \Gamma r^{\alpha_i}}{\Delta_1 C_i} \right) 
               \mathcal{L}^{(s)}_{I_i^N}\left(  \frac{\mu \Gamma r^{\alpha_i}}{\Delta_1 C_i} \right) \bar{f}^{(s)}_i(r)\mathrm{d}r.
               \label{eq:sinrfinal}
              \end{equation}
          } 

          \def \inside{
              \begin{align}
              \mathcal{L}^{(s)}_{I_i^j}(t)& \triangleq \mathbf{E}_{\Phi_j,h,\Delta_I}\Big[e^{-tI}\Big] = \mathbf{E}_{\Phi_j,h,\Delta_I} \left[ \exp \left( -t \sum_{k\in \Phi_j} |h_k|^2 \ell_j(r_k) \Delta_{I_k} \right)\right]  \notag \\
              &\stackrel{(a)}{=}  \mathbf{E}_{\Phi_j} \left\lbrace \prod_{k\in\Phi_j} \mathbf{E}_{h,\Delta_I} \left[ \exp \left( -t |h_k|^2 r_k^{-\alpha_j}C_j\Delta_{I_k} \right)\right] \right\rbrace\notag  \\
              &\stackrel{(b)}{=}   \exp \left( -2\lambda_b \int_{ \left( \frac{C_{j}}{C_i} r^{\alpha_{i}} \right)^{\frac{1}{\alpha_{j}}}}^\infty \left\lbrace 1-\mathbf{E}_{h_k,\Delta_{I_k}}\left[ \exp\Big( -t |h_k|^2 v^{-\alpha_j}C_j\Delta_{I_k} \Big) \right] \right\rbrace p_j^{(s)}(v) \mathrm{d}v  \right),  
              \label{eq:L_first}
              \end{align}
          }

\title{Coverage \hspace*{-0.3cm} and \hspace*{-0.3cm} Connectivity \hspace*{-0.3cm} Analysis \hspace*{-0.3cm} of  \hspace*{-0.3cm} \\ Millimeter \hspace*{-0.3cm} Wave \hspace*{-0.3cm} Vehicular \hspace*{-0.3cm} Networks}

\author{{{ Marco Giordani}, { Mattia Rebato}, { Andrea Zanella}, { Michele Zorzi} }\\
\normalsize Department of Information Engineering (DEI), University of Padova, Italy \\
\small{$\{$\texttt{giordani}, \texttt{rebatoma}, \texttt{zanella}, \texttt{zorzi}$\}$\texttt{@dei.unipd.it}
}
\thanks{A preliminary version of this paper was
presented at the  \emph{6th International Conference on Modern Circuits and Systems Technologies (MOCAST)}, Thessaloniki, Greece, May 2017 \cite{MOCAST_2017}.}}

\allowdisplaybreaks

\begin{document}
\maketitle

\tikzstyle{startstop} = [rectangle, rounded corners, minimum width=2cm, minimum height=0.5cm,text centered, draw=black]
\tikzstyle{io} = [trapezium, trapezium left angle=70, trapezium right angle=110, minimum width=3cm, minimum height=1cm, text centered, draw=black]
\tikzstyle{process} = [rectangle, minimum width=2cm, minimum height=0.5cm, text centered, draw=black, align=center]
\tikzstyle{decision} = [ellipse, minimum width=2cm, minimum height=1cm, text centered, draw=black]
\tikzstyle{arrow} = [thick,<->,>=stealth]
\tikzstyle{line} = [thick,>=stealth]
\tikzstyle{darrow} = [thick,<->,>=stealth,dashed]
\tikzstyle{sarrow} = [thick,->,>=stealth]

\begin{abstract}  
The next generations of vehicles will require data transmission rates in the order of terabytes per driving hour, to support advanced automotive services. 
This unprecedented amount of data to be exchanged goes beyond the capabilities of existing communication technologies for vehicular communication and calls for new solutions.
A possible answer to this growing demand for ultra-high transmission speeds can be found in the millimeter-wave (mmWave) bands which, however,  are subject to high signal attenuation and challenging propagation characteristics.
In particular, mmWave links are typically directional, to benefit from the resulting beamforming gain, and require precise alignment of the transmitter and the
receiver beams, an operation which may increase the latency of the communication and lead to deafness due to beam misalignment.
In this paper, we propose a stochastic model for characterizing the beam coverage and connectivity probability in mmWave automotive networks.
The purpose is to exemplify some of the complex and interesting tradeoffs
that have to be considered when designing solutions for vehicular  scenarios based on mmWave links. 
The results show that  the performance of the automotive nodes in highly mobile mmWave systems strictly depends on the specific environment in which the vehicles are deployed, and must account for several automotive-specific features such as the nodes speed, the beam alignment periodicity, the base stations density and the antenna geometry.

\end{abstract}


\section{Introduction}
In recent years, vehicle-to-vehicle (V2V) and vehicle-to-infrastructure (V2I) communications, which are collectively referred to as vehicle-to-everything (V2X) communications, have been investigated as a means to support emerging automotive applications ranging from safety services to infotainment \cite{bookV2X}. 
The standard V2V communication protocol is the so-called dedicated short-range communication  (DSRC)  transmission service, which provides a nominal coverage range of about 1 km, with achievable data rates in the order of 2-6 Mbps \cite{DSRC}. V2I communication, instead, exploits the 4G-LTE connectivity below 6 GHz, enabling a data rate of up to 100 Mbps in high mobility scenarios \cite{3GPP_LTE}.  
However, the next generation of automotive systems will include advanced services based on sophisticated sensors to support enhanced automated driving applications and is expected to require very high data rates (in the order of terabytes per driving hour) that cannot be provided by current V2X technologies~\cite{Heath_surveyV2X}.
A possible answer to this growing demand for ultra-high transmission speeds can be found in next-generation radio technologies  and interfaces, such as the millimeter-wave (mmWave) bands between 10 and 300 GHz \cite{rappaportmillimeter}.\footnote{Although strictly speaking mmWave bands include frequencies between
30 and 300 GHz, the industry has loosely defined it to include any frequency
above 10 GHz.}
Besides the extremely large bandwidths available at such frequencies,
the small size of antennas
at mmWaves makes it possible to build complex
antenna arrays  and  obtain high
 gains by beamforming (BF), thus further increasing the transmission rates.
In addition, the inherent security of communication is also
improved because of the
relatively narrow beamwidth that can be achieved \cite{magazine_IA}.
However,  there are many concerns about the transmission characteristics at these frequencies. 
The path loss is indeed very large and the communication range is quite limited. Raindrops are roughly the same size as the radio wavelengths and cause scattering of the radio signal \cite{keyelement_mmWave}. Moreover, mmWave signals do not pass through most solid materials, and movements of obstacles and reflectors cause the channel to rapidly appear and disappear \cite{Yamamoto_08}. 
Additionally,  mmWave links are typically directional and require precise alignment of the transmitter and receiver beams to maintain
connectivity, an operation that resembles handover in cellular
systems \cite{JSAC_2017}.
Those  limitations pose new challenges
for the design of vehicular protocols
and exemplify how  the
connectivity performance of the automotive nodes operating at mmWaves is heavily influenced by the specific features of the environment in which the vehicles are deployed.

\subsection{Related Work}


Given the simplicity of their topology and their high level of automation, highway scenarios have been heavily investigated in the literature for evaluating the connectivity performance of moving nodes in vehicular networks \cite{TCP_DSRC,Highway_DSRC,throughput_infrastructure_chen}.
In particular, \cite{TCP_DSRC} analyzes the performance of multi-hop  transport protocols in a multi-lane highway environment, with particular emphasis on the effect (in terms of 
throughput and latency) of tuning the transmission power.
In \cite{Highway_DSRC}, the authors conducted a realistic analysis of the vehicular ad hoc network topology by integrating realistic microscopic mobility traces and real database traffic demand with realistic
channel models, taking into account the effect of vehicles on
the received signal power.
The article in \cite{throughput_infrastructure_chen} provides  a closed form expression of the achievable throughput of infrastructure-based vehicular networks under a cooperative communication
strategy,  exploring the combined use of V2I and V2V
communications to facilitate the data transmission.
However, such analyses strictly deal with DSRC systems operating at 5.8~GHz, whose propagation characteristics are completely different from those 
of mmWave channels.
Furthermore, in conventional vehicular systems, transmissions are mostly omnidirectional (though beamforming or other directional transmissions
can be performed after a physical link between the nodes has
been established). 
These solutions are therefore unsuitable for a mmWave scenario, which requires highly directional transmission schemes instead.

The potential of the mmWave technology as a means to enable future Intelligent Transportation System (ITS) communications has been first  acknowledged  in  \cite{Heath_V2X_magazine}, which makes the case that the mmWave band is the only viable approach to handle the massive data rates that can be generated in next-generation vehicles.
A non-exhaustive list of relevant works regarding  V2X communication systems operating at mmWaves includes articles \cite{Heath_automotive_radar, Heath_high_speed_train, Heath_surveyV2X, MOCAST_2017}. 
However, the presented results were not analytically investigated nor validated, and suffer from scalability issues.

In this context, stochastic geometry has emerged as a tractable approach to model and analyze the performance of wireless systems via spatial processes, such as the Poisson Point Process~(PPP)~\cite{Bacelli_book, Andrews_tractable_approach_stochastic}.
 In \cite{stochastic_highway_DSRC, stochastic_highway_DSRC_tong,stochastic_highway_DSRC_Farooq}, the authors exploit stochastic geometry and queuing theory to develop
tractable and accurate modeling frameworks to characterize
and analyze the performance of traditional vehicular networks in a multi-lane
highway setup. 
However, it is not possible to directly apply those results to mmWave automotive scenarios due to the specific features of this type of communication 
In this respect, several literature works, including \cite{Coverage_rate_analysis, on_the_feasibility, rebato17}, provide general  schemes to stochastically evaluate the coverage and rate performance in mmWave 2-D cellular networks.
However, it is not easy to translate such studies into the context of mmWave systems for automotive  scenarios, due to the more challenging propagation characteristics of highly mobile vehicular nodes (VNs).
Finally, to the best of our knowledge, paper \cite{Tassi17_Highway} is the only available contribution that models a highway communication network operating at mmWave frequencies and characterizes its fundamental metrics. 
However,  it does not consider some important automotive-specific features, e.g., the vehicles speed or the beam alignment probability, and adopts an approximated path loss model in which the Line-of-Sight (LOS) and Non-Line-of-Sight (NLOS) probabilities are independent of the distance and distribution of the nodes. 
Furthermore, it does not investigate the connectivity performance of the vehicular nodes when modeling a dynamic~environment.

\subsection{Contributions}

The above discussion makes it apparent that next-generation  mmWave
automotive networks should support a mechanism by
which the vehicles and the infrastructure can quickly determine the best directions to establish the mmWave link, an operation which may increase the latency and the overhead of the communication and have a substantial impact on the connectivity of  vehicular nodes.
With this in mind, as an extension of our work \cite{MOCAST_2017}, in this paper we provide the first analytical model to evaluate the  coverage,  connectivity and  throughput performance of a dynamic V2X network operating at mmWaves.
We therefore consider a typical unidimensional multi-lane highway setup based on a V2I communication scenario, in which cars exchange data with mmWave Base Stations (BSs) deployed on both sides of the road.
The original contributions of this paper can be synthesized  as follows:
\begin{itemize}
\item We develop a novel tractable framework  based on
stochastic geometry to evaluate both the coverage and the connectivity performance of an automotive node in a dynamic mmWave vehicular environment, based on a realistic measurement-based distance-dependent path loss model.
In particular, this is the first contribution in which an analytical expression for the beam alignment probability and connection stability (i.e.,  the probability that the vehicle does not disconnect from its serving infrastructure over time) is evaluated considering a dynamic  scenario.
\item  We prove that  the performance of the automotive nodes in highly mobile mmWave systems strictly depends on the specific environment in which the vehicles are deployed, i.e., on the nodes speed, the  beam alignment periodicity, the base stations density and the antenna geometry.
\item We show that an optimal value of  throughput can be associated with a density threshold above which the deployment of more BSs results in a considerable increase of the system complexity while leading to worse communication performance. 
\item We evaluate and compare the connectivity capabilities of the V2X network adopting both a \textit{rural }path loss model, in which the communication between the endpoints is impaired by large vehicles acting as~blockages, and a distance-dependent \textit{urban} path loss implementation, based on real-world measurements, in which environmental obstructions (i.e., urban buildings) can occlude the path between the transceiver.
The results prove that, although the two models are intrinsically remarkably different, they yield comparable results in terms of connectivity~performance.
\end{itemize}
Overall, the purpose is to exemplify some of the complex and interesting tradeoffs to be considered when designing solutions for next-generation automotive scenarios operating at~mmWaves.

The remainder of this paper is organized as follows.
The system
model is described in Sec.~\ref{sec:system_model}.
In Sec.~\ref{sec:analysis} we introduce the association rule for the vehicular nodes and we  derive the expressions of the coverage and connectivity probabilities and the achievable throughput in a general  scenario.
In Sec.~\ref{sec:results}, we  validate our theoretical framework through simulations and we present our main findings and results. 
Finally, conclusions and suggestions for future work are provided in Sec.~\ref{sec:concl}.


\section{System Model}
\label{sec:system_model}
In this section we present the system model for evaluating the coverage and connectivity  performance of a mmWave vehicular network.
The notation and the system parameters that will be used throughout this manuscript are summarized in Tab. \ref{tab:notation}, while their values will be detailed in Sec. \ref{sec:sim_params}.

\begin{table}[t!]
\renewcommand{\arraystretch}{0.8}
\centering
\caption{Notation and meaning of the main system parameters.}
\label{tab:notation}
\begin{adjustbox}{max width=\columnwidth}
\centering
\begin{tabular}{c|P{22cm}  |P{0.1cm} | P{0.1cm}}
\hline
\textbf{                  Parameter} &  \multicolumn{3}{c}{\textbf{Meaning}}\\ \hline \hline
                  $\Phi_b$, $\Phi_o$, $\Phi_{L}$, $\Phi_{N}$ &  \multicolumn{3}{c}{PPP of BSs, obstacles, LOS BSs, NLOS BSs}\\ \hline
$\lambda_{L}$, $\lambda_{N}$ & \multicolumn{3}{c}{Density of LOS and NLOS BSs}  \\ \hline
$p_L$, $p_N$ & \multicolumn{3}{c}{Probability of a BS being in LOS (or NLOS) w.r.t. the test VN} \\ \hline
$\ell_i(r)$ & \multicolumn{3}{c}{Path loss component of BS $\in \Phi_i$, for $i\in\{L,\,N\}$, at distance $r$ from the VN}\\ \hline
$\Delta_1=G_b\cdot G_{\rm VN}$ & \multicolumn{3}{c}{Overall antenna gain (assuming perfect beam alignment) }\\\hline 
$\Delta_{I_j}$ & \multicolumn{3}{c}{Antenna gain between the test VN and interfering BS $j$ }\\\hline 
$|h_i|^2 \sim \text{Exp}(1/\mu)$ & \multicolumn{3}{c}{Small scale fading component of the $i$-th BS }\\\hline  
\rule{0pt}{0.4cm}$\bar{f}_i(r)$ & \multicolumn{3}{c}{ PDF of the distance $r$ from the test VN to the serving  BS  $\in \Phi_i$, for $i\in\{L,\,N\}$}\\\hline  
$P_i(r)$ &\multicolumn{3}{c}{ Probability that the test VN connects to a  BS  $\in \Phi_i$, for $i\in\{L,\,N\}$ }\\\hline  
$P_{ cov}$   & \multicolumn{3}{c}{SINR Coverage probability} \\\hline  
$P_{ C}$   & \multicolumn{3}{c}{Connectivity probability} \\\hline
$B$   & \multicolumn{3}{c}{Achievable throughput within one slot} \\\hline	
\end{tabular}
\vspace{0.6cm}
\end{adjustbox}
\end{table} 

\begin{table}[t!]
\renewcommand{\arraystretch}{0.8}
\centering
\caption{System parameters. Their values will be detailed in Sec. \ref{sec:sim_params}.}
\begin{adjustbox}{max width=\columnwidth}
\begin{tabular}{c|P{6cm}  |P{3.5cm} | P{3.5cm}}
\hline
\multirow{2}{*}{\textbf{Parameter}}  & \multirow{2}{*}{\textbf{Meaning}} & \multicolumn{2}{c}{\textbf{Value}} \\ \cline{3-4} 
                  &          & \textbf{Rural Model $\sim$ \cite{Tassi17_Highway}}         & \textbf{HIghway Model $\sim$ \cite{Mustafa}}\\ \hline \hline
             $a_{\rm LOS}$ &  LOS parameter & Not applicable & $0.0149$ m$^{-1}$ \\ \hline
$\alpha_{L}$ & LOS path loss exponent & $2.8$ & $2$  \\ \hline
$\alpha_{N}$ & NLOS path loss exponent & $4$ & $2.92$  \\ \hline
\rule{0pt}{0.4cm}  $C_L$ &\begin{tabular} {@{}c @{}}LOS path loss gain  (unit distance)\end{tabular} & $10^{-6.1}$ dB& $10^{-7.2}$ dB\\ \hline
\rule{0pt}{0.4cm}$C_N$ & \begin{tabular} {@{}c @{}}NLOS path loss gain  (unit distance)\end{tabular} & $10^{-6.1}$ dB & $10^{-6.14}$ dB\\ \hline
$\lambda_o$ & Density of obstacles& $20$ obs/km & Not applicable  \\ \hline
$\tau_o$ & Length of obstacle &$11.1$ m & Not applicable     \\ \hline
$\lambda_b$ & Density of BSs& \multicolumn{2}{c}{$\{0\dots45\}$ BS/km}  \\ \hline
$W_{\rm tot}$                    & Total bandwidth&  \multicolumn{2}{c}{$1$ GHz}     \\ \hline
$f_{C}$                    &  Carrier frequency & \multicolumn{2}{c}{$28$ GHz}     \\ \hline
$P_{TX}$& Downlink transmission power & \multicolumn{2}{c}{$27$ dBm}     \\ \hline
\rule{0pt}{0.4cm}$\sigma^2$ &  Thermal noise & \multicolumn{2}{c}{$10\log_{10}(1.381\cdot10^{-23} \cdot 290 \cdot W_{\rm tot}\cdot 10^3)$ dBm} \\\hline
$\mu$& Rayleigh channel parameter & \multicolumn{2}{c}{1}     \\ \hline
$w$ & Road lane width & \multicolumn{2}{c}{$3.7$ m}     \\ \hline
$N_l$ & Number of lanes & \multicolumn{2}{c}{$4$}     \\ \hline
$\Gamma$ & SINR threshold & \multicolumn{2}{c}{$-5$ dB}     \\ \hline
$2W, \, L$ & Total road width and length &  \multicolumn{2}{c}{$2W = 14.8$ m, $L=50$ km} \\ \hline
$T_S$ & Slot duration & \multicolumn{2}{c}{$\{0.1,0.3,0.5,1\}$ s.} \\\hline
$V$ & Vehicle Speed & \multicolumn{2}{c}{$\{30,60,90,100,130\}$ km/h} \\\hline
$\phi$ & Beamwidth of  VN main  lobe & \multicolumn{2}{c}{$60^\circ$} \\\hline
$\psi$ & Beamwidth of  BS main  lobe & \multicolumn{2}{c}{$\{30^\circ,60^\circ,90^\circ\}$} \\\hline
$G_{b_i}$ & BF gain of  main lobe of the $i$-th BS  & \multicolumn{2}{c}{$\{20, 12, 6\}$ dB} \\\hline
$G_{\rm VN}$& BF gain of  main lobe of the VN  & \multicolumn{2}{c}{$12$ dB} \\\hline
$g_{b_i}$, $g_{\rm VN}$ &  BF gain of side lobes  of the $i$-th BS  and VN & \multicolumn{2}{c}{$-10$ dB} \\\hline
$N_b \times M_b$ & Number of antenna elements of the BS  UPA & \multicolumn{2}{c}{$\{4,16,64\}$} \\\hline
$N_{\rm VN} \times M_{\rm VN}$ & Number of antenna elements of the VN UPA& \multicolumn{2}{c}{$16$} \\\hline
\end{tabular}
\end{adjustbox}
\end{table} 

\subsection{Network Model}
\label{sec:network_model}

\begin{figure}[t!]
\centering
\vspace*{-0.3cm}
 \includegraphics[trim= 0cm 0cm 0cm 0cm , clip=true, width= 0.95\textwidth]{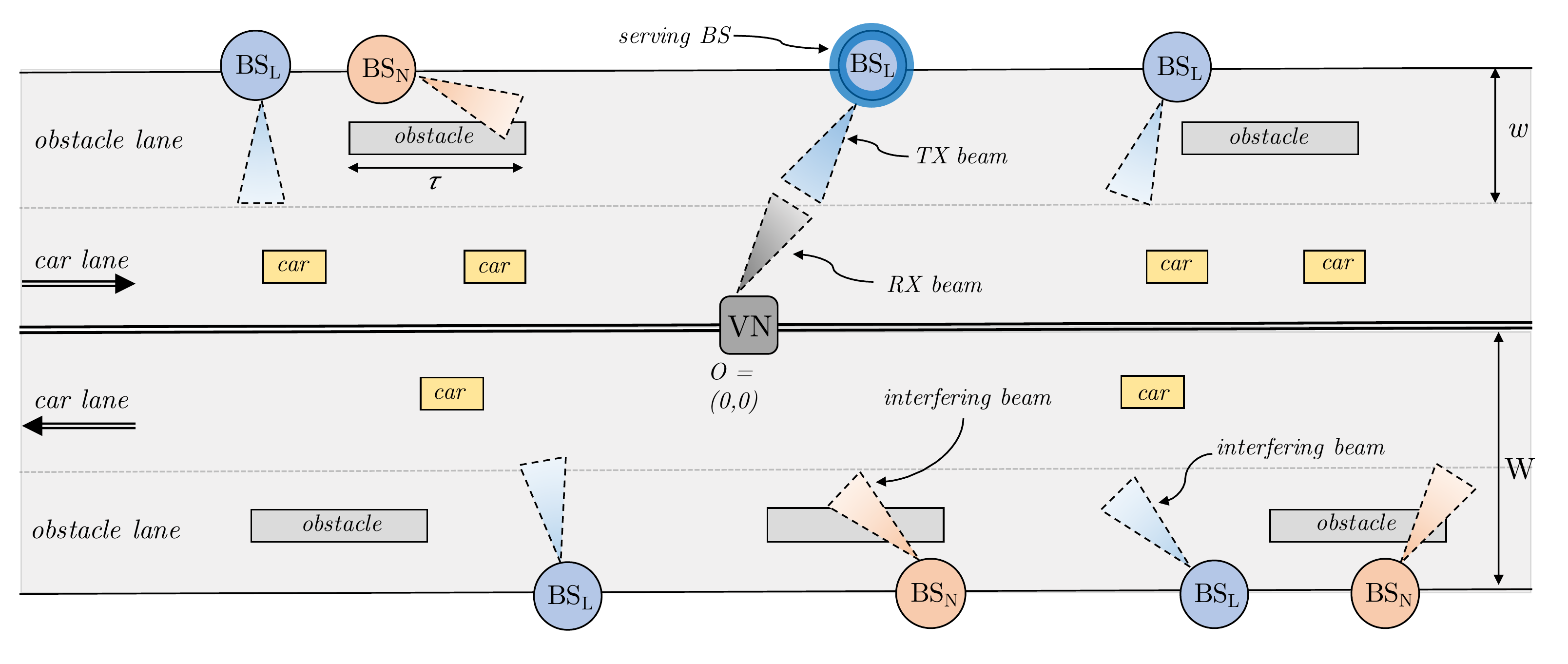}
 \caption{Illustration of the considered highway system model, composed of $N_l=4$ lanes of width $w$, with one car lane and one obstacle lane in each traffic direction. The target VN is placed at the center of the scenario, while LOS (NLOS) BSs follow a PPP $\Phi_{L}$ ($\Phi_{N}$) of density $\lambda_{L}$ ($\lambda_{N}$). }
 \label{fig:highway_section}
\end{figure}

The network model consists of VNs and BSs which are deployed over a section of a highway, as depicted in Fig. \ref{fig:highway_section}.
We  consider two possible scenarios: 

\begin{itemize}
 \item  A \emph{rural section of highway}, denoted with superscript $(R)$ throughout the paper, in which the communication can be blocked by large vehicles. The communication follows the model presented in \cite{Tassi17_Highway}.
\item A \emph{urban section of highway }, denoted with superscript $(U)$ throughout the paper, in which environmental obstructions (i.e., urban buildings) can occlude the path (via either reflections or scattering) from the transmitter to the receiver.  The communication  follows the model presented in \cite{Mustafa}.
\end{itemize}

 We assume that the highway is composed of $N_l=4$ infinitely long parallel lanes of width $w$, making $2W=N_l\cdot w$ be the total road width. 
 In each driving direction, a \emph{car lane} (closer to the innermost part of the road) is used by fast vehicles (e.g., cars, motorcycles) whose data traffic performance and behavior will be analyzed in this work. An \emph{obstacle lane} (closer to the outermost part of the road) is instead reserved for large and slow vehicles (i.e., trucks, buses), which are therefore assumed to act as communication blockages (if they obstruct the LOS of the vehicles in the car~lane).
Moreover, without loss of generality, we assume that the target VN is deployed at the origin of a coordinate system centered on the point $O=(0,0)$,   on the mid line of the road.

We assume that the BSs form a one-dimensional homogeneous PPP $\Phi_b$  with density $\lambda_b$. 
Note that the BSs can be located along  either  the upper or the lower side of the road with equal probability.

Furthermore, the  vehicles in the obstacle lane form a one-dimensional homogeneous PPP $\Phi_o$ with density $\lambda_o$. 
Each blockage node potentially obstructing the  LOS of the vehicles in the car lane is associated with a segment of length $\tau_o$, while obstacles widths and heights are not part of our model.
When considering an urban scenario, we assume that blockages (typically buildings in urban areas) form a process of random shapes on a plane. The distribution of the blockage process is assumed to be stationary and isotropic (i.e., invariant to the motions of translation and rotation)~\cite{Coverage_rate_analysis}.

\subsection{Path Loss Model}
\label{sec:PL_model}

The path loss characterization follows either the model presented in \cite{Tassi17_Highway} (if a rural environment is considered) or that in~\cite{Mustafa} (if an urban scenario is considered). 
Since vehicles in the obstacle lane or urban buildings can block the link connecting the test VN to its serving BS, it is necessary to distinguish between LOS and NLOS BSs, respectively denoted with subscripts $L$ and $N$ throughout the paper.\footnote{For the tractability of the analysis, in this work we neglect the outage condition induced by severe attenuation and incorporated~in~\cite{Mustafa}.}

\begin{itemize}
\item \emph{Rural path loss model.}
 When considering a rural highway scenario, BS $n$ is assumed to be in LOS (with probability $p_{n,L}^{(R)}$) if the ideal segment connecting the target VN and BS $n$ does not intersect with any of the segments of length $\tau_o$ describing the footprint of the vehicles in the obstacle lane. According to \cite{Tassi17_Highway}, this probability is independent of the distance from BS $n$ to $O$ and can be approximated  as
$
p_L^{(R)} \simeq e^{-\lambda_o \tau_o},
$
for any $n$.
Accordingly, a BS will be in NLOS with probability $p_N^{(R)} = 1-p_L^{(R)}$.

\item \emph{Distance-dependent urban path loss model.}
When considering an urban highway scenario, BS $n$ is assumed to be in LOS (with probability $p_{n,L}^{(U)}(r)$) if the ideal segment connecting the test VN and BS $n$ (at distance $r$)  does not intersect with any building. 
According to \cite{Mustafa}, $p_{n,L}^{(U)}(r)$ is independent among different links and is a non-increasing function of $r$: the longer the link, the more likely to intersect with one or more blockages. When obstacles are modeled as random shapes, it follows~that
$
p_L^{(U)}(r) = e^{-a_{\rm LOS} r},
$
where the parameter $a_{\rm LOS} = 0.0149$ m$^{-1}$ is derived from the measurement campaign conducted in \cite{Mustafa}. Again, a BS is in NLOS with probability $p_N^{(U)}(r) = 1-p_L^{(U)}(r)$.
\end{itemize}
By the thinning theorem of PPP \cite{Bacelli_book}, the PPPs of the LOS BSs $\Phi_{L}\subseteq \Phi_{b}$ and the NLOS BSs $\Phi_{N}\subseteq \Phi_{b}$ are independent and have density $\lambda_{L}^{(s)} = p_L^{(s)} \lambda_b$ and $\lambda_{N}^{(s)} = p_N^{(s)} \lambda_b$, respectively, with $s\in \{{R, U}\}$.
Therefore, the path loss component $\ell_i(r)$ affecting the propagation of the test VN, at distance $r$ from a BS $\in \Phi_i$, for $i\in \{L,N\}$,  is defined~as
\beq
\ell_i(r) =  C_i r^{-\alpha_i} ,
\label{eq:PL_component}
\eeq
where  $\alpha_i$ is the path loss exponent and $C_i$ is the path loss gain at unit distance. 

\subsection{Antenna Model and Beam Tracking}
\label{sec:antenna_model}

As mentioned, isotropic transmission at mmWave frequencies incurs severe path loss. 
To overcome this problem, antenna arrays are typically deployed at both the BSs and the VNs, to perform directional beamforming and benefit from the resulting antenna gain~\cite{MOCAST_2017}.

 For the tractability of the analysis, following the model proposed in prior literature work (e.g.,  \cite{Coverage_rate_analysis, coverage_het}), the actual antenna array patterns are approximated by a sectored antenna model. 
 We therefore assume that BSs are equipped with a Uniform Planar Array (UPA) of $N_b \times M_b$ elements, allowing to steer beams consisting of a main lobe with beamwidth $\psi$ and a side lobe that covers the remainder of the antenna radiation pattern. Similarly, VNs are equipped with a UPA of $N_{\rm VN}\times M_{\rm VN}$ elements,  allowing to steer beams consisting of a main lobe with beamwidth $\phi$ and a side lobe that covers the remainder of the antenna radiation~pattern.

Moreover, we let $G_{b_n}$ and $g_{b_n}$ be the main lobe directivity gain (assumed constant for all angles in the main lobe) and the side lobe gain of the $n$-th BS antenna, respectively.
Similarly, we let $G_{\rm VN}$ and $g_{\rm VN}$ be the main and side lobe gains of the VN antenna.
Then, we define $\Delta_1 = G_b \cdot G_{\rm VN}$ as  the overall antenna gain in case of perfect beam alignment between the test VN and its serving BS.

Also, the beam direction of an interfering link is modeled as a uniform random variable in $[0, \, 2\pi]$. Therefore, the effective interference antenna gain between an interfering BS $j$ and the test VN is a discrete random variable $\Delta_{I_j}$ described as
\beq
\Delta_{I_j}=\begin{cases} G_{b_j} \cdot G_{\rm VN}, & \mbox{with probability }\mbox{ $\theta_b/\pi$} \\ g_{b_j}\cdot g_{\rm VN}, & \mbox{with probability }\mbox{ $1-(\theta_b/\pi)$},
\end{cases}
\eeq
where  $\theta_b$ is defined as the half beamwidth of the aggregate antenna radiation pattern.\\

As far as beam tracking is concerned, according to the procedure described in \cite{MedHoc2016,TWC2017,JSAC_2017}, we assume that measurement reports are periodically exchanged
among the nodes so that, at the beginning of every slot of
duration $T_S$, VNs and BSs identify the optimal directions for
their respective beams. Such configuration is kept fixed for
the whole slot duration, during which nodes may lose the
alignment due to  mobility. 
In case the connectivity
is lost during a slot, it can only be recovered at the beginning
of the subsequent slot, when the beam tracking procedure
is performed again \cite{MOCAST_2017}. 
We also assume that, if the  VN connects to BS $n$, at the beginning of the slot the main lobes of the BS's and VN's transmit beams are perfectly aligned \cite{rebato17}. 
This guarantees that, at every slot, the maximum gain $\Delta_1$ is experienced between the VN and its serving~BS.

\subsection{Channel Model}
\label{sec:channel}

Available measurements at mmWaves in the V2X
context are  very limited and realistic scenarios are
indeed hard to simulate. In fact, the reflectivity
and scattering from common objects and the poor diffraction
and penetration capabilities of mmWaves are the main factors
preventing the reuse of existing lower frequency channel models for an automotive mmWave scenario. Moreover,
current models for mmWave cellular systems (e.g., \cite{Mustafa})
present many limitations for their applicability to a V2X context, due to the more challenging propagation characteristics
of highly mobile VNs \cite{MOCAST_2017}. 
It is thus necessary to adopt conservative assumptions on signal~propagation.

 The channel between the test VN and its serving BS is described as a Rayleigh channel model\footnote{It has been observed in previous works (e.g., \cite{on_the_feasibility,Park_channelModel}) that considering a general fading model such as Nakagami-m may not provide significant design insights and performance improvements compared to Rayleigh fading, while complicating the analysis significantly. Therefore, as a first step, in this paper we consider only Rayleigh fading, and leave extensions to more general channel models as future work.}
 with mean $\mu$, i.e., $|h_1|^2 \sim \text{Exp}(1/\mu)$~\cite{Andrews_tractable_approach_stochastic}.
Similarly, to capture the clustering of interfering communications, the channels between the interfering BSs and the test VN are modeled as independent and identically distributed (i.i.d.) exponential  random variables with mean $\mu$ \cite{Tassi17_Highway}.

We define the Signal-to-Interference-plus-Noise Ratio (SINR) measured between the test VN, attached to a BS $\in \Phi_i$, for $i\in\{L,\,N\}$, at distance $r_1$, as
\beq
\text{SINR}_i = \frac{ |h_1|^2\Delta_1\ell_i{(r_1)}}{( I_L + I_N ) + \sigma^2} =\frac{ |h_1|^2\Delta_1\ell_i{(r_1)}}{\left(\sum\limits_{k\in \Phi_L} |h_k|^2 \Delta_{I_k}\ell_L(r_k) + \sum\limits_{k\in \Phi_N} |h_k|^2 \Delta_{I_k}\ell_N(r_k)\right) + \sigma^2},
\label{eq:SINR}
\eeq
where $|h_n|^2$ and $\Delta_n$  are the small scale fading components and the overall antenna gains measured between  BS $n$ and the test VN (at distance $r_n$), respectively, while $\ell_i(r_n)$ is the path loss component affecting the propagation between   BS $n\in \Phi_i$ and the VN, as given in \eqref{eq:PL_component}.  $I_L$ and $I_N$ represent the interference produced to the test VN by BSs in $\Phi_L$ and $ \Phi_N$, respectively. Finally, $\sigma^2$ represents the thermal noise power, normalized with respect to the transmission power~$P_{\rm TX}$, which is assumed equal for all nodes.

\section{Coverage and Connectivity Analysis}
\label{sec:analysis}
In this section, we analyze the coverage  and the  connectivity of a moving VN in the proposed scenario.
 The purpose is to exemplify some of the complex and interesting tradeoffs that have to be considered when designing solutions for mmWave automotive scenarios.
First, we present the  association rule for the VN and we derive the expression of the \emph{probability density function} (PDF) of the distance $r$ between the VN and its serving BS  (LOS or NLOS). 
Second, we derive the expressions for the SINR coverage probability and  the probability of the moving VN to stay in the communication range of its serving BS during one slot and, consequently, to keep connectivity.
Finally,  we analytically determine an expression for the average achievable throughput as a function of  the vehicle speed $V$, the slot duration (or beam tracking periodicity) $T_S$ and the beamwidth $\psi$.

\subsection{Association Rule}
According to the system model presented in Sec. \ref{sec:system_model}, every \ts  both the VN and the BSs estimate the surrounding channels and then adjust their antenna orientation accordingly, to exploit the maximum beamforming gain. We also assume that the measured channel information is perfect, neglecting any estimation error. Therefore, letting $r_n$ be the distance between the VN and  BS $n$, the  VN always connects to  BS $n^*\in \Phi_i$,  $i\in\{L,N\}$, that provides the maximum average received power (i.e., the minimum path loss):
\beq
n^* = \argmax_{\forall i,\: \forall n} \{\ell_i (r_n)\},
\eeq
where $\ell_i(r_n)$ is as in Eq. \eqref{eq:PL_component}.

\begin{lemma} \label{lemma:f_L} The  probability density function of the distance $r$ between the VN and the nearest LOS (NLOS) BS is
\beq
f_i^{(s)}(r) = \frac{\partial }{ \partial r}\left( 1-\exp \left( -2\lambda_b \int_0^{b(r)} p_i^{(s)}(x) \mathrm{d}x \right)  \right),
\label{eq:f_L}
\eeq
where $r$ is larger than the road width $W$, i.e., $r>W$, by construction, $s\in \{{R, U}\}$ (according to the simulated scenario\footnote{When considering a rural highway scenario, the path loss state (i.e., the probability of LOS/NLOS conditions) does not depend on the distance $r$ between the VN and the BS, therefore the expression in Eq. \eqref{eq:f_L} simplifies to $ f_i^{(R)}(r)=\frac{2\lambda^{(R)}_i r}{b(r)} \exp\Big(-2\lambda^{(R)}_i b(r)\Big)$, $i\in \{{L, N}\}$. Similar simplifications apply for all the results referred to a rural  environment presented throughout the paper.}), $i\in \{{L, N}\}$ (according to the path loss state of the nearest BS)  and $b(r) = \sqrt{r^2-W^2}$.
\end{lemma}

\emph{Proof:} See Appendix \ref{appendix:f_L}.
\hfill $\blacksquare$

%

However, the test VN may not always perform association to the closest BS, especially when considering a very dense urban environment in which the nearest infrastructure may be NLOS. On the contrary, the serving BS can be either the nearest BS in $\Phi_L$ or the nearest one in $\Phi_N$ \cite{Coverage_rate_analysis}. 
Assuming that the test VN connects to a LOS (NLOS) BS, there must be no NLOS (LOS) BSs at distance greater than or equal to $A_N(r)$ ($A_L(r)$), which is defined~as
\beq
A_i(r) = \left( \frac{C_{i^*}}{C_i} r^{\alpha_{i}} \right)^{\frac{1}{\alpha_{i^*}}},
\label{eq:A}
\eeq
where $i^*$ indicates the  LOS/NLOS state other than $i$.

We can therefore compute the probability that the test VN is associated with either a LOS or NLOS BS as follows.
\begin{lemma} \label{lemma:P_L}
The test VN connects to a   BS $\in \Phi_i$, for $i\in\{L,\,N\}$, with probability
\beq
P^{(s)}_i = \int_W^\infty  \exp\left( -2\lambda_b \int_0^{b(A_{i}(r))} p^{(s)}_{i^*}(x) \mathrm{d}x \right) f^{(s)}_i(r) \mathrm{d}r
\label{eq:P_L}
\eeq
where $s\in \{{R, U}\}$ and  $b(A_{i}(r)) = \sqrt{A_i(r)^2-W^2}$.
\end{lemma}

\emph{Proof:} See Appendix \ref{appendix:P_L}.
\hfill $\blacksquare$

\begin{lemma} \label{lemma:f_r}
Given that the test VN connects to a  BS $\in \Phi_i$, for $i\in\{L,\,N\}$, the PDF $\bar{f}^{(s)}_i(r)$, $s\in\{R,\,U\}$, of the distance $r$ between the vehicular node and the  serving BS is given by
\beq
\bar{f}^{(s)}_i(r) =  \exp\left( -2\lambda_b \int_0^{b(A_{i}(r))} p^{(s)}_{i^*}(x) dx \right) f^{(s)}_i(r).
\label{eq:bar_f}
\eeq
\end{lemma}

\emph{Proof:} The proof is based on the same rationale used to prove Lemma \ref{lemma:P_L}, and is therefore omitted here.
\hfill $\blacksquare$

\subsection{SINR Coverage Analysis}
The SINR coverage probability $P^{(s)}_{cov}(\Gamma)$ is defined as the probability that the target VN experiences an SINR larger than a predefined threshold $\Gamma>0$, i.e., $P^{(s)}_{cov}(\Gamma) = \mathbb{P}[\text{SINR}>\Gamma]$. 
By using the law of total probability, we get
\beq
P^{(s)}_{cov}(\Gamma) = \mathbb{P} [\underbrace{ \text{SINR} > \Gamma,\, n^* \in \Phi_{L}}_{\text{SINR}_L > \Gamma}  ]  +\mathbb{P} [\underbrace{  \text{SINR} > \Gamma, \, n^* \in \Phi_{N} }_{\text{SINR}_N > \Gamma}]
\label{eq:def_P_cov}
\eeq
where $n^*$ is the serving BS referred to the target VN.
Based on the lemmas and the assumptions introduced in the previous sections, we present the main theorem on the SINR coverage probability.

\begin{theorem} \label{th:P_cov} The  coverage probability $P^{(s)}_{cov}(\Gamma)$ for a target SINR threshold $\Gamma>0$ is given~by
\beq
P^{(s)}_{cov}(\Gamma) = \sum_{i\in\{L,\,N\}} \int_W^\infty \exp\left( \frac{-\mu \sigma^2 \Gamma r^{\alpha_i}}{\Delta_1 C_i} \right)
\mathcal{L}^{(s)}_{I_i^L}\left(  \frac{\mu \Gamma r^{\alpha_i}}{\Delta_1 C_i} \right) 
\mathcal{L}^{(s)}_{I_i^N}\left(  \frac{\mu \Gamma r^{\alpha_i}}{\Delta_1 C_i} \right) 
\bar{f}^{(s)}_i(r) \mathrm{d}r,
\label{eq:P_cov}
\eeq
where $\mathcal{L}^{(s)}_{I_i^j}(t)$ is the Laplace functional of the interference from BSs $\in \Phi_j$, for $j\in\{L,\,N\}$, to the test VN, and is expressed as
\medmuskip=0mu
\thickmuskip=0mu
 \eq_text
\end{theorem}

\emph{Proof:} See Appendix \ref{appendix:P_cov}.
\hfill $\blacksquare$

\subsection{Connectivity  Analysis}
\begin{figure}[t!]
     \centering
     \vspace*{-0.5cm}
     \includegraphics[trim= 0cm 0cm 0cm 0cm , clip=true, width= 1\textwidth]{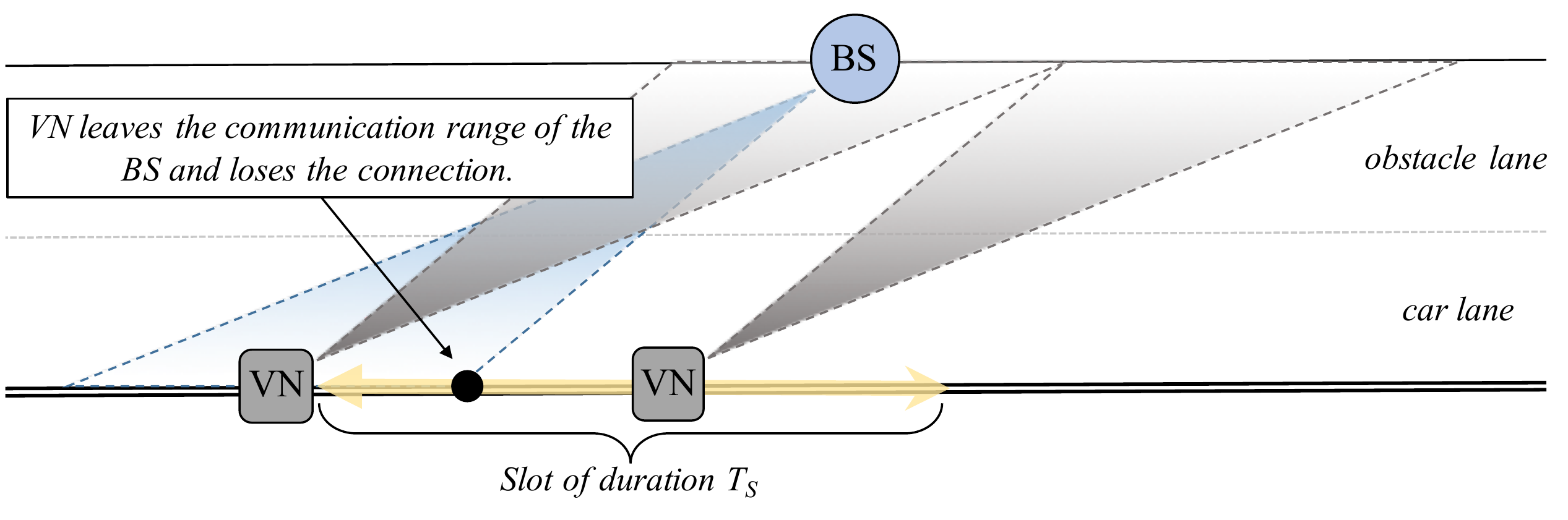}
\caption{At the beginning of the slot of duration $T_S$, the VN is connected and aligned to its serving BS. When moving at constant speed $V$ during the slot, the VN leaves the communication range of the BS. Since the beam direction cannot
be updated during the slot, the link between the VN and the BS will be lost
until the beginning of the next slot.}
\label{fig:beam_alignment}
\end{figure}

As we pointed out in Sec. \ref{sec:antenna_model}, a directional beam pair
needs to be determined to enable the transmission between two nodes, thus beam tracking heavily affects the connectivity
performance of a V2X mmWave system.
Assuming that perfect beam alignment is obtained at the beginning of every slot of duration $T_S$, the vehicle can be  in either a connected (C) or an idle (I) state, depending on whether the  endpoints experience an SINR larger than a predefined threshold $\Gamma$.
Starting from state C, the  VN can either
maintain connectivity to the serving BS for the whole slot
duration, or lose the beam alignment and get disconnected. This second event is illustrated in Fig. \ref{fig:beam_alignment}.
Starting from state I,
instead, the vehicle can either remain out-of-range for the whole
slot, or enter the coverage range of a new BS
within \ts (\emph{catch-up}). Even in this second case, however, the
connection will be established only at the beginning
of the following slot, when the beam alignment procedure
will be performed.
Therefore, preservation of the connectivity
during a slot requires that the VN is within the coverage range
of the BS at the beginning of the slot, with sufficient signal quality, and does not lose beam
alignment in the slot period $T_S$.

In this subsection, we analytically derive the expression of the probability $P_{C}$ of the moving VN not to disconnect from its serving infrastructure during a slot of duration $T_S$.
\begin{theorem} \label{th:P_C} 
For $s\in\{R,U\}$, the probability that the VN is in the connected state C for the whole duration of a slot is given by
\beq
P^{(s)}_{C} = P^{(s)}_{cov}(\Gamma) \cdot P^{(s)}_{NL},
\label{eq:P_C}
\eeq 
where $P^{(s)}_{cov}(\Gamma)$ is the probability of being in state C at the beginning of the slot and $P^{(s)}_{\rm NL} = \mathbb{P}[T_L > T_S]$ represents the  probability that the VN does not leave the communication range of the serving BS during the slot.\footnote{Notice that, according to \eqref{eq:P_C}, $P^{(s)}_{\rm NL}$ is to be interpreted as a \emph{conditional} probability, i.e., the probability that the VN remains connected during a slot \emph{given} that it is connected at the beginning (otherwise the endpoints would not be able to determine the optimal directions for their beams and no communication would be possible).}
This can be expressed as a function of $r$ as
\beq
P^{(s)}_{\rm NL}  = \mathbb{P}[T_L > T_S] = \mathbb{P}\left[ r > \frac{VT_S}{\sin(\psi/2)} \left( \frac{W}{r}\sin(\eta) + \sqrt{1-\left(\frac{W}{r}\right)^2}\cos(\eta)\right) \right],
\label{eq:P_T_L}
\eeq
where $\eta=\pi/2 - \psi/2$  and $\psi$ is the beamwidth of the main lobe of the BS. 
\end{theorem}

\emph{Proof:} See Appendix \ref{appendix:P_C}.
\hfill $\blacksquare$

The last expression  can be easily solved via numerical computation by determining the value~$r^*$ for which the inequality in \eqref{eq:P_T_L} is satisfied as an equality.
Considering that the right-hand side of the inequality in \eqref{eq:P_T_L} is monotonically decreasing in $r$, we hence have
\beq
P^{(s)}_{\rm NL} =\mathbb{P}[T_L > T_S] =  \sum_{i\in\{L,\,N\}} \int_{r^*}^\infty   \bar{f}^{(s)}_i(r) \mathrm{d}r,
\eeq
where $ \bar{f}^{(s)}_i(r) $ is as in Eq. \eqref{eq:bar_f}.

\subsection{Throughput Analysis}
\label{sec:throughput_analysis}

In this subsection, we determine the  expression for the average  throughput $B$ experienced by the  target VN moving across the considered mmWave vehicular scenario.
In particular, let $\mathbb{E}[T_{\rm comm}]\in \{0,T_S\}$ represent the average time (i.e., the portion of slot) during which the VN is  within the coverage range of its serving infrastructure and properly aligned with it.
In this case, the nodes are able to exchange data, on average, at a rate $\mathbb{E}[\gamma(r)]$ that depends on their distance~$r$.
The average achievable throughput during one slot is therefore defined as  
\beq
B^{(s)}(r) = \mathbb{E}[\gamma(r)]\cdot \frac{\mathbb{E}[T_{\rm comm}]}{T_S}, \quad s\in\{R,U\}.
\label{eq:B}
\eeq
Notice that $B(r) = 0$ if the VN is disconnected at the beginning of the slot, while $B(r) = \mathbb{E}[\gamma(r)]$ if the VN is in the connected state for the whole duration of a slot (i.e., $\mathbb{E}[T_{\rm comm}]=T_S$).
The average rate $\mathbb{E}[\gamma(r)]$ in Eq.~\eqref{eq:B} can be computed using  Lemma \ref{lemma:E_R}, while the average communication duration $\mathbb{E}[T_{\rm comm}] $ is evaluated through Theorem \ref{th:B}.
\begin{lemma}
Given the SINR coverage probability $P_{cov}(\Gamma)$, the average achievable rate experienced by the target VN, at distance $r$ from its serving BS, is given by the following expression:
\begin{equation}
\mathbb{E}[\gamma(r)] = W_{\rm tot}\mathbb{E}\{\log_2(1+\text{SINR})\} = \frac{W_{\rm tot}}{\log(2)} \int_0^\infty\mathbb{P}[\text{SINR} > e^t-1]\mathrm{d}t = \frac{W_{\rm tot}}{\log(2)} \int_0^\infty P_{cov}( e^t-1)\mathrm{d}t.
\end{equation}
\label{lemma:E_R}
\end{lemma}
\emph{Proof:} 
See \cite[Theorem 3]{Andrews_tractable_approach_stochastic} and \cite[Section V]{akoum2012coverage}.
\hfill $\blacksquare$

\begin{theorem}
\label{th:B}
Being $d(r)$  the maximum distance the VN can cover before leaving the communication range of its serving BS and being $VT_S$ the total distance covered by the VN, moving at speed $V$, within the slot of duration $T_S$, the  average time (i.e., the portion of slot) in which the VN experiences a non-zero throughput is expressed as:
\begin{align}
\mathbb{E}[T_{\rm comm}] &= \left(1-P^{}_{\rm NL}\right)\cdot\mathbb{E}[T_{\rm comm,L}] + P^{}_{\rm NL}\cdot\mathbb{E}[T_{\rm comm,NL}] \notag \\
&=\left(1-P^{}_{\rm NL}\right)\cdot\left(\frac{1}{V}\cdot \frac{\int_0^{VT_S}F_d(VT_S)-F_d(u)\mathrm{d}u}{F_d(VT_S)}\right)+ P^{}_{\rm NL}\cdot{T_S},
\end{align}
where $f_d(x)$ and $F_d(x)$ represent the PDF and CDF of the distance $d(r)$, respectively, and $P^{}_{\rm NL}$ is as in Eq.~\eqref{eq:P_T_L}.
\label{th:B}
\end{theorem}

\emph{Proof:} 
According to the analysis we developed in Appendix \ref{appendix:P_C}, if the VN does not disconnect during the slot (with probability $P^{(s)}_{\rm NL}$, i.e., with probability $\mathbb{P}[d(r) > VT_S]$) then $\mathbb{E}[T_{\rm comm,NL}] = T_S$. Otherwise, the VN experiences a non-zero throughput only during the portion of the slot in which the alignment with the serving infrastructure is maintained,  and therefore:
\begin{align}
\mathbb{E}[T_{\rm comm,L}] &= \frac{\mathbb{E}\Big[d(r) \mid d(r) < VT_S\Big] }{V} \stackrel{(a)}{=} \frac{1}{V}\cdot\frac{\int_0^{VT_S}xf_d(x)\mathrm{d}x}{\mathbb{P}[d<VT_S]}\notag\\
&=\frac{1}{V}\cdot\frac{\int_0^{VT_S}\mathrm{d}x\int_0^uf_d(x)\mathrm{d}u}{\mathbb{P}[d<VT_S]}\notag\\
&\stackrel{(b)}{=}\frac{1}{V}\cdot\frac{\int_0^{VT_S}\mathrm{d}u\int_0^{VT_S}f_d(x)\mathrm{d}x}{\mathbb{P}[d<VT_S]}\notag\\
&=\frac{1}{V}\cdot \frac{\int_0^{VT_S}F_d(VT_S)-F_d(u)\mathrm{d}u}{F_d(VT_S)},
\end{align}
where step $(a)$ derives from the definition of conditional expectation and from the fact that $d(r)<VT_S$,
while step $(b)$ has been obtained by changing the order of integration for the integrals in $\mathrm{d}x$ and $\mathrm{d}u$.
\hfill $\blacksquare$

\section{Numerical Results}
\label{sec:results}
In this section, after introducing our main
simulation parameters, we provide some numerical results based on the analysis presented in Section \ref{sec:analysis}, with the following objectives.
\begin{itemize}
\item[ (i)]  Comparing the coverage and  connectivity performance of vehicles  considering both~a rural and an urban path loss characterization, following the models of  \cite{Tassi17_Highway}~and~\cite{Mustafa},~respectively.
\item[(ii)] Evaluating the throughput performance of the VNs in a mmWave vehicular network. 
The validity of  the proposed theoretical model will be assessed by comparing the analytical results representing Eq.~\eqref{eq:B} with simulation outcomes.
\item[ (iii)] Providing insights on the impact on  the  performance of V2X nodes in highly mobile mmWave networks of several automotive-specific features, e.g., the vehicles speed, the beam tracking periodicity, the nodes density, the antenna configuration.
\end{itemize}

\subsection{System  Parameters}
\label{sec:sim_params}
The parameters used to derive the results are based on realistic system design considerations
and are summarized in Tab.~\ref{tab:notation}(b).
In particular, we assume that the mmWave network is operated at ${f_C = 28}$ GHz, and the total available bandwidth  is $W_{\rm tot} = 1$ GHz.
In addition, the mmWave channel follows the model presented in Sec. \ref{sec:channel} where the Rayleigh parameter $\mu$ is set to $1$.
The BSs are equipped with an antenna array of $N_b\times M_b = \{4,16,64\}$ elements. 
The resulting main lobe width $\psi$ and BF gain $G_b$ are proportional to the array size, since narrower beams can be steered and larger gains can be achieved when considering larger-scale  arrays\cite{tse_book}. 
Therefore, we assume $\psi\simeq \{90^\circ,60^\circ,30^\circ\}$ and $G_b\simeq \{20,12,6\}$ dB, according to the respective array dimension.
On the other hand, the VN is equipped with $N_{\rm VN} \times M_{\rm VN}  = 16$ elements, steering beams of width $\phi\simeq 60^\circ$ and producing a gain $G_{\rm VN} \simeq 12$ dB. Finally, the side lobe gain of both base stations and vehicular nodes is set to $g_b=g_{\rm VN}=-10$ dB.
The BSs are positioned uniformly at random on both sides of
the road  according to a PPP  with density $\lambda_L$ and $\lambda_N$ for LOS and NLOS BSs, respectively.
The road width is ${2W=14.8}$ m, while the length is ${L=50}$ km. 
As introduced in Sec. \ref{sec:antenna_model}, perfect alignment between the VN, moving at constant speed $V=\{30,\,60,\,90,\,100,\,130\}$ km/h, and its serving infrastructure is guaranteed every $T_S=\{0.1,\,0.3,\,0.5,\,1\}$ s. Conversely, interfering BSs steer their beams through random angle configurations.

\subsection{Results}
\begin{figure}[t!]
     \centering
     		\setlength{\belowcaptionskip}{-0.5cm}
     		             \begin{subfigure}[t!]{0.45\textwidth}
      		\setlength{\belowcaptionskip}{0cm}
	\setlength{\belowcaptionskip}{0cm}
	\setlength\fwidth{0.95\columnwidth}
	\setlength\fheight{0.9\columnwidth}
%
%
\definecolor{mycolor1}{rgb}{0.00000,0.44700,0.74100}%
\definecolor{mycolor2}{rgb}{0.85000,0.32500,0.09800}%
\definecolor{mycolor3}{rgb}{0.92900,0.69400,0.12500}%
\definecolor{mycolor4}{rgb}{0.49400,0.18400,0.55600}%
\definecolor{mycolor5}{rgb}{0.46600,0.67400,0.18800}%
\definecolor{mycolor6}{rgb}{0.30100,0.74500,0.93300}%
\pgfplotsset{
tick label style={font=\footnotesize},
label style={font=\footnotesize},
legend  style={font=\footnotesize}
}

\begin{tikzpicture}

\begin{axis}[%
width=\fwidth,
height=\fheight,
at={(0\fwidth,0\fheight)},
xmin=0,
xmax=45,
xlabel style={font=\color{white!15!black}},
xlabel={$\lambda_b$ [BS/km]},
ymin=0,
ymax=1,
ylabel style={font=\color{white!15!black}},
ylabel={$P^{(U)}_{NL}$},
axis background/.style={fill=white},
label style={font=\footnotesize},
legend columns={1},
legend style={at={(0.97,0.97)},legend cell align=left, align=center, draw=white!15!black},
xmajorgrids,
ymajorgrids]
\addplot [ blue,mark=triangle*, mark options={fill=blue, scale = 1.5}]
  table[row sep=crcr]{%
0.2	0.991070791710024\\
2.82105263157895	0.881219197532395\\
5.4421052631579	0.783638044928052\\
8.06315789473684	0.696944803580098\\
10.6842105263158	0.619914199752603\\
13.3052631578947	0.551460145340653\\
15.9263157894737	0.490619767258278\\
18.5473684210526	0.436539288442064\\
21.1684210526316	0.388461544902708\\
23.7894736842105	0.345714944766223\\
26.4105263157895	0.307703703163784\\
29.0315789473684	0.273899202510937\\
31.6526315789474	0.243832347461742\\
34.2736842105263	0.217086799595136\\
36.8947368421053	0.193292987564122\\
39.5157894736842	0.172122805323296\\
42.1368421052632	0.153284917022534\\
44.7578947368421	0.136520598200944\\
};
\addlegendentry{ $\psi = 30^\circ$}

\addplot   [ green,mark=square*, mark options={fill=green, scale = 1.2}]
  table[row sep=crcr]{%
0.2	0.993427832077114\\
2.82105263157895	0.911193452344912\\
5.4421052631579	0.835778761612487\\
8.06315789473684	0.766617060753205\\
10.6842105263158	0.703188837211299\\
13.3052631578947	0.645017827966112\\
15.9263157894737	0.591667411790497\\
18.5473684210526	0.542737303896425\\
21.1684210526316	0.497860525749934\\
23.7894736842105	0.456700629080459\\
26.4105263157895	0.418949149979946\\
29.0315789473684	0.384323282663042\\
31.6526315789474	0.352563730404251\\
34.2736842105263	0.323432757560435\\
36.8947368421053	0.296712388362041\\
39.5157894736842	0.27220276205695\\
42.1368421052632	0.249720624588128\\
44.7578947368421	0.229097944528977\\
};
\addlegendentry{ $ \psi = 60^\circ$}

\addplot   [ red,mark=diamond*, mark options={fill=red, scale = 1.5}]
  table[row sep=crcr]{%
0.2	0.994719792468107\\
2.82105263157895	0.928044434600351\\
5.4421052631579	0.865838863463869\\
8.06315789473684	0.807803356305565\\
10.6842105263158	0.753658287377147\\
13.3052631578947	0.703142781052078\\
15.9263157894737	0.656013454910739\\
18.5473684210526	0.612043247457377\\
21.1684210526316	0.571020324368821\\
23.7894736842105	0.532747057791932\\
26.4105263157895	0.497039073853088\\
29.0315789473684	0.463724368562667\\
31.6526315789474	0.432642468936403\\
34.2736842105263	0.403643669957004\\
36.8947368421053	0.376588307606021\\
39.5157894736842	0.351346086694258\\
42.1368421052632	0.327795452108856\\
44.7578947368421	0.305823002586628\\
};
\addlegendentry{  $\psi = 90^\circ$}

\end{axis}
\end{tikzpicture}%
    \caption{\footnotesize  $P^{(U)}_{NL}$,  with parameters $T_S=300$ ms, $V=100$ km/h, for different antenna configurations. }
    \label{fig:NYU_P_T_psi}
      \end{subfigure} \qquad \quad 
                   \begin{subfigure}[t!]{0.45\textwidth}
             		\setlength{\belowcaptionskip}{0cm}
	\setlength{\belowcaptionskip}{0cm}
	\setlength\fwidth{0.93\columnwidth}
	\setlength\fheight{0.9\columnwidth}
%
%
\definecolor{mycolor1}{rgb}{0.00000,0.44700,0.74100}%
\definecolor{mycolor2}{rgb}{0.85000,0.32500,0.09800}%
\definecolor{mycolor3}{rgb}{0.92900,0.69400,0.12500}%
\definecolor{mycolor4}{rgb}{0.49400,0.18400,0.55600}%
\definecolor{mycolor5}{rgb}{0.46600,0.67400,0.18800}%
\definecolor{mycolor6}{rgb}{0.30100,0.74500,0.93300}%
\definecolor{mycolor7}{rgb}{0.63500,0.07800,0.18400}%

\pgfplotsset{
tick label style={font=\footnotesize},
label style={font=\footnotesize},
legend  style={font=\footnotesize}
}
\begin{tikzpicture}

\begin{axis}[%
width=\fwidth,
height=\fheight,
at={(0\fwidth,0\fheight)},
xmin=0,
xmax=45,
xlabel style={font=\color{white!15!black}},
xlabel={$\lambda_b$ [BS/km]},
ymin=0,
ymax=1,
ylabel style={font=\color{white!15!black}},
ylabel={$P^{(U)}_{NL}$},
axis background/.style={fill=white},
label style={font=\footnotesize},
legend columns={1},
legend style={at={(0.97,0.97)},legend cell align=left, align=center, draw=white!15!black},
xmajorgrids,
ymajorgrids]
\addplot [only marks, blue,mark=triangle*, mark options={fill=blue, scale = 1.5}]
  table[row sep=crcr]{%
0.2	0.996399610652271\\
2.82105263157895	0.950398482071096\\
5.4421052631579	0.906524745279001\\
8.06315789473684	0.864679867259624\\
10.6842105263158	0.824769884742726\\
13.3052631578947	0.786705191954601\\
15.9263157894737	0.750400338465047\\
18.5473684210526	0.715773836102691\\
21.1684210526316	0.682747975094641\\
23.7894736842105	0.651248648641076\\
26.4105263157895	0.6212051856681\\
29.0315789473684	0.592550191349426\\
31.6526315789474	0.5652193950384\\
34.2736842105263	0.539151505265767\\
36.8947368421053	0.514288071474599\\
39.5157894736842	0.490573352505296\\
42.1368421052632	0.467954189471019\\
44.7578947368421	0.446379888156759\\
};
\addlegendentry{$T_S = 0.1$ s}

\addplot[only marks, green,mark=square*, mark options={fill=green, scale = 1.2}]
  table[row sep=crcr]{%
0.2	0.991070791726652\\
2.82105263157895	0.881219202453336\\
5.4421052631579	0.783638070766736\\
8.06315789473684	0.696944879542231\\
10.6842105263158	0.619914369969346\\
13.3052631578947	0.551460468947589\\
15.9263157894737	0.490620316302098\\
18.5473684210526	0.436540144693359\\
21.1684210526316	0.388462795299132\\
23.7894736842105	0.345716677401388\\
26.4105263157895	0.307706003099507\\
29.0315789473684	0.273902148219578\\
31.6526315789474	0.243836008102274\\
34.2736842105263	0.217091232436392\\
36.8947368421053	0.193298236944609\\
39.5157894736842	0.172128901743083\\
42.1368421052632	0.153291876930767\\
44.7578947368421	0.136528424185606\\
};
\addlegendentry{$T_S = 0.3$ s}

\addplot  [  red,mark=diamond*, mark options={fill=red, scale = 1.5}]
  table[row sep=crcr]{%
0.2	0.98704000688406\\
2.82105263157895	0.832262617356999\\
5.4421052631579	0.702254926229663\\
8.06315789473684	0.592962511688894\\
10.6842105263158	0.501010633890932\\
13.3052631578947	0.423587515653719\\
15.9263157894737	0.358348136399703\\
18.5473684210526	0.303334856534452\\
21.1684210526316	0.256911863665892\\
23.7894736842105	0.217710976006025\\
26.4105263157895	0.184586816124218\\
29.0315789473684	0.156579624832137\\
31.6526315789474	0.132884497329445\\
34.2736842105263	0.11282579752455\\
36.8947368421053	0.0958359198695566\\
39.5157894736842	0.0814376226691809\\
42.1368421052632	0.0692293203998401\\
44.7578947368421	0.0588728266778577\\
};
\addlegendentry{$T_S = 0.5$ s}

\addplot [  orange!100!black,mark=*, mark options={fill=orange!100!black, scale = 1.2}]
  table[row sep=crcr]{%
0.2	0.978312996501961\\
2.82105263157895	0.735754206986591\\
5.4421052631579	0.555748506847818\\
8.06315789473684	0.421530080413596\\
10.6842105263158	0.320986279256394\\
13.3052631578947	0.245327746326148\\
15.9263157894737	0.188147522487088\\
18.5473684210526	0.144753368571537\\
21.1684210526316	0.111692422366562\\
23.7894736842105	0.0864115914189293\\
26.4105263157895	0.0670140025993506\\
29.0315789473684	0.0520836342626072\\
31.6526315789474	0.0405585050574878\\
34.2736842105263	0.0316385659537638\\
36.8947368421053	0.0247184894069049\\
39.5157894736842	0.0193383932250627\\
42.1368421052632	0.0151475392424829\\
44.7578947368421	0.0118774607615241\\
};
\addlegendentry{$T_S = 1$ s}

\end{axis}
\end{tikzpicture}%
    \caption{\footnotesize $P^{(U)}_{NL}$, with parameters $\psi=30^\circ$, $V=100$ km/h, for different values of the slot duration.}
      \end{subfigure}
             \begin{subfigure}[t!]{0.45\textwidth}
             		\setlength{\belowcaptionskip}{0cm}
	\setlength{\belowcaptionskip}{0cm}
	\setlength\fwidth{0.93\columnwidth}
	\setlength\fheight{0.9\columnwidth}
%
%
\definecolor{mycolor1}{rgb}{0.00000,0.44700,0.74100}%
\definecolor{mycolor2}{rgb}{0.85000,0.32500,0.09800}%
\definecolor{mycolor3}{rgb}{0.92900,0.69400,0.12500}%
\definecolor{mycolor4}{rgb}{0.49400,0.18400,0.55600}%
\definecolor{mycolor5}{rgb}{0.46600,0.67400,0.18800}%
\definecolor{mycolor6}{rgb}{0.30100,0.74500,0.93300}%
\definecolor{mycolor7}{rgb}{0.63500,0.07800,0.18400}%

\pgfplotsset{
tick label style={font=\footnotesize},
label style={font=\footnotesize},
legend  style={font=\footnotesize}
}

\begin{tikzpicture}

\begin{axis}[%
width=\fwidth,
height=\fheight,
at={(0\fwidth,0\fheight)},
xmin=0,
xmax=45,
xlabel style={font=\color{white!15!black}},
xlabel={$\lambda_b$ [BS/km]},
ymin=0,
ymax=1,
ylabel style={font=\color{white!15!black}},
ylabel={$P^{(U)}_{NL}$},
axis background/.style={fill=white},
label style={font=\footnotesize},
legend columns={2},
legend style={at={(1.02,0.3)},legend cell align=left, align=center, draw=white!15!black},
xmajorgrids,
ymajorgrids]
\addplot [ blue,mark=triangle*, mark options={fill=blue, scale = 1.5}]
  table[row sep=crcr]{%
0.2	0.997795449433559\\
2.82105263157895	0.969357819691701\\
5.4421052631579	0.94174594043639\\
8.06315789473684	0.914935653986459\\
10.6842105263158	0.888903511682144\\
13.3052631578947	0.863626752970767\\
15.9263157894737	0.839083285353865\\
18.5473684210526	0.815251664575929\\
21.1684210526316	0.79211107566385\\
23.7894736842105	0.7696413144066\\
26.4105263157895	0.747822769407878\\
29.0315789473684	0.726636405085209\\
31.6526315789474	0.706063742651204\\
34.2736842105263	0.686086847917895\\
36.8947368421053	0.666688311149599\\
39.5157894736842	0.647851234053454\\
42.1368421052632	0.62955921287544\\
44.7578947368421	0.611796326202337\\
};
\addlegendentry{$V = 30$ km/h}

\addplot [ green,mark=square*, mark options={fill=green, scale = 1.2}]
  table[row sep=crcr]{%
0.2	0.995107696496371\\
2.82105263157895	0.933161890861349\\
5.4421052631579	0.875072243842876\\
8.06315789473684	0.820598706924877\\
10.6842105263158	0.769516174820634\\
13.3052631578947	0.721613555177263\\
15.9263157894737	0.676692896408735\\
18.5473684210526	0.634568569516535\\
21.1684210526316	0.595066501076651\\
23.7894736842105	0.558023453872306\\
26.4105263157895	0.523286352332276\\
29.0315789473684	0.490711649958766\\
31.6526315789474	0.460164736133549\\
34.2736842105263	0.431519379851066\\
36.8947368421053	0.404657208079778\\
39.5157894736842	0.379467216596098\\
42.1368421052632	0.355845311473004\\
44.7578947368421	0.333693878015102\\
};
\addlegendentry{$V = 60$ km/h}

\addplot  [ red,mark=diamond*, mark options={fill=red, scale = 1.5}]
  table[row sep=crcr]{%
0.2	0.993216635011626\\
2.82105263157895	0.90846713687985\\
5.4421052631579	0.830965406553835\\
8.06315789473684	0.760090105030739\\
10.6842105263158	0.695273299906053\\
13.3052631578947	0.635995861816673\\
15.9263157894737	0.581783259091843\\
18.5473684210526	0.532201715603854\\
21.1684210526316	0.486854700810087\\
23.7894736842105	0.445379722892887\\
26.4105263157895	0.407445397772337\\
29.0315789473684	0.372748777485811\\
31.6526315789474	0.341012893509981\\
34.2736842105263	0.311984532563962\\
36.8947368421053	0.285432189091351\\
39.5157894736842	0.261144200267033\\
42.1368421052632	0.238927040967167\\
44.7578947368421	0.218603764727174\\
};
\addlegendentry{$V = 90$ km/h}

\addplot   [ orange,mark=*, mark options={fill=orange, scale = 1.5}]
  table[row sep=crcr]{%
0.2	0.992651490542881\\
2.82105263157895	0.901210937966952\\
5.4421052631579	0.81822299813982\\
8.06315789473684	0.742903334277229\\
10.6842105263158	0.674540693853333\\
13.3052631578947	0.612490067823418\\
15.9263157894737	0.55616649343408\\
18.5473684210526	0.505039439324162\\
21.1684210526316	0.45862771836715\\
23.7894736842105	0.416494878131294\\
26.4105263157895	0.378245022915115\\
29.0315789473684	0.343519034039597\\
31.6526315789474	0.311991128919823\\
34.2736842105263	0.283365763047743\\
36.8947368421053	0.257374807046456\\
39.5157894736842	0.233774993932443\\
42.1368421052632	0.212345604427945\\
44.7578947368421	0.192886367840777\\
};
\addlegendentry{$V = 100$ km/h}

\addplot [ purple,mark=star, mark options={fill=purple, scale = 1.5}]
  table[row sep=crcr]{%
0.2	0.991070791726652\\
2.82105263157895	0.881219202453336\\
5.4421052631579	0.783638070766736\\
8.06315789473684	0.696944879542231\\
10.6842105263158	0.619914369969346\\
13.3052631578947	0.551460468947589\\
15.9263157894737	0.490620316302098\\
18.5473684210526	0.436540144693359\\
21.1684210526316	0.388462795299132\\
23.7894736842105	0.345716677401388\\
26.4105263157895	0.307706003099507\\
29.0315789473684	0.273902148219578\\
31.6526315789474	0.243836008102274\\
34.2736842105263	0.217091232436392\\
36.8947368421053	0.193298236944609\\
39.5157894736842	0.172128901743083\\
42.1368421052632	0.153291876930767\\
44.7578947368421	0.136528424185606\\
};
\addlegendentry{$V = 130$ km/h}

\end{axis}
\end{tikzpicture}%
    \caption{\footnotesize  $P^{(U)}_{NL}$,   with parameters $\psi=30^\circ$, $T_S=300$ ms, for different values of the VN speed.}
      \end{subfigure}
\caption{Probability of not leaving the communication range of the serving BS ($P^{(U)}_{NL} = \mathbb{P}[T_L>T_S]$) within a slot of duration $T_S$, when varying the BSs density $\lambda_b$. An urban path loss model is considered. The curves are analytically obtained from Eq.~\eqref{eq:P_T_L}.}
\label{fig:NYU_P_T}
\end{figure}
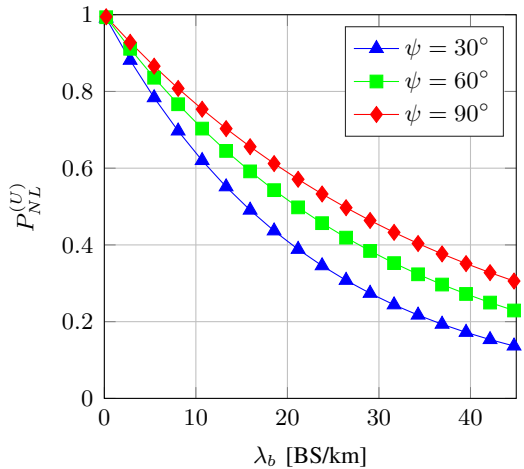
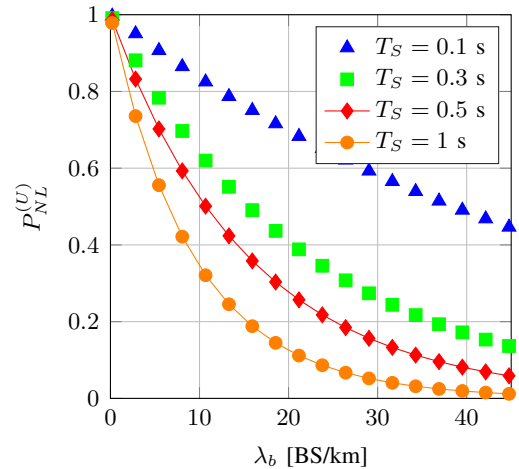
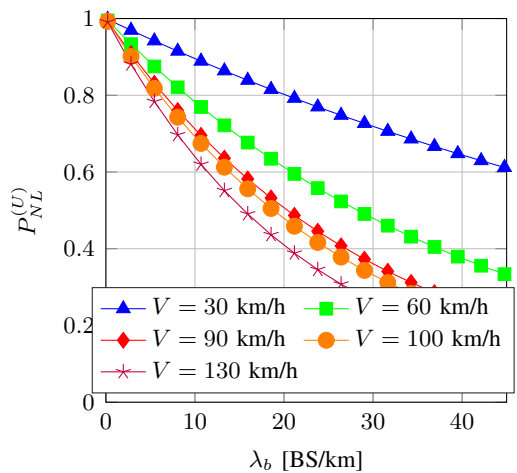

\textbf{Coverage and Connectivity (Urban Path Loss Model).}
Considering an urban path loss characterization,  Fig. \ref{fig:NYU_P_T} illustrates the probability $P^{(U)}_{NL}$ that the VN does not leave the communication range of its serving BS within a slot of duration $T_S$, as expressed by Eq. \eqref{eq:P_T_L}. We vary the BSs density $\lambda_b$ and consider different system configurations.
We observe that $P^{(U)}_{NL}$ decreases with $\lambda_b$. 
The reason is that, when considering denser networks, the distance $r$ to the serving BS decreases, thereby geometrically tightening the region shaped by the projection of the BS's beam onto the road surface and consequently reducing the maximum distance  that the VN can cover before leaving the communication range of its serving cell (see also Fig. \ref{fig:scenario_d}).

Furthermore, from Fig. \ref{fig:NYU_P_T}(a), we notice that $P^{(U)}_{NL}$ increases with $\psi$, since wider beams enlarge the area in which the VN can benefit from the coverage of its serving BS and therefore guarantee a more durable alignment within the slot.

Finally, Figs. \ref{fig:NYU_P_T}(b) and (c) show that $P^{(U)}_{NL}$ increases for shorter slots and/or lower vehicle speeds because of the shorter distance covered by the VN during one slot.

   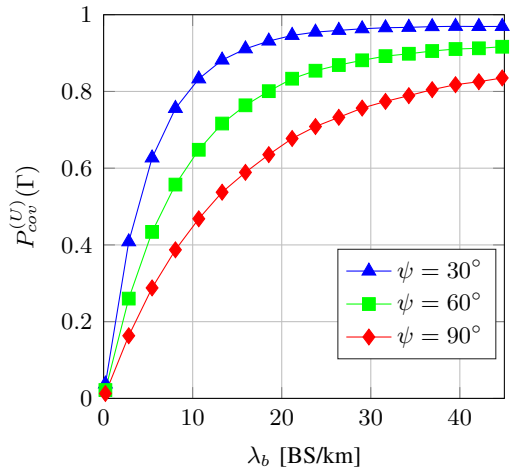
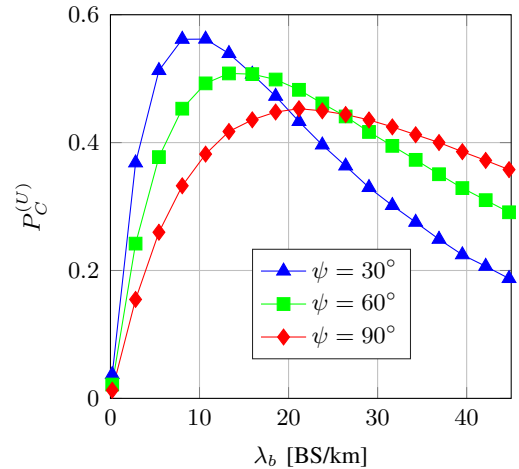
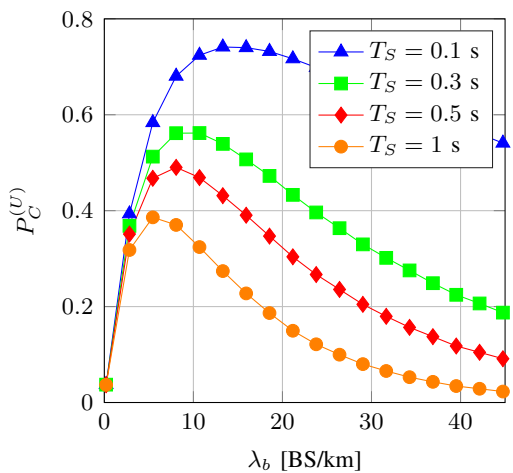
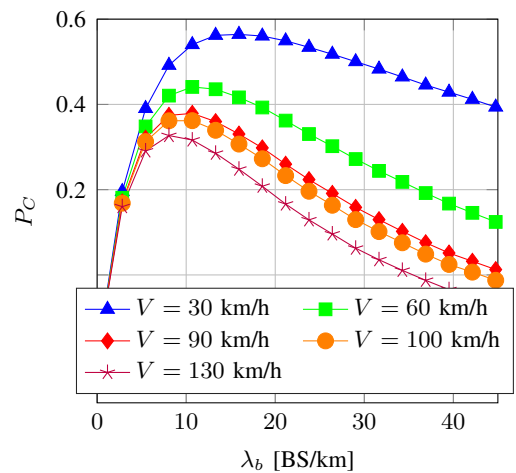
\begin{figure}[t!]
     \centering
     		\setlength{\belowcaptionskip}{-0.5cm}
     		             \begin{subfigure}[t!]{0.45\textwidth}
      		\setlength{\belowcaptionskip}{0cm}
	\setlength{\belowcaptionskip}{0cm}
	\setlength\fwidth{0.93\columnwidth}
	\setlength\fheight{0.90\columnwidth}
%
%
\definecolor{mycolor1}{rgb}{0.00000,0.44700,0.74100}%
\definecolor{mycolor2}{rgb}{0.85000,0.32500,0.09800}%
\definecolor{mycolor3}{rgb}{0.92900,0.69400,0.12500}%
\definecolor{mycolor4}{rgb}{0.49400,0.18400,0.55600}%
\definecolor{mycolor5}{rgb}{0.46600,0.67400,0.18800}%
\definecolor{mycolor6}{rgb}{0.30100,0.74500,0.93300}%
\pgfplotsset{
tick label style={font=\footnotesize},
label style={font=\footnotesize},
legend  style={font=\footnotesize}
}

\begin{tikzpicture}

\begin{axis}[%
width=\fwidth,
height=\fheight,
at={(0\fwidth,0\fheight)},
xmin=0,
xmax=45,
xlabel style={font=\color{white!15!black}},
xlabel={$\lambda_b$ [BS/km]},
ymin=0,
ymax=1,
ylabel style={font=\color{white!15!black}},
ylabel={$P^{(U)}_{cov}(\Gamma)$},
axis background/.style={fill=white},
label style={font=\footnotesize},
legend columns={1},
legend style={at={(0.98,0.39)},legend cell align=left, align=center, draw=white!15!black},
xmajorgrids,
ymajorgrids]
\addplot [  blue,mark=triangle*, mark options={fill=blue, scale = 1.5}]
  table[row sep=crcr]{%
0.2	0.03751\\
2.82105263157895	0.408145\\
5.4421052631579	0.62667\\
8.06315789473684	0.755745\\
10.6842105263158	0.83275\\
13.3052631578947	0.881805\\
15.9263157894737	0.911075\\
18.5473684210526	0.93133\\
21.1684210526316	0.945995\\
23.7894736842105	0.95461\\
26.4105263157895	0.95913\\
29.0315789473684	0.963345\\
31.6526315789474	0.96601\\
34.2736842105263	0.967055\\
36.8947368421053	0.968915\\
39.5157894736842	0.96967\\
42.1368421052632	0.969265\\
44.7578947368421	0.969425\\
};
\addlegendentry{$\psi= 30^\circ$}

\addplot [  green,mark=square*, mark options={fill=green, scale = 1.2}]
  table[row sep=crcr]{%
0.2	0.02177\\
2.82105263157895	0.26009\\
5.4421052631579	0.433995\\
8.06315789473684	0.557275\\
10.6842105263158	0.647995\\
13.3052631578947	0.716355\\
15.9263157894737	0.76386\\
18.5473684210526	0.800705\\
21.1684210526316	0.833225\\
23.7894736842105	0.85403\\
26.4105263157895	0.86879\\
29.0315789473684	0.88117\\
31.6526315789474	0.89203\\
34.2736842105263	0.898225\\
36.8947368421053	0.905345\\
39.5157894736842	0.910535\\
42.1368421052632	0.91245\\
44.7578947368421	0.916495\\
};
\addlegendentry{$\psi = 60^\circ$}

\addplot [  red,mark=diamond*, mark options={fill=red, scale = 1.5}]
  table[row sep=crcr]{%
0.2	0.012835\\
2.82105263157895	0.163245\\
5.4421052631579	0.28795\\
8.06315789473684	0.387205\\
10.6842105263158	0.46812\\
13.3052631578947	0.53734\\
15.9263157894737	0.58906\\
18.5473684210526	0.635385\\
21.1684210526316	0.677365\\
23.7894736842105	0.70857\\
26.4105263157895	0.73254\\
29.0315789473684	0.75662\\
31.6526315789474	0.77399\\
34.2736842105263	0.78906\\
36.8947368421053	0.804695\\
39.5157894736842	0.817615\\
42.1368421052632	0.825415\\
44.7578947368421	0.83515\\
};
\addlegendentry{$\psi = 90^\circ$}

\end{axis}
\end{tikzpicture}%
    \caption{\footnotesize \footnotesize  $P^{(U)}_{cov}$, for different antenna configurations. SINR threshold $\Gamma=-5$ dB.}
      \end{subfigure} \qquad \quad 
             \begin{subfigure}[t!]{0.45\textwidth}
             		\setlength{\belowcaptionskip}{0cm}
	\setlength{\belowcaptionskip}{0cm}
	\setlength\fwidth{0.93\columnwidth}
	\setlength\fheight{0.9\columnwidth}
%
%
\definecolor{mycolor1}{rgb}{0.00000,0.44700,0.74100}%
\definecolor{mycolor2}{rgb}{0.85000,0.32500,0.09800}%
\definecolor{mycolor3}{rgb}{0.92900,0.69400,0.12500}%
\definecolor{mycolor4}{rgb}{0.49400,0.18400,0.55600}%
\definecolor{mycolor5}{rgb}{0.46600,0.67400,0.18800}%
\definecolor{mycolor6}{rgb}{0.30100,0.74500,0.93300}%

\pgfplotsset{
tick label style={font=\footnotesize},
label style={font=\footnotesize},
legend  style={font=\footnotesize}
}

\begin{tikzpicture}

\begin{axis}[%
width=\fwidth,
height=\fheight,
at={(0\fwidth,0\fheight)},
xmin=0,
xmax=45,
xlabel style={font=\color{white!15!black}},
xlabel={$\lambda_b$ [BS/km]},
ymin=0,
ymax=0.6,
ylabel style={font=\color{white!15!black}},
ylabel={$P^{(U)}_{C}$},
axis background/.style={fill=white},
label style={font=\footnotesize},
legend columns={1},
legend style={at={(0.75,0.39)},legend cell align=left, align=center, draw=white!15!black},
xmajorgrids,
ymajorgrids]
\addplot [  blue,mark=triangle*, mark options={fill=blue, scale = 1.5}]
  table[row sep=crcr]{%
0.2	0.03724611715\\
2.82105263157895	0.368291681475\\
5.4421052631579	0.51270692715\\
8.06315789473684	0.5615638797\\
10.6842105263158	0.5618480975\\
13.3052631578947	0.539289892875\\
15.9263157894737	0.506926685375\\
18.5473684210526	0.47248699225\\
21.1684210526316	0.432986641475\\
23.7894736842105	0.3962967954\\
26.4105263157895	0.3634623135\\
29.0315789473684	0.32974336005\\
31.6526315789474	0.30164145255\\
34.2736842105263	0.27542693455\\
36.8947368421053	0.24878830455\\
39.5157894736842	0.2245852687\\
42.1368421052632	0.20654067885\\
44.7578947368421	0.1874286295\\
};
\addlegendentry{ $\psi = 30^\circ$}

\addplot   [  green,mark=square*, mark options={fill=green, scale = 1.2}]
  table[row sep=crcr]{%
0.2	0.02166365355\\
2.82105263157895	0.242039754\\
5.4421052631579	0.377287043325\\
8.06315789473684	0.453067361375\\
10.6842105263158	0.4926317188\\
13.3052631578947	0.508164328125\\
15.9263157894737	0.5070731838\\
18.5473684210526	0.4986470458\\
21.1684210526316	0.482574757125\\
23.7894736842105	0.4612530627\\
26.4105263157895	0.44088920525\\
29.0315789473684	0.41667445205\\
31.6526315789474	0.3949195216\\
34.2736842105263	0.373055298125\\
36.8947368421053	0.350536003825\\
39.5157894736842	0.3288670313\\
42.1368421052632	0.31000032525\\
44.7578947368421	0.29105131715\\
};
\addlegendentry{ $ \psi = 60^\circ$}

\addplot   [  red,mark=diamond*, mark options={fill=red, scale = 1.5}]
  table[row sep=crcr]{%
0.2	0.0127885373\\
2.82105263157895	0.154742384175\\
5.4421052631579	0.259794249\\
8.06315789473684	0.33248131735\\
10.6842105263158	0.3820257102\\
13.3052631578947	0.4173761583\\
15.9263157894737	0.4356216512\\
18.5473684210526	0.4478447634\\
21.1684210526316	0.452998004225\\
23.7894736842105	0.44970457905\\
26.4105263157895	0.4439119146\\
29.0315789473684	0.4354499424\\
31.6526315789474	0.42459930415\\
34.2736842105263	0.4123193577\\
36.8947368421053	0.400130565275\\
39.5157894736842	0.385803901975\\
42.1368421052632	0.37210533615\\
44.7578947368421	0.357761557\\
};
\addlegendentry{  $\psi = 90^\circ$}

\end{axis}
\end{tikzpicture}%
    \caption{\footnotesize  $P^{(U)}_{C}$ with parameters $T_S=300$ ms, $V=100$ km/h, for different antenna configurations.}
      \end{subfigure}
      \begin{subfigure}[t!]{0.45\textwidth}
      		\setlength{\belowcaptionskip}{0cm}
	\setlength{\belowcaptionskip}{0cm}
	\setlength\fwidth{0.93\columnwidth}
	\setlength\fheight{0.9\columnwidth}
%
%
\definecolor{mycolor1}{rgb}{0.00000,0.44700,0.74100}%
\definecolor{mycolor2}{rgb}{0.85000,0.32500,0.09800}%
\definecolor{mycolor3}{rgb}{0.92900,0.69400,0.12500}%
\definecolor{mycolor4}{rgb}{0.49400,0.18400,0.55600}%
\definecolor{mycolor5}{rgb}{0.46600,0.67400,0.18800}%
\definecolor{mycolor6}{rgb}{0.30100,0.74500,0.93300}%
\definecolor{mycolor7}{rgb}{0.63500,0.07800,0.18400}%

\pgfplotsset{
tick label style={font=\footnotesize},
label style={font=\footnotesize},
legend  style={font=\footnotesize}
}
\begin{tikzpicture}

\begin{axis}[%
width=\fwidth,
height=\fheight,
at={(0\fwidth,0\fheight)},
xmin=0,
xmax=45,
xlabel style={font=\color{white!15!black}},
xlabel={$\lambda_b$ [BS/km]},
ymin=0,
ymax=0.8,
ylabel style={font=\color{white!15!black}},
ylabel={$P^{(U)}_{C}$},
axis background/.style={fill=white},
label style={font=\footnotesize},
legend columns={1},
legend style={at={(0.97,0.97)},legend cell align=left, align=center, draw=white!15!black},
xmajorgrids,
ymajorgrids]
\addplot [  blue,mark=triangle*, mark options={fill=blue, scale = 1.5}]
  table[row sep=crcr]{%
0.2	0.03741303665\\
2.82105263157895	0.393408924775\\
5.4421052631579	0.58364597115\\
8.06315789473684	0.680416117125\\
10.6842105263158	0.72408861625\\
13.3052631578947	0.741276146175\\
15.9263157894737	0.73994778275\\
18.5473684210526	0.73208591645\\
21.1684210526316	0.71673311175\\
23.7894736842105	0.6984594987\\
26.4105263157895	0.67904006175\\
29.0315789473684	0.658903896375\\
31.6526315789474	0.63878860265\\
34.2736842105263	0.6183156259\\
36.8947368421053	0.59872164595\\
39.5157894736842	0.57915964925\\
42.1368421052632	0.56026424795\\
44.7578947368421	0.540885831625\\
};
\addlegendentry{$T_S = 0.1$ s}

\addplot[  green,mark=square*, mark options={fill=green, scale = 1.2}]
  table[row sep=crcr]{
0.2	0.03724611715\\
2.82105263157895	0.368291681475\\
5.4421052631579	0.51270692715\\
8.06315789473684	0.5615638797\\
10.6842105263158	0.5618480975\\
13.3052631578947	0.539289892875\\
15.9263157894737	0.506926685375\\
18.5473684210526	0.47248699225\\
21.1684210526316	0.432986641475\\
23.7894736842105	0.3962967954\\
26.4105263157895	0.3634623135\\
29.0315789473684	0.32974336005\\
31.6526315789474	0.30164145255\\
34.2736842105263	0.27542693455\\
36.8947368421053	0.24878830455\\
39.5157894736842	0.2245852687\\
42.1368421052632	0.20654067885\\
44.7578947368421	0.1874286295\\
};
\addlegendentry{$T_S = 0.3$ s}

\addplot  [  red,mark=diamond*, mark options={fill=red, scale = 1.5}]
  table[row sep=crcr]{%
0.2	0.0371180205\\
2.82105263157895	0.351231220475\\
5.4421052631579	0.4675083534\\
8.06315789473684	0.490285790025\\
10.6842105263158	0.46900063625\\
13.3052631578947	0.4312908255\\
15.9263157894737	0.39047763425\\
18.5473684210526	0.3469763048\\
21.1684210526316	0.30382521415\\
23.7894736842105	0.2668516794\\
26.4105263157895	0.23574935835\\
29.0315789473684	0.20462411145\\
31.6526315789474	0.1796585398\\
34.2736842105263	0.156638733625\\
36.8947368421053	0.13734370125\\
39.5157894736842	0.11779066325\\
42.1368421052632	0.104675773675\\
44.7578947368421	0.09129075225\\
};
\addlegendentry{$T_S = 0.5$ s}

\addplot [  orange!100!black,mark=*, mark options={fill=orange!100!black, scale = 1.2}]
  table[row sep=crcr]{%
0.2	0.0368490738\\
2.82105263157895	0.317995973125\\
5.4421052631579	0.3861039204\\
8.06315789473684	0.37029237765\\
10.6842105263158	0.32411879125\\
13.3052631578947	0.27403853985\\
15.9263157894737	0.227490872125\\
18.5473684210526	0.18653142905\\
21.1684210526316	0.14960910925\\
23.7894736842105	0.1214072998\\
26.4105263157895	0.09972554175\\
29.0315789473684	0.080068419675\\
31.6526315789474	0.06572249035\\
34.2736842105263	0.052815708825\\
36.8947368421053	0.0430973392\\
39.5157894736842	0.03433116635\\
42.1368421052632	0.028685397675\\
44.7578947368421	0.022863888625\\
};
\addlegendentry{$T_S = 1$ s}

\end{axis}
\end{tikzpicture}%
    \caption{\footnotesize \footnotesize  $P^{(U)}_{C}$ with parameters $\psi=30^\circ$, $V=100$ km/h, for different values of the slot duration. }
      \end{subfigure} \qquad \quad
       \begin{subfigure}[t!]{0.45\textwidth}
      		\setlength{\belowcaptionskip}{0cm}
	\setlength{\belowcaptionskip}{0cm}
	\setlength\fwidth{0.93\columnwidth}
	\setlength\fheight{0.9\columnwidth}
%
%
\definecolor{mycolor1}{rgb}{0.00000,0.44700,0.74100}%
\definecolor{mycolor2}{rgb}{0.85000,0.32500,0.09800}%
\definecolor{mycolor3}{rgb}{0.92900,0.69400,0.12500}%
\definecolor{mycolor4}{rgb}{0.49400,0.18400,0.55600}%
\definecolor{mycolor5}{rgb}{0.46600,0.67400,0.18800}%
\definecolor{mycolor6}{rgb}{0.30100,0.74500,0.93300}%
\definecolor{mycolor7}{rgb}{0.63500,0.07800,0.18400}%

\pgfplotsset{
tick label style={font=\footnotesize},
label style={font=\footnotesize},
legend  style={font=\footnotesize}
}

\begin{tikzpicture}

\begin{axis}[%
width=\fwidth,
height=\fheight,
at={(0\fwidth,0\fheight)},
xmin=0,
xmax=45,
xlabel style={font=\color{white!15!black}},
xlabel={$\lambda_b$ [BS/km]},
ymin=-0.1,
ymax=0.8,
ylabel style={font=\color{white!15!black}},
ylabel={$P_{C}$},
yticklabels={, , ,0.2,0.4,0.6,0.8},
axis background/.style={fill=white},
label style={font=\footnotesize},
legend columns={2},
legend cell align={center},
legend style={at={(1.04,0.3)},legend cell align=left, align=center, draw=white!15!black,
/tikz/every even column/.append style={column sep=0.2cm}},
xmajorgrids,
ymajorgrids]
\addplot [ blue,mark=triangle*, mark options={fill=blue, scale = 1.5}]
  table[row sep=crcr]{%
0.2	0.037427478\\
2.82105263157895	0.3956476001\\
5.4421052631579	0.5903670069\\
8.06315789473684	0.691691832525\\
10.6842105263158	0.74029393125\\
13.3052631578947	0.7620029727\\
15.9263157894737	0.764287151375\\
18.5473684210526	0.76032384205\\
21.1684210526316	0.74912398055\\
23.7894736842105	0.7339423524\\
26.4105263157895	0.717716979\\
29.0315789473684	0.70026511395\\
31.6526315789474	0.68285797885\\
34.2736842105263	0.664371620275\\
36.8947368421053	0.645437882675\\
39.5157894736842	0.6286758478\\
42.1368421052632	0.6119163798\\
44.7578947368421	0.5936952585\\
};
\addlegendentry{$V = 30$ km/h}

\addplot [ green,mark=square*, mark options={fill=green, scale = 1.2}]
  table[row sep=crcr]{%
0.2	0.0373344532\\
2.82105263157895	0.38107274215\\
5.4421052631579	0.5483487834\\
8.06315789473684	0.620561113125\\
10.6842105263158	0.64101347625\\
13.3052631578947	0.635574180825\\
15.9263157894737	0.616483454125\\
18.5473684210526	0.5927636051\\
21.1684210526316	0.5620345494\\
23.7894736842105	0.5304958692\\
26.4105263157895	0.5019702768\\
29.0315789473684	0.471822297375\\
31.6526315789474	0.4440651369\\
34.2736842105263	0.418207770025\\
36.8947368421053	0.392056921025\\
39.5157894736842	0.36733523775\\
42.1368421052632	0.345644745325\\
44.7578947368421	0.3243114395\\
};
\addlegendentry{$V = 60$ km/h}

\addplot  [ red,mark=diamond*, mark options={fill=red, scale = 1.5}]
  table[row sep=crcr]{%
0.2	0.03726637255\\
2.82105263157895	0.371234406925\\
5.4421052631579	0.5207878368\\
8.06315789473684	0.574339748925\\
10.6842105263158	0.5789194725\\
13.3052631578947	0.55963754325\\
15.9263157894737	0.530405088125\\
18.5473684210526	0.4979914643\\
21.1684210526316	0.45975357\\
23.7894736842105	0.42363205275\\
26.4105263157895	0.3911811705\\
29.0315789473684	0.3584414076\\
31.6526315789474	0.32971370315\\
34.2736842105263	0.302199852225\\
36.8947368421053	0.27556911515\\
39.5157894736842	0.2514257343\\
42.1368421052632	0.23201296305\\
44.7578947368421	0.212531889875\\
};
\addlegendentry{$V = 90$ km/h}

\addplot   [ orange,mark=*, mark options={fill=orange, scale = 1.5}]
  table[row sep=crcr]{%
0.2	0.03724611715\\
2.82105263157895	0.368291681475\\
5.4421052631579	0.51270692715\\
8.06315789473684	0.5615638797\\
10.6842105263158	0.5618480975\\
13.3052631578947	0.539289892875\\
15.9263157894737	0.506926685375\\
18.5473684210526	0.47248699225\\
21.1684210526316	0.432986641475\\
23.7894736842105	0.3962967954\\
26.4105263157895	0.3634623135\\
29.0315789473684	0.32974336005\\
31.6526315789474	0.30164145255\\
34.2736842105263	0.27542693455\\
36.8947368421053	0.24878830455\\
39.5157894736842	0.2245852687\\
42.1368421052632	0.20654067885\\
44.7578947368421	0.1874286295\\
};
\addlegendentry{$V = 100$ km/h}

\addplot [ purple,mark=star, mark options={fill=purple, scale = 1.5}]
  table[row sep=crcr]{%
0.2	0.0371810373\\
2.82105263157895	0.360104292775\\
5.4421052631579	0.49084241085\\
8.06315789473684	0.52695831615\\
10.6842105263158	0.516454895\\
13.3052631578947	0.4853983803\\
15.9263157894737	0.448062129625\\
18.5473684210526	0.40802032965\\
21.1684210526316	0.366246694225\\
23.7894736842105	0.3291399819\\
26.4105263157895	0.29612659185\\
29.0315789473684	0.2626849146\\
31.6526315789474	0.23558568875\\
34.2736842105263	0.21073095505\\
36.8947368421053	0.18732033695\\
39.5157894736842	0.1654741855\\
42.1368421052632	0.149455816675\\
44.7578947368421	0.132660964125\\
};
\addlegendentry{$V = 130$ km/h}

\end{axis}
\end{tikzpicture}%
    \caption{\footnotesize \footnotesize  $P^{(U)}_{C}$ with parameters $\psi=30^\circ$, $T_S=300$ ms, for different values of the VN speed.}
      \end{subfigure}
\caption{Coverage and connectivity probabilities ($P^{(U)}_{cov}$ and $P^{(U)}_{C}$, respectively) within a slot of duration $T_S$, when varying the BSs density $\lambda_b$. An urban path loss model is considered. The curves  are analytically obtained from Eqs. \eqref{eq:P_cov} and \eqref{eq:P_C}.}
\label{fig:NYU_P_C}
\end{figure}

Nevertheless, according to Theorem \ref{th:P_C}, the preservation of the connectivity during a slot requires  both accurate alignment  between the endpoints and satisfactory signal quality.
The value of $P^{(U)}_{NL}$  therefore becomes particularly meaningful if weighted by and constrained with the VN's coverage probability at the beginning of the slot.
In Fig. \ref{fig:NYU_P_C}(a), we thus plot the SINR coverage probability, i.e., the probability that the VN is connected at the beginning of a synchronization slot, as expressed by Eq.~\eqref{eq:P_cov}, for different transmit antenna configurations.
We note that  $P^{(U)}_{cov}(\Gamma)$ increases with the BSs density $\lambda_b$ because of the higher probability of having a BS at shorter distance which can offer better signal quality.
However, the gain progressively reduces with $\lambda_b$ because of the increasing impact of the interference from the surrounding BSs.
Anyway, the increasing trend shown by $P^{(U)}_{cov}(\Gamma)$ proves that the reduction of the attachment distance  is dominant over the increased interference.
Finally, the figure shows that   narrower beams  result in higher SINR (and higher $P^{(U)}_{cov}(\Gamma)$) due to the higher gain achieved by beamforming, as expected.

In Figs. \ref{fig:NYU_P_C}(b), (c), (d) we report the connectivity probability $P^{(U)}_C$ representing the probability, as expressed by Eq.~\eqref{eq:P_C}, that the VN is connected during an entire slot of duration $T_{S}$, i.e., the VN is still connected at the end of one slot given that it was connected at the beginning of the same slot, as a function of $\psi$, $V$ and $T_S$.
We observe that ${P^{(U)}_C= P^{(U)}_{cov}(\Gamma) \cdot P^{(U)}_{NL}}$ exhibits a maximum for a given  density $\lambda_b^*$.
In detail, we notice that $P^{(U)}_C$  increases with $\lambda_b$ for sparse networks.
In this region, the reduction of
the attachment distance $r$ to the serving BS is more significant than
the increase of the interference coming from the neighboring
BSs. Moreover, $r$ is still
sufficiently large to allow for a loose beam alignment (thanks
to the widening of the beam's projection on the road's surface with the distance), so that the connectivity between the endpoints is maintained for a
relatively large number of slots. 
After a certain value of $\lambda_b$, $P^{(U)}_{C}$ starts decreasing. 
In this range, the coverage probability does not increase significantly, as depicted in Fig. \ref{fig:NYU_P_C}(a), while   $r$ keeps reducing and the resulting smaller beam projected on the highway lanes contributes to increasing the risk of losing connectivity during a slot, as illustrated in Fig. \ref{fig:NYU_P_T}.

Finally, as mentioned above, Figs. \ref{fig:NYU_P_C}(c) and (d) emphasize how more durable connectivity capabilities are guaranteed for smaller values of $T_S$ and $V$,~respectively.\\

\begin{figure}[t!]
     \centering
     		\setlength{\belowcaptionskip}{-0.5cm}
     		             \begin{subfigure}[t!]{0.45\textwidth}
      		\setlength{\belowcaptionskip}{0cm}
	\setlength{\belowcaptionskip}{0cm}
	\setlength\fwidth{0.93\columnwidth}
	\setlength\fheight{0.9\columnwidth}
%
%
\definecolor{mycolor1}{rgb}{0.00000,0.44700,0.74100}%
\definecolor{mycolor2}{rgb}{0.85000,0.32500,0.09800}%
\definecolor{mycolor3}{rgb}{0.92900,0.69400,0.12500}%
\definecolor{mycolor4}{rgb}{0.49400,0.18400,0.55600}%
\definecolor{mycolor5}{rgb}{0.46600,0.67400,0.18800}%
\definecolor{mycolor6}{rgb}{0.30100,0.74500,0.93300}%
\pgfplotsset{
tick label style={font=\footnotesize},
label style={font=\footnotesize},
legend  style={font=\footnotesize}
}

\begin{tikzpicture}

\begin{axis}[%
width=\fwidth,
height=\fheight,
at={(0\fwidth,0\fheight)},
xmin=0,
xmax=45,
xlabel style={font=\color{white!15!black}},
xlabel={$\lambda_b$ [BS/km]},
ymin=0,
ymax=2,
ylabel style={font=\color{white!15!black}},
ylabel={$B^{(U)}$ [Gbps]},
axis background/.style={fill=white},
label style={font=\footnotesize},
legend columns={2},
legend style={at={(0.99,0.22)},legend cell align=left, align=center, draw=white!15!black},
xmajorgrids,
ymajorgrids]
\addplot [only marks, blue,mark=triangle*, mark options={fill=blue, scale = 1.5}]
  table[row sep=crcr]{%
0.2	0.0537365485349919\\
2.82105263157895	0.593472195652716\\
5.4421052631579	0.961582428564448\\
8.06315789473684	1.21452474923663\\
10.6842105263158	1.40049240563193\\
13.3052631578947	1.53524330586648\\
15.9263157894737	1.60122827410759\\
18.5473684210526	1.66426800306715\\
21.1684210526316	1.70633477511376\\
23.7894736842105	1.72837388551604\\
26.4105263157895	1.75016030264221\\
29.0315789473684	1.74890033807294\\
31.6526315789474	1.73361766736327\\
34.2736842105263	1.72653397448476\\
36.8947368421053	1.69907143980583\\
39.5157894736842	1.68705545827722\\
42.1368421052632	1.65708149783734\\
44.7578947368421	1.62411791858463\\
47.3789473684211	1.59946760518255\\
50	1.57306640105923\\
};
\addlegendentry{ $\psi = 30^\circ$}

\addplot   [only marks,green,mark=square*, mark options={fill=green, scale = 1.2}]
  table[row sep=crcr]{%
0.2	0.0281916602602655\\
2.82105263157895	0.347053586819619\\
5.4421052631579	0.587831117724639\\
8.06315789473684	0.776619182698531\\
10.6842105263158	0.924549327658808\\
13.3052631578947	1.04711810692336\\
15.9263157894737	1.12696467787488\\
18.5473684210526	1.19153633536385\\
21.1684210526316	1.250452072025\\
23.7894736842105	1.28860707053785\\
26.4105263157895	1.32662052032973\\
29.0315789473684	1.35271431296196\\
31.6526315789474	1.36829020941559\\
34.2736842105263	1.37987067399179\\
36.8947368421053	1.37775689872026\\
39.5157894736842	1.39904748133483\\
42.1368421052632	1.38759100514973\\
44.7578947368421	1.37862746057413\\
47.3789473684211	1.3760732854853\\
50	1.37209674997682\\
};
\addlegendentry{ $ \psi = 60^\circ$}

\addplot   [only marks,red,mark=diamond*, mark options={fill=red, scale = 1.5}]
  table[row sep=crcr]{%
0.2	0.0164848286474305\\
2.82105263157895	0.209686967478141\\
5.4421052631579	0.368533265651197\\
8.06315789473684	0.500410635490789\\
10.6842105263158	0.61179800702134\\
13.3052631578947	0.70904694785489\\
15.9263157894737	0.781573781697945\\
18.5473684210526	0.845858218564306\\
21.1684210526316	0.908317028514498\\
23.7894736842105	0.954449153253118\\
26.4105263157895	0.999507955628739\\
29.0315789473684	1.03810424247374\\
31.6526315789474	1.06734631994481\\
34.2736842105263	1.09877531674643\\
36.8947368421053	1.11322532604426\\
39.5157894736842	1.14561718566275\\
42.1368421052632	1.15844455418229\\
44.7578947368421	1.16657297225606\\
47.3789473684211	1.17785325285236\\
50	1.19138374596571\\
};
\addlegendentry{ $ \psi = 90^\circ$}

\addplot[color=black, dashed, line width = 0.25 mm, forget plot]
  table[row sep=crcr]{
0.2	0.0522937983027348\\
2.82105263157895	0.597650089705572\\
5.4421052631579	0.964986921583883\\
8.06315789473684	1.2205640528491\\
10.6842105263158	1.4008693041818\\
13.3052631578947	1.52833806588538\\
15.9263157894737	1.61760923199863\\
18.5473684210526	1.67864128658832\\
21.1684210526316	1.71843242865098\\
23.7894736842105	1.74203321873534\\
26.4105263157895	1.75317099503488\\
29.0315789473684	1.75464919699453\\
31.6526315789474	1.74861110102668\\
34.2736842105263	1.73671805251374\\
36.8947368421053	1.72027307904459\\
39.5157894736842	1.70030860602663\\
42.1368421052632	1.67764938823786\\
44.7578947368421	1.65295843734815\\
47.3789473684211	1.62677188924498\\
50	1.59952489427186\\
};
\addplot[color=black, dashed, line width = 0.25 mm, forget plot]
  table[row sep=crcr]{
0.2	0.0287192232324177\\
2.82105263157895	0.35035226560478\\
5.4421052631579	0.594586700960701\\
8.06315789473684	0.783842289576677\\
10.6842105263158	0.932253364350468\\
13.3052631578947	1.04946853560232\\
15.9263157894737	1.14236764353789\\
18.5473684210526	1.21600822216979\\
21.1684210526316	1.27419795633978\\
23.7894736842105	1.3198599806235\\
26.4105263157895	1.35527468994272\\
29.0315789473684	1.38224714855671\\
31.6526315789474	1.40222231898782\\
34.2736842105263	1.41636963726226\\
36.8947368421053	1.42564437614268\\
39.5157894736842	1.43083306786603\\
42.1368421052632	1.43258810576593\\
44.7578947368421	1.43145383660362\\
47.3789473684211	1.42788782902155\\
50	1.42227418615344\\
};
\addplot[color=black, dashed, line width = 0.25 mm]
  table[row sep=crcr]{
0.2	0.0172553623319388\\
2.82105263157895	0.218993694291496\\
5.4421052631579	0.384396640754856\\
8.06315789473684	0.522233673625299\\
10.6842105263158	0.638375494270526\\
13.3052631578947	0.73704090596417\\
15.9263157894737	0.821382474810419\\
18.5473684210526	0.893821720919296\\
21.1684210526316	0.956258724027269\\
23.7894736842105	1.01021007454728\\
26.4105263157895	1.05690429386487\\
29.0315789473684	1.09734874447726\\
31.6526315789474	1.13237795674629\\
34.2736842105263	1.16269025058339\\
36.8947368421053	1.18887448367508\\
39.5157894736842	1.21143076755799\\
42.1368421052632	1.23078708910016\\
44.7578947368421	1.24731128579181\\
47.3789473684211	1.26132145503808\\
50	1.27309398344939\\
};
\addlegendentry{Simulation}
\end{axis}
\end{tikzpicture}%
    \caption{\footnotesize $B^{(U)}$ with parameters  $T_S=300$ ms, $V=130$ km/h, for different antenna configurations.}
      \end{subfigure} \qquad \quad 
             \begin{subfigure}[t!]{0.45\textwidth}
             		\setlength{\belowcaptionskip}{0cm}
	\setlength{\belowcaptionskip}{0cm}
	\setlength\fwidth{0.93\columnwidth}
	\setlength\fheight{0.9\columnwidth}
%
%
\definecolor{mycolor1}{rgb}{0.00000,0.44700,0.74100}%
\definecolor{mycolor2}{rgb}{0.85000,0.32500,0.09800}%
\definecolor{mycolor3}{rgb}{0.92900,0.69400,0.12500}%
\definecolor{mycolor4}{rgb}{0.49400,0.18400,0.55600}%
\definecolor{mycolor5}{rgb}{0.46600,0.67400,0.18800}%
\definecolor{mycolor6}{rgb}{0.30100,0.74500,0.93300}%
\definecolor{mycolor7}{rgb}{0.63500,0.07800,0.18400}%
\pgfplotsset{
tick label style={font=\footnotesize},
label style={font=\footnotesize},
legend  style={font=\footnotesize}
}
\begin{tikzpicture}

\begin{axis}[%
width=\fwidth,
height=\fheight,
at={(0\fwidth,0\fheight)},
xmin=0,
xmax=45,
xlabel style={font=\color{white!15!black}},
xlabel={$\lambda_b$ [BS/km]},
ymin=0,
ymax=3.5,
ylabel style={font=\color{white!15!black}},
ylabel={$B^{(U)}$ [Gbps]},
axis background/.style={fill=white},
label style={font=\footnotesize},
legend columns={1},
legend style={at={(0.47,0.98)},legend cell align=left, align=center, draw=white!15!black},
xmajorgrids,
ymajorgrids]
\addplot [only marks,blue,mark=triangle*, mark options={fill=blue, scale = 1.5}]
  table[row sep=crcr]{%
0.2	0.0726968574201499\\
2.82105263157895	0.86222760233505\\
5.4421052631579	1.42234021732383\\
8.06315789473684	1.8182290026815\\
10.6842105263158	2.13907691948752\\
13.3052631578947	2.37758041850113\\
15.9263157894737	2.5395275443375\\
18.5473684210526	2.66611316797409\\
21.1684210526316	2.77805213188939\\
23.7894736842105	2.86486833447186\\
26.4105263157895	2.92960946268982\\
29.0315789473684	2.98248515163951\\
31.6526315789474	3.00983790741079\\
34.2736842105263	3.04573935926732\\
36.8947368421053	3.05249772295629\\
39.5157894736842	3.0810085266451\\
42.1368421052632	3.08237362004994\\
44.7578947368421	3.05823044297727\\
47.3789473684211	3.06198883098916\\
50	3.05574910750322\\
};
\addlegendentry{$T_S = 0.1$ s}

\addplot[only marks,green,mark=square*, mark options={fill=green, scale = 1.2}]
  table[row sep=crcr]{
0.2	0.0537365485349919\\
2.82105263157895	0.593472195652716\\
5.4421052631579	0.961582428564448\\
8.06315789473684	1.21452474923663\\
10.6842105263158	1.40049240563193\\
13.3052631578947	1.53524330586648\\
15.9263157894737	1.60122827410759\\
18.5473684210526	1.66426800306715\\
21.1684210526316	1.70633477511376\\
23.7894736842105	1.72837388551604\\
26.4105263157895	1.75016030264221\\
29.0315789473684	1.74890033807294\\
31.6526315789474	1.73361766736327\\
34.2736842105263	1.72653397448476\\
36.8947368421053	1.69907143980583\\
39.5157894736842	1.68705545827722\\
42.1368421052632	1.65708149783734\\
44.7578947368421	1.62411791858463\\
47.3789473684211	1.59946760518255\\
50	1.57306640105923\\
};
\addlegendentry{$T_S = 0.3$ s}

\addplot  [only marks,red,mark=diamond*, mark options={fill=red, scale = 1.5}]
  table[row sep=crcr]{%
0.2	0.0501751808807657\\
2.82105263157895	0.550991179766295\\
5.4421052631579	0.879601293323525\\
8.06315789473684	1.08549378687531\\
10.6842105263158	1.24257897161598\\
13.3052631578947	1.3423673550278\\
15.9263157894737	1.38433056425734\\
18.5473684210526	1.40893821836392\\
21.1684210526316	1.43265388765256\\
23.7894736842105	1.41826128747417\\
26.4105263157895	1.4119519980068\\
29.0315789473684	1.39247752481399\\
31.6526315789474	1.36079779783662\\
34.2736842105263	1.34820341980584\\
36.8947368421053	1.29583868420497\\
39.5157894736842	1.26732542065305\\
42.1368421052632	1.23596204795434\\
44.7578947368421	1.18904124086665\\
47.3789473684211	1.15869991571868\\
50	1.13158162130616\\
};
\addlegendentry{$T_S = 0.5$ s}

\addplot [only marks,orange!100!black,mark=*, mark options={fill=orange!100!black, scale = 1.2}]
  table[row sep=crcr]{%
0.2	0.0477223364253884\\
2.82105263157895	0.514992653301016\\
5.4421052631579	0.802211931385355\\
8.06315789473684	0.966037043694431\\
10.6842105263158	1.07640007202475\\
13.3052631578947	1.11572108636254\\
15.9263157894737	1.11991137514241\\
18.5473684210526	1.10830537298428\\
21.1684210526316	1.0838555392323\\
23.7894736842105	1.04434267812594\\
26.4105263157895	1.00538228581592\\
29.0315789473684	0.971501835035907\\
31.6526315789474	0.917270235381098\\
34.2736842105263	0.88577087689141\\
36.8947368421053	0.840795459052997\\
39.5157894736842	0.792905930723791\\
42.1368421052632	0.757502738765143\\
44.7578947368421	0.720432801320302\\
47.3789473684211	0.687391389471251\\
50	0.658955766858236\\
};
\addlegendentry{$T_S=1$ s}

\addplot[color=black, dashed, line width = 0.25 mm]
  table[row sep=crcr]{
0.2	0.0750205003922906\\
2.82105263157895	0.870887342344244\\
5.4421052631579	1.42853688267393\\
8.06315789473684	1.83588691704706\\
10.6842105263158	2.14115797216522\\
13.3052631578947	2.37397701262276\\
15.9263157894737	2.55368067431715\\
18.5473684210526	2.69343256396236\\
21.1684210526316	2.8024933613491\\
23.7894736842105	2.88755651166726\\
26.4105263157895	2.95357201596799\\
29.0315789473684	3.0042740799409\\
31.6526315789474	3.04252853470108\\
34.2736842105263	3.07056943922918\\
36.8947368421053	3.09016161244511\\
39.5157894736842	3.10271651992633\\
42.1368421052632	3.10937506407774\\
44.7578947368421	3.11106870239579\\
47.3789473684211	3.10856408932649\\
50	3.10249763676277\\
};
\addplot[color=black, dashed, line width = 0.25 mm]
  table[row sep=crcr]{
0.2	0.0522937983027348\\
2.82105263157895	0.597650089705572\\
5.4421052631579	0.964986921583883\\
8.06315789473684	1.2205640528491\\
10.6842105263158	1.4008693041818\\
13.3052631578947	1.52833806588538\\
15.9263157894737	1.61760923199863\\
18.5473684210526	1.67864128658832\\
21.1684210526316	1.71843242865098\\
23.7894736842105	1.74203321873534\\
26.4105263157895	1.75317099503488\\
29.0315789473684	1.75464919699453\\
31.6526315789474	1.74861110102668\\
34.2736842105263	1.73671805251374\\
36.8947368421053	1.72027307904459\\
39.5157894736842	1.70030860602663\\
42.1368421052632	1.67764938823786\\
44.7578947368421	1.65295843734815\\
47.3789473684211	1.62677188924498\\
50	1.59952489427186\\
};
\addplot[color=black, dashed, line width = 0.25 mm]
  table[row sep=crcr]{
0.2	0.0491860115557257\\
2.82105263157895	0.554600554586886\\
5.4421052631579	0.883220766815852\\
8.06315789473684	1.10158358822812\\
10.6842105263158	1.24646968092185\\
13.3052631578947	1.34051888952522\\
15.9263157894737	1.39850138625561\\
18.5473684210526	1.43045583766149\\
21.1684210526316	1.44341778232576\\
23.7894736842105	1.44244352515133\\
26.4105263157895	1.43124008185072\\
29.0315789473684	1.41256884769116\\
31.6526315789474	1.38851285861786\\
34.2736842105263	1.36065898116831\\
36.8947368421053	1.33022338060121\\
39.5157894736842	1.29814203009492\\
42.1368421052632	1.26513658231495\\
44.7578947368421	1.23176204735617\\
47.3789473684211	1.19844327123424\\
50	1.16550325105515\\
};
\addplot[color=black, dashed, line width = 0.25 mm]
  table[row sep=crcr]{
0.2	0.0479827069236464\\
2.82105263157895	0.525454007863108\\
5.4421052631579	0.811845372728615\\
8.06315789473684	0.981637453651263\\
10.6842105263158	1.07638139359548\\
13.3052631578947	1.12167298729832\\
15.9263157894737	1.13410301517033\\
18.5473684210526	1.12477574034132\\
21.1684210526316	1.10127966629176\\
23.7894736842105	1.06886686017231\\
26.4105263157895	1.03119468721831\\
29.0315789473684	0.990810029753496\\
31.6526315789474	0.949475365602846\\
34.2736842105263	0.908394814431267\\
36.8947368421053	0.868371992235416\\
39.5157894736842	0.829923774345135\\
42.1368421052632	0.793361927646404\\
44.7578947368421	0.758852124125333\\
47.3789473684211	0.726457396638529\\
50	0.696169007698609\\
};
\addlegendentry{Simulation}
\end{axis}
\end{tikzpicture}%
    \caption{\footnotesize $B^{(U)}$ with parameters $\psi=30^\circ$, $V=130$ km/h, for different values of the slot duration.}
      \end{subfigure}
      \begin{subfigure}[t!]{0.45\textwidth}
      		\setlength{\belowcaptionskip}{0cm}
	\setlength{\belowcaptionskip}{0cm}
	\setlength\fwidth{0.93\columnwidth}
	\setlength\fheight{0.94\columnwidth}
%
%
\definecolor{mycolor1}{rgb}{0.00000,0.44700,0.74100}%
\definecolor{mycolor2}{rgb}{0.85000,0.32500,0.09800}%
\definecolor{mycolor3}{rgb}{0.92900,0.69400,0.12500}%
\definecolor{mycolor4}{rgb}{0.49400,0.18400,0.55600}%
\definecolor{mycolor5}{rgb}{0.46600,0.67400,0.18800}%
\definecolor{mycolor6}{rgb}{0.30100,0.74500,0.93300}%
\definecolor{mycolor7}{rgb}{0.63500,0.07800,0.18400}%
\pgfplotsset{
tick label style={font=\footnotesize},
label style={font=\footnotesize},
legend  style={font=\footnotesize}
}

\begin{tikzpicture}

\begin{axis}[%
width=\fwidth,
height=\fheight,
at={(0\fwidth,0\fheight)},
xmin=0,
xmax=45,
xlabel style={font=\color{white!15!black}},
xlabel={$\lambda_b$ [BS/km]},
ymin=0,
ymax=4,
ylabel style={font=\color{white!15!black}},
ylabel={$B^{(U)}$ [bps]},
axis background/.style={fill=white},
label style={font=\footnotesize},
legend columns={2},
legend style={at={(0.99,0.3)},legend cell align=left, align=center, draw=white!15!black},
xmajorgrids,
ymajorgrids]
\addplot [only marks,blue,mark=triangle*, mark options={fill=blue, scale = 1.5}]
  table[row sep=crcr]{%
0.2	0.0900408060677211\\
2.82105263157895	1.0606887133234\\
5.4421052631579	1.74884340426923\\
8.06315789473684	2.24370058679102\\
10.6842105263158	2.64426834842246\\
13.3052631578947	2.94154124427111\\
15.9263157894737	3.15152416049252\\
18.5473684210526	3.31433231870394\\
21.1684210526316	3.46387838222685\\
23.7894736842105	3.57935280150588\\
26.4105263157895	3.67443676510714\\
29.0315789473684	3.74511834320231\\
31.6526315789474	3.78043480468155\\
34.2736842105263	3.83874915541013\\
36.8947368421053	3.85039301351941\\
39.5157894736842	3.89637238193118\\
42.1368421052632	3.90039941313932\\
44.7578947368421	3.8932570585285\\
47.3789473684211	3.89671959399189\\
50	3.89145249362918\\
};
\addlegendentry{$V = 30$ km/h}

\addplot [only marks,green,mark=square*, mark options={fill=green, scale = 1.2}]
  table[row sep=crcr]{%
0.2	0.0631983663831961\\
2.82105263157895	0.748255050395669\\
5.4421052631579	1.22465994824042\\
8.06315789473684	1.56598466247919\\
10.6842105263158	1.82216572905426\\
13.3052631578947	2.03538609858055\\
15.9263157894737	2.16091847401684\\
18.5473684210526	2.25711210582626\\
21.1684210526316	2.35356144731339\\
23.7894736842105	2.41012782294973\\
26.4105263157895	2.45955865954237\\
29.0315789473684	2.50559440631969\\
31.6526315789474	2.51481253732463\\
34.2736842105263	2.54609750231552\\
36.8947368421053	2.52740593102524\\
39.5157894736842	2.55035408017492\\
42.1368421052632	2.53785179979293\\
44.7578947368421	2.51451430306801\\
47.3789473684211	2.50836232288764\\
50	2.50005212510041\\
};
\addlegendentry{$V = 60$ km/h}

\addplot  [only marks, red,mark=diamond*, mark options={fill=red, scale = 1.5}]
  table[row sep=crcr]{%
0.2	0.0568691159769731\\
2.82105263157895	0.651095177043413\\
5.4421052631579	1.05827987518791\\
8.06315789473684	1.3567276952021\\
10.6842105263158	1.5724859526656\\
13.3052631578947	1.73737055799003\\
15.9263157894737	1.83218820224671\\
18.5473684210526	1.90179350075329\\
21.1684210526316	1.96977462712708\\
23.7894736842105	2.00316983983123\\
26.4105263157895	2.03424486887643\\
29.0315789473684	2.05788778030879\\
31.6526315789474	2.05801665261571\\
34.2736842105263	2.05667054330638\\
36.8947368421053	2.03901494607476\\
39.5157894736842	2.05149128097088\\
42.1368421052632	2.02344769730198\\
44.7578947368421	1.99580923544636\\
47.3789473684211	1.9804163936501\\
50	1.96208528447374\\
};
\addlegendentry{$V = 90$ km/h}

\addplot   [only marks,orange,mark=*, mark options={fill=orange, scale = 1.5}]
  table[row sep=crcr]{%
0.2	0.0560843363853609\\
2.82105263157895	0.633842160773509\\
5.4421052631579	1.02706913273571\\
8.06315789473684	1.30983681645035\\
10.6842105263158	1.51355701529347\\
13.3052631578947	1.67064483874118\\
15.9263157894737	1.75974597430512\\
18.5473684210526	1.82558594882971\\
21.1684210526316	1.88345561697812\\
23.7894736842105	1.91849666207663\\
26.4105263157895	1.94563847688589\\
29.0315789473684	1.96140404424336\\
31.6526315789474	1.96175563837636\\
34.2736842105263	1.9565835281952\\
36.8947368421053	1.93540020726122\\
39.5157894736842	1.9397431869101\\
42.1368421052632	1.91287850972179\\
44.7578947368421	1.88211423271638\\
47.3789473684211	1.8657547918941\\
50	1.84301060996797\\
};
\addlegendentry{$V = 100$ km/h}

\addplot [only marks,purple,mark=star, mark options={fill=purple, scale = 1.5}]
  table[row sep=crcr]{%
0.2	0.0537365485349919\\
2.82105263157895	0.593472195652716\\
5.4421052631579	0.961582428564448\\
8.06315789473684	1.21452474923663\\
10.6842105263158	1.40049240563193\\
13.3052631578947	1.53524330586648\\
15.9263157894737	1.60122827410759\\
18.5473684210526	1.66426800306715\\
21.1684210526316	1.70633477511376\\
23.7894736842105	1.72837388551604\\
26.4105263157895	1.75016030264221\\
29.0315789473684	1.74890033807294\\
31.6526315789474	1.73361766736327\\
34.2736842105263	1.72653397448476\\
36.8947368421053	1.69907143980583\\
39.5157894736842	1.68705545827722\\
42.1368421052632	1.65708149783734\\
44.7578947368421	1.62411791858463\\
47.3789473684211	1.59946760518255\\
50	1.57306640105923\\
};
\addlegendentry{$V = 130$ km/h}

\addplot[color=black, dashed, line width = 0.25 mm]
  table[row sep=crcr]{
0.2	0.0915831148480798\\
2.82105263157895	1.06532529832275\\
5.4421052631579	1.75103656101053\\
8.06315789473684	2.25492225322456\\
10.6842105263158	2.63520502639453\\
13.3052631578947	2.92765680243897\\
15.9263157894737	3.15562787452995\\
18.5473684210526	3.33501805092188\\
21.1684210526316	3.47701581487079\\
23.7894736842105	3.5897103550167\\
26.4105263157895	3.67908561576123\\
29.0315789473684	3.74965757004321\\
31.6526315789474	3.80489394008143\\
34.2736842105263	3.84749887424124\\
36.8947368421053	3.87961009664669\\
39.5157894736842	3.90293828258302\\
42.1368421052632	3.91886739214448\\
44.7578947368421	3.92852806664\\
47.3789473684211	3.93285221891169\\
50	3.93261410952694\\
};
\addplot[color=black, dashed, line width = 0.25 mm]
  table[row sep=crcr]{
0.2	0.0649706881849419\\
2.82105263157895	0.751965372880411\\
5.4421052631579	1.22976802504984\\
8.06315789473684	1.57569384033293\\
10.6842105263158	1.83217948741731\\
13.3052631578947	2.02529954722604\\
15.9263157894737	2.17207001427815\\
18.5473684210526	2.2840703493671\\
21.1684210526316	2.36944553663553\\
23.7894736842105	2.43408019070629\\
26.4105263157895	2.48232379524155\\
29.0315789473684	2.51745741229561\\
31.6526315789474	2.54199662125258\\
34.2736842105263	2.55790116007975\\
36.8947368421053	2.56671828975852\\
39.5157894736842	2.56968389110585\\
42.1368421052632	2.5677958469891\\
44.7578947368421	2.56186758324815\\
47.3789473684211	2.55256767503493\\
50	2.54044994218256\\
};
\addplot[color=black, dashed, line width = 0.25 mm]
  table[row sep=crcr]{
0.2	0.0564901299131941\\
2.82105263157895	0.649991758703115\\
5.4421052631579	1.05672729804413\\
8.06315789473684	1.34591953777975\\
10.6842105263158	1.55561826048443\\
13.3052631578947	1.70920929945182\\
15.9263157894737	1.8219513145773\\
18.5473684210526	1.90422628156957\\
21.1684210526316	1.96333281394226\\
23.7894736842105	2.00454288680919\\
26.4105263157895	2.03175047372498\\
29.0315789473684	2.04788951077816\\
31.6526315789474	2.05520848163686\\
34.2736842105263	2.05545637366108\\
36.8947368421053	2.05001126674326\\
39.5157894736842	2.03997151336174\\
42.1368421052632	2.02622111452337\\
44.7578947368421	2.00947841497616\\
47.3789473684211	1.99033153187297\\
50	1.96926257031912\\
};
\addplot[color=black, dashed, line width = 0.25 mm]
  table[row sep=crcr]{
0.2	0.0552482198586007\\
2.82105263157895	0.634720571316362\\
5.4421052631579	1.03029406493591\\
8.06315789473684	1.31019687374539\\
10.6842105263158	1.51194819485772\\
13.3052631578947	1.65860959252686\\
15.9263157894737	1.76522767459306\\
18.5473684210526	1.84203961789772\\
21.1684210526316	1.89624253497609\\
23.7894736842105	1.933031377934\\
26.4105263157895	1.9562447071651\\
29.0315789473684	1.96877341206185\\
31.6526315789474	1.97283172703147\\
34.2736842105263	1.9701405593366\\
36.8947368421053	1.9620550660073\\
39.5157894736842	1.94965349857126\\
42.1368421052632	1.9338029933295\\
44.7578947368421	1.91520617941717\\
47.3789473684211	1.89443654924216\\
50	1.87196510848854\\
};
\addplot[color=black, dashed, line width = 0.25 mm]
  table[row sep=crcr]{%
0.2	0.0522937983027348\\
2.82105263157895	0.597650089705572\\
5.4421052631579	0.964986921583883\\
8.06315789473684	1.2205640528491\\
10.6842105263158	1.4008693041818\\
13.3052631578947	1.52833806588538\\
15.9263157894737	1.61760923199863\\
18.5473684210526	1.67864128658832\\
21.1684210526316	1.71843242865098\\
23.7894736842105	1.74203321873534\\
26.4105263157895	1.75317099503488\\
29.0315789473684	1.75464919699453\\
31.6526315789474	1.74861110102668\\
34.2736842105263	1.73671805251374\\
36.8947368421053	1.72027307904459\\
39.5157894736842	1.70030860602663\\
42.1368421052632	1.67764938823786\\
44.7578947368421	1.65295843734815\\
47.3789473684211	1.62677188924498\\
50	1.59952489427186\\
};
\addlegendentry{Simulation}
\end{axis}
\end{tikzpicture}%
    \caption{\footnotesize $B^{(U)}$ with parameters $\psi=30^\circ$, $T_S=300$ ms, for different values of the VN speed.}
      \end{subfigure} 
\caption{Average throughput  ($B$) experienced within a slot of duration $T_S$, when varying the BSs density $\lambda_b$. An urban path loss model is considered.  The dashed lines are drawn from Monte Carlo simulations, the markers are obtained from Eq.~\eqref{eq:B}.}
\label{fig:NYU_B}
\end{figure}
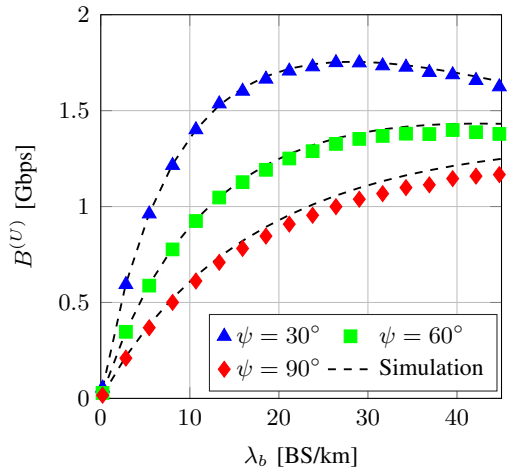
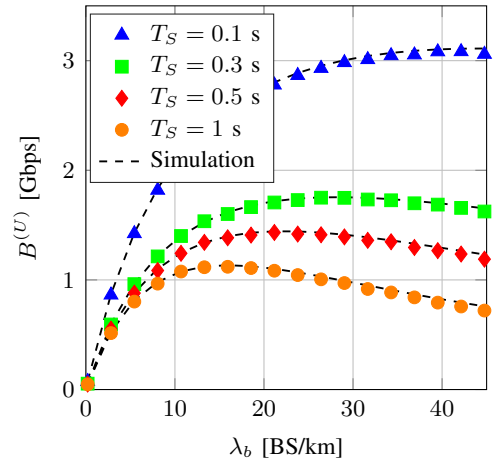
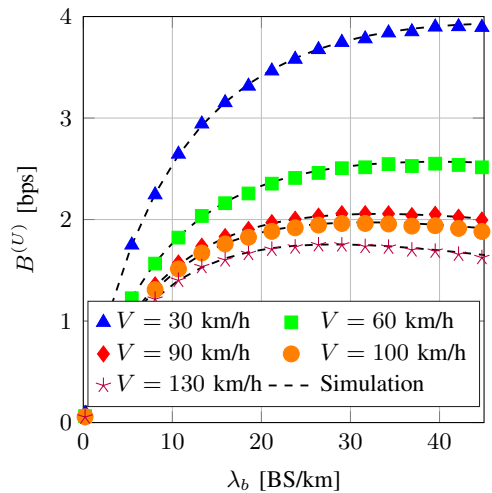

%
%
%

\textbf{Throughput Results (Urban Path Loss Model).}
As mentioned in Sec.~\ref{sec:throughput_analysis}, a non-zero throughput $B$ is experienced  when the vehicle
is within the coverage of its serving infrastructure and properly aligned with it.
In Fig. \ref{fig:NYU_B} we plot the  results  representing the average throughput measured by the target VN within a slot of duration $T_S$.
While the analytical connectivity results presented in the previous paragraphs were exact, these throughput curves were obtained from an approximate theoretical framework. To assess the accuracy of the approximation, we report in the graphs also the simulation results obtained through a Monte Carlo approach, where multiple independent simulations are repeated to get different statistical quantities of interest.
At each iteration, the simulator computes the path loss, according to the urban characterization proposed in \cite{Mustafa}, from each BS to the test VN and (i) makes the optimal association decision according to a max-path loss policy, (ii) measures the SINR from the VN to its serving cell, and (iii) computes the data rate, according to the Shannon formula, for the fraction of time in which the nodes are properly aligned.
Finally, the throughput is estimated by averaging over the total number of repetitions.

\begin{figure}[t!]
 \centering
 \includegraphics[trim= 0cm 0cm 0cm 0cm , clip=true, width= 0.95\textwidth]{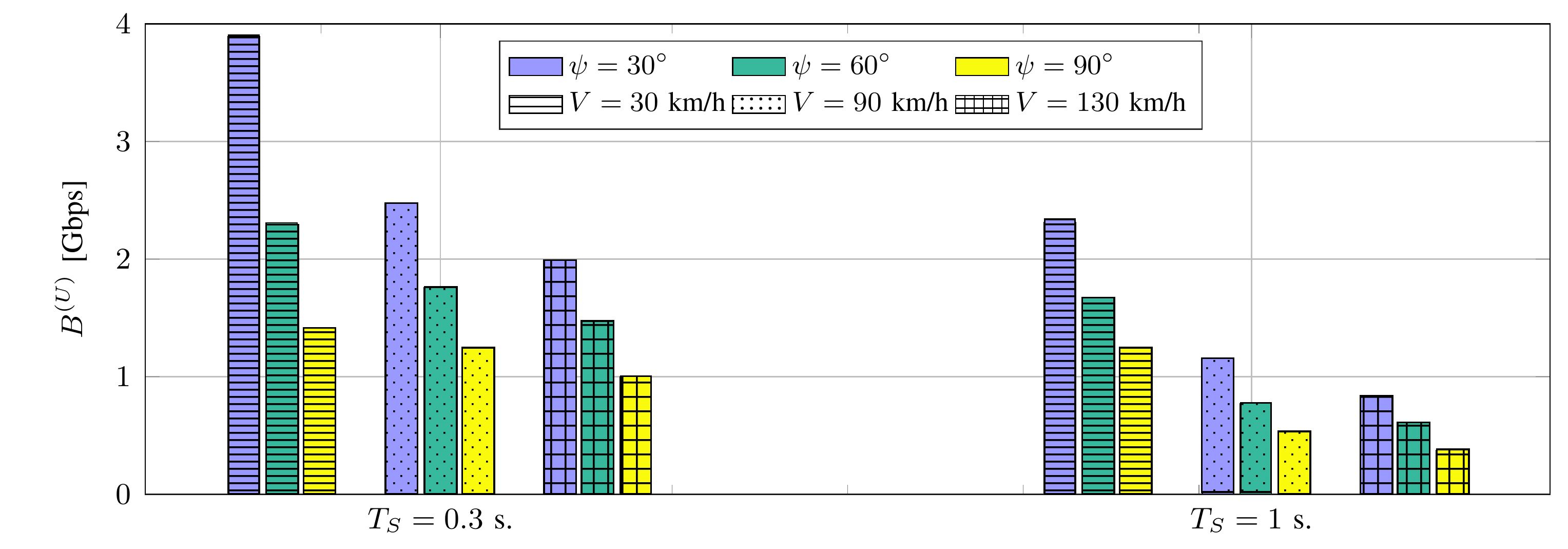}
    \caption{\footnotesize Histogram of the average throughput  ($B$) experienced  for different values of slot duration $T_S$, speed $V$ and BS's beamwidth $\psi$. An urban path loss model and a fixed BS density  $\lambda_b=25$ BS/km are considered.}
    \label{fig:NYU_B_hist}
      \end{figure}

First, we observe that the numerical results closely follow the analytic curves representing Eq.~\eqref{eq:B}, thereby validating our theoretical framework.
Moreover, it is  interesting to observe that, in all considered configurations, the throughput  exhibits a similar trend when varying the BSs density $\lambda_b$ and, most importantly, follows the behavior of the connectivity curves presented in Fig.~\ref{fig:NYU_P_C}. 
An optimal value of $B^{(s)}$ can therefore be identified, meaning that there exists a density threshold $\lambda_b^*$ above which the deployment of more BSs results in a considerable increase of the system complexity while leading to worse communication performance.
Moreover, Fig.~\ref{fig:NYU_B_hist} exemplifies how $B$ grows as $V$ decreases, because of the higher probability of remaining connected  during a slot.
Similarly, $B$ grows
as $T_S$ decreases, because the beam alignment is repeated more frequently, thus reducing the disconnection time. 
However,
the overhead  (which is not accounted for in this 
analysis) would also increase, thus limiting or even nullifying such
a gain if $T_S$ drops below a certain threshold.\\

\begin{figure}[t!]
     \centering
     		\setlength{\belowcaptionskip}{-0.5cm}
     		             \begin{subfigure}[t!]{0.45\textwidth}
      		\setlength{\belowcaptionskip}{0cm}
	\setlength{\belowcaptionskip}{0cm}
	\setlength\fwidth{0.99\columnwidth}
	\setlength\fheight{0.95\columnwidth}
	\includegraphics[ width= 0.99\columnwidth, height = 0.89\columnwidth]{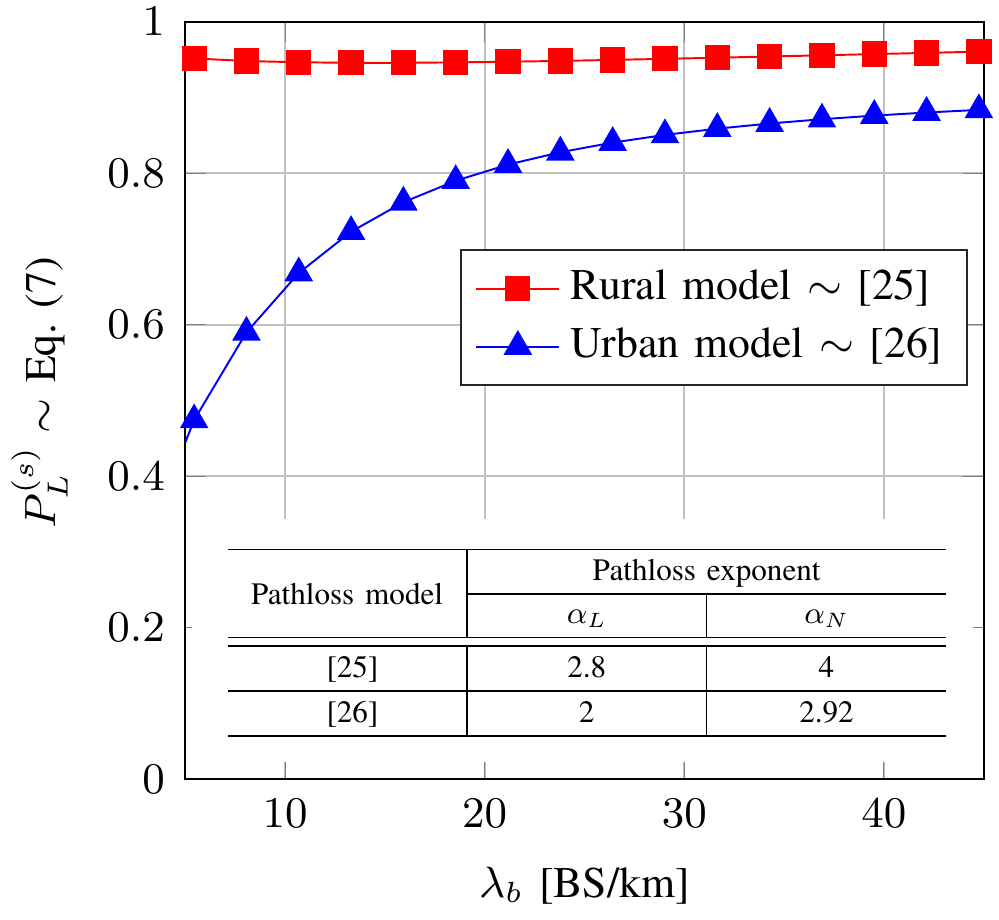}
    \caption{\footnotesize  LOS connection probability $P^{(s)}_{L}$ between the target VN and its serving BS. }
      \end{subfigure} \qquad \quad 
             \begin{subfigure}[t!]{0.45\textwidth}
             		\setlength{\belowcaptionskip}{0cm}
	\setlength{\belowcaptionskip}{0cm}
	\setlength\fwidth{0.99\columnwidth}
	\setlength\fheight{0.95\columnwidth}
%
%
\definecolor{mycolor1}{rgb}{0.00000,0.44700,0.74100}%
\definecolor{mycolor2}{rgb}{0.85000,0.32500,0.09800}%
\pgfplotsset{
tick label style={font=\footnotesize},
label style={font=\footnotesize},
legend  style={font=\footnotesize}
}
\begin{tikzpicture}

\begin{axis}[%
width=\fwidth,
height=\fheight,
at={(0\fwidth,0\fheight)},
xmin=0,
xmax=45,
xlabel style={font=\color{white!15!black}},
xlabel={$\lambda_b$ [BS/km]},
ymin=0,
ymax=0.6,
ylabel style={font=\color{white!15!black}},
ylabel={$P^{(s)}_{C}$},
axis background/.style={fill=white},
label style={font=\footnotesize},
legend columns={1},
legend style={at={(0.7,0.27)},legend cell align=left, align=center, draw=white!15!black},
xmajorgrids,
ymajorgrids]

\addplot[red,mark=square*, mark options={fill=red, scale = 1.2}]
  table[row sep=crcr]{
0.2	0.0334344665896892\\
2.82105263157895	0.335944975448609\\
5.4421052631579	0.47601175218823\\
8.06315789473684	0.531904440128491\\
10.6842105263158	0.543734705056194\\
13.3052631578947	0.532568162926696\\
15.9263157894737	0.509705130238364\\
18.5473684210526	0.481310309824911\\
21.1684210526316	0.450784495157431\\
23.7894736842105	0.420008949284662\\
26.4105263157895	0.390013867750441\\
29.0315789473684	0.361345309535974\\
31.6526315789474	0.334270705143923\\
34.2736842105263	0.308896095071725\\
36.8947368421053	0.285234123570698\\
39.5157894736842	0.26324404556739\\
42.1368421052632	0.242855536514941\\
44.7578947368421	0.223982988211259\\
};
\addlegendentry{\begin{NoHyper} Rural model $\sim$ \cite{Tassi17_Highway} \end{NoHyper}}

\addplot [blue,mark=triangle*, mark options={fill=blue, scale = 1.5}]
  table[row sep=crcr]{%
0.2	0.0374908821244268\\
2.82105263157895	0.369104763849123\\
5.4421052631579	0.512246058628286\\
8.06315789473684	0.56080304845788\\
10.6842105263158	0.562045649194926\\
13.3052631578947	0.540148726972374\\
15.9263157894737	0.50763271945305\\
18.5473684210526	0.471031496045371\\
21.1684210526316	0.433751805470879\\
23.7894736842105	0.397542052616507\\
26.4105263157895	0.363260144593812\\
29.0315789473684	0.331281732908154\\
31.6526315789474	0.30172077474995\\
34.2736842105263	0.274550450900904\\
36.8947368421053	0.249670328742723\\
39.5157894736842	0.226943969669926\\
42.1368421052632	0.206220120872714\\
44.7578947368421	0.187344615853642\\
};
\addlegendentry{\begin{NoHyper} Urban model $\sim$ \cite{Mustafa}\end{NoHyper}}

\end{axis}
\end{tikzpicture}%
    \caption{\footnotesize Connectivity probability $P^{(s)}_{C}$ with parameters $\psi=30^\circ$, $T_S=300$ ms and $V=100$ km/h.}
      \end{subfigure}
\caption{Comparison between the rural and the urban path loss models, denoted with superscript $s\in\{R,\,U\}$ respectively, in terms of (a) LOS attachment probability, as given by Eq. \eqref{eq:P_L}, and (b) connectivity probability, as given by Eq. \eqref{eq:P_C},   as a function of  $\lambda_b$.}
\label{fig:PL_comparison}
\end{figure}
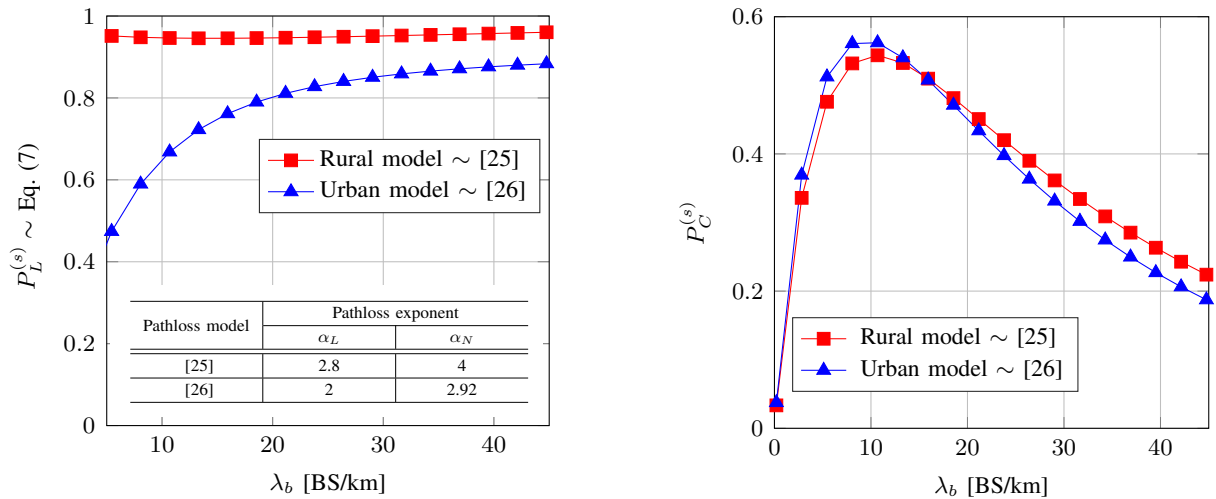

\textbf{Path Loss Models Comparison.}
In Fig. \ref{fig:PL_comparison}, we compare the connectivity performance of the V2X network adopting either a rural  or an urban highway path loss model.
It is interesting to notice that the two models are intrinsically different and capture the characteristics of two clearly distinct environments.
In particular, for  $s\in\{R,\,U\}$, Fig.~\ref{fig:PL_comparison}(a) illustrates the probability $P^{(s)}_L$  that the target VN is attached to a LOS BS, as given by Eq.~\eqref{eq:P_L}. For the rural case,  $P^{(R)}_L$  does not vary much with the density $\lambda_b$, due to the assumption that the LOS probability $p^{(R)}_L$ is independent of the distance from the VN to the BS. 
This result is perfectly in line with the simulation outcomes presented in \cite{Tassi17_Highway}.
On the other hand, for the urban case,  $P^{(U)}_L$ monotonically increases with $\lambda_b$ because the probability of LOS conditions to one of the BSs increases.
Moreover, from the table superimposed to Fig. \ref{fig:PL_comparison}(a), we realize how the path loss parameters (i.e., $\alpha_L$ and $\alpha_N$) of the two models are remarkably different. More specifically, the path loss characterized by \cite{Tassi17_Highway} is much more severe than its counterpart.
It is interesting to note that, although referred to different scenarios and channel characterizations, all connectivity curves displayed in Fig.~\ref{fig:PL_comparison}(b) show similar trends.
Finally, notice that the rural characterization has a significant impact mainly in sparse networks (i.e., $\lambda_b < 15$ BS/km, with the  settings of Fig. \ref{fig:PL_comparison}), showing the worst connectivity performance, while the urban path loss model affects the connectivity behavior of denser networks~instead.
 
\section{Final Remarks and Open Challenges}
\label{sec:concl}




In this work, we proposed a  stochastic geometry framework to characterize the coverage (i.e., downlink coverage) and the connectivity (i.e., beam alignment probability) performance of a dynamic mmWave vehicular network deployed along a multi-lane highway section. 
The key point is that, in order to
compensate for the increased isotropic path loss experienced
at high frequencies, next-generation mmWave automotive
communication must provide mechanisms by which the vehicles and the infrastructure determine suitable directions
of transmission to exchange data. 
In this context, our model characterizes the base stations as independent homogeneous LOS and NLOS one-dimensional  Poisson processes and implements both a rural and a distance-dependent urban path loss model, in which the communication between the endpoints is impaired by large vehicles acting as~blockages or environmental obstructions, respectively.
We derived expressions for the  SINR coverage probability, the probability that the vehicle does not disconnect from its serving cell over time, and the achievable throughput, as a function of the infrastructure density.
We  showed that:
\begin{itemize}
\item[(i)] The preservation of the connectivity requires both accurate alignment  between the endpoints and satisfactory signal quality. 
\item[(ii)] For sparse networks, the connectivity probability grows with the BS density.
In this region, the connectivity can  be improved by considering narrower beams due to the resulting higher gain achieved by beamforming.
\item[(iii)] For dense networks, the connectivity probability presents a decreasing behavior when increasing the number of  BSs. In this region, larger beams should be produced to generate larger connectivity regions and  ensure a more robust alignment between the  nodes.
\item[(iv)] Better connectivity and throughput performance is achieved by considering frequent beam re-alignment operations (which allow the endpoints to reduce the disconnection time) and slower cars (because of the resulting higher probability of remaining connected during one~slot).
\item[(v)] Consistent connectivity trends have been observed under intrinsically diverse path loss models (i.e., a rural model characterized by stringent path loss parameters and an urban model featuring moderate LOS attachment probability values).
\end{itemize}

Most vehicular-related challenges  are still largely unexplored, so that additional research is needed. As part of our future work, we aim at further validating the presented results considering innovative and original channel models specifically tailored to a next-generation V2X context.
Moreover, it would be interesting to formally characterize space and time dependent target performance metrics to assess
the connectivity performance of the nodes in the rapidly time-varying mmWave environment, while investigating  realistic scenarios and models.

\appendices
\section{Proof of Lemma \ref{lemma:f_L}}
\label{appendix:f_L}

\begin{figure}[t!]
     \centering
             \begin{subfigure}[t!]{0.45\textwidth}
             		 \includegraphics[trim= 0cm 0cm 0cm 0cm , clip=true, width= 0.95\textwidth]{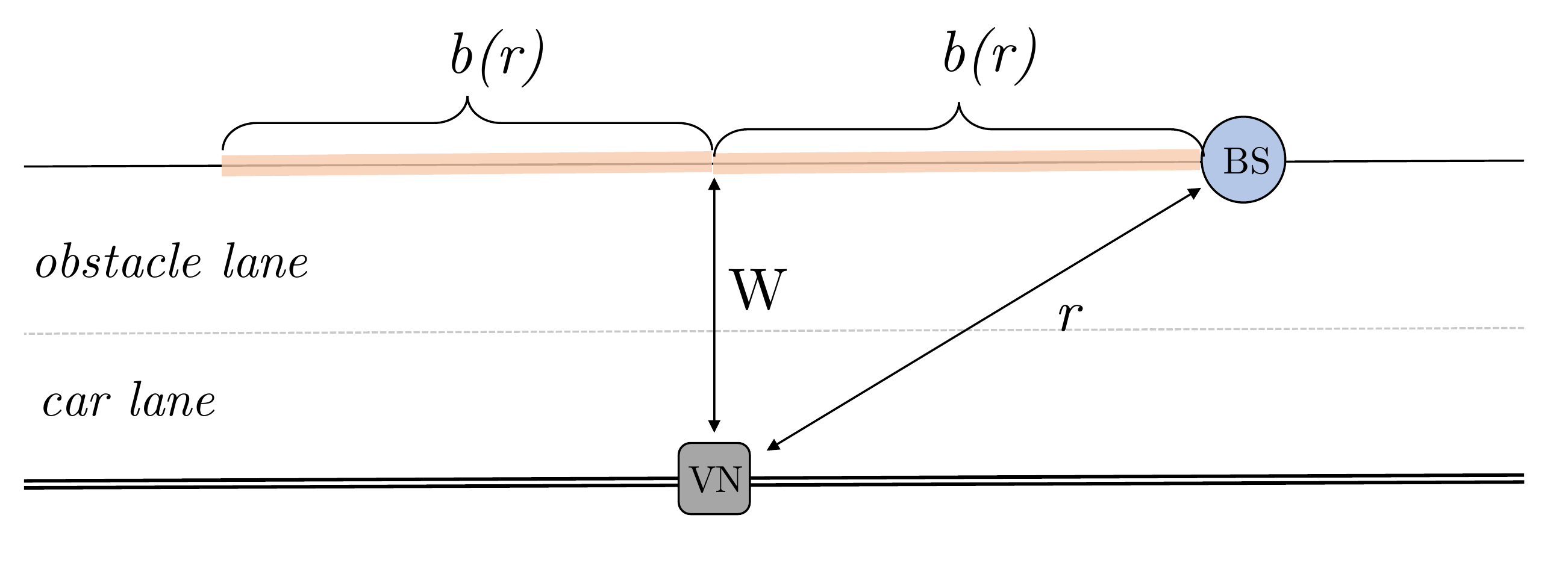}
    \caption{The test VN is at distance $r$ from the closest LOS (or NLOS) BS, and ${b(r)=\sqrt{r^2-W^2}}$.}
      \end{subfigure} \qquad \quad
           \begin{subfigure}[t!]{0.45\textwidth}
       \includegraphics[trim= 0cm 0cm 0cm 0cm , clip=true, width= 0.95\textwidth]{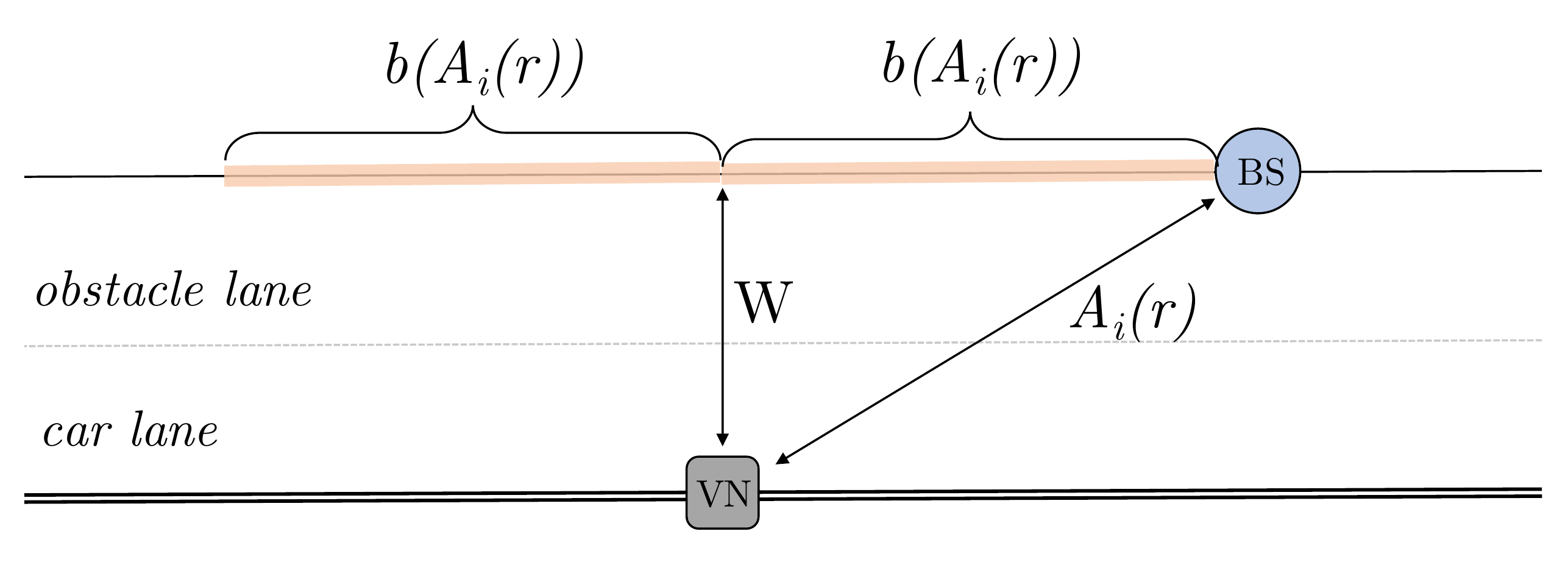}
    \caption{The test VN is at distance $A_i(r)$ from a BS $\in \Phi_i$, for $i\in\{L,\,N\}$, and $b(A_i(r))=\sqrt{A_i(r)^2-W^2}$.}
      \end{subfigure}
\caption{Illustration of a half section of highway of width $W$, as a support to the proof of Lemma  \ref{lemma:f_L} and Lemma \ref{lemma:P_L}.}
\label{fig:scenario_scheme}
\end{figure}

Considering the LOS case, the VN is at distance $r$ from the closest LOS BS if no other LOS BSs lie at distance closer than $r$. 
Considering the highway scenario described in  Sec. \ref{sec:network_model} and taking Fig. \ref{fig:scenario_scheme}(a) as a reference, there must be no other LOS BSs within the interval $[-b(r), \, b(r)]$, with $b(r)=\sqrt{r^2-W^2}$.
Since the spatial distribution of the LOS BSs is modeled as a 1-D Poisson process $\Phi_{L}$  with density $\lambda_L^{(s)}(r)$, we have

 \begin{align}
 F^{(s)}_L(r)=&\mathbb{P}\Big[\text{No LOS BSs within the interval }[-b(r), \, b(r)]\Big] \notag  \\ 
& = \exp \left(-\int_{-b(r)}^{b(r)}\lambda_L^{(s)}(x) \mathrm{d}x \right)\notag  \\\notag
&\stackrel{(a)}{=} \exp \left(-2\int_{0}^{b(r)}\lambda_L^{(s)}(x) \mathrm{d}x  \right)  \\
& \stackrel{(b)}{=} \exp \left(-2\lambda_b\int_{0}^{b(r)}p_L^{(s)}(x) \mathrm{d}x  \right) \label{eq:inside_proof_lemma_f_L}
  \end{align}
with $s\in \{R,\,U\}$. 
Step $(a)$ follows from the symmetry of the scenario and $(b)$ from the fact that  $\lambda_{L}^{(s)}(x) = p_L^{(s)}(x) \lambda_b$. 
The PDF of $r$  can be computed as
$$
f_L^{(s)}(r) = \frac{\partial }{ \partial r} \Big(1- F^{(s)}_L(r)\Big) = \frac{\partial }{ \partial r}\left( 1-\exp \left( -2\lambda_b \int_0^{b(r)} p_L^{(s)}(x) \mathrm{d}x \right)  \right),
$$
which gives  Eq. \eqref{eq:f_L} for the LOS case. 
With a similar reasoning, it is also possible to prove the lemma for the NLOS case.
\section{Proof of Lemma \ref{lemma:P_L}}
\label{appendix:P_L}

Let $r_L$ and $r_N$ be the random variables expressing the distance to the closest LOS and NLOS BSs, respectively.
 For $s\in\{R,\,U\}$, consider the event in which the test VN connects to a LOS BS, at distance $r_L$. 
Such an event requires the LOS BS to have smaller path loss than that of the nearest NLOS BS, at distance $r_N$. The probability $P^{(s)}_L$ of connecting to the LOS BS can therefore be expressed as
\begin{align}
P^{(s)}_L&= \mathbb{P}\Big[ C_Lr_L^{-\alpha_L} > C_Nr_N^{-\alpha_N}\Big] =\mathbb{P}\Big[r_N^{-\alpha_N} < \Big(C_L/C_N\Big)r_L^{-\alpha_L}\Big] \notag \\
 &= \mathbb{P}\Big[ r_N^{\alpha_N} > \Big(C_N/C_L\Big) r_L^{\alpha_L}  \Big] = \mathbb{P}\Big[ r_N > \left( (C_N/C_L) r_L^{\alpha_L} \Big)^{\frac{1}{\alpha_N}} \right]  \notag\\
  &{=} \int_W^\infty  \mathbb{P}\Big[r_N > A_L(r)\Big]f^{(s)}_L(r) \mathrm{d}r,
  \label{eq:lemma_p_N}
\end{align}
where the last step follows from the fact that $r>W$ by construction and that $f^{(s)}_L(r) $ is the PDF of $r_L$, as per Lemma \ref{lemma:f_L}.
From Lemma~\ref{lemma:f_L} and considering the highway scenario represented in Fig.~\ref{fig:scenario_scheme}(b), $\mathbb{P}[r_N > A_L(r)]$ can be regarded as the probability that there are no NLOS BSs within the interval $[-b(A_L(r)), \, b(A_L(r))]$, with $b(r)=\sqrt{b(A_L(r))^2-W^2}$, and can be written as
\beq
\mathbb{P}[r_N > A_L(r)] = \exp \left(-2\lambda_b \int_0^{b(A_L(r))}p_N^{(s)}(x) \mathrm{d}x\right).
\label{eq:lemma_p_N_2}
\eeq
By  substituting Eq. \eqref{eq:lemma_p_N_2} into Eq. \eqref{eq:lemma_p_N}, we get Lemma \ref{lemma:P_L} for the LOS scenario.
The proof for the NLOS case follows the same line of reasoning.

\section{Proof of Theorem \ref{th:P_cov}}
\label{appendix:P_cov}

Let $r_i^{n^*}$  be the random variable expressing the distance to the serving BS $n^* \in \Phi_i$, for $i\in\{L,N\}$.
The joint probability ${\mathbb{P}[ \text{SINR} > \Gamma,\, n^* \in \Phi_{i} ] }$ in Eq. \eqref{eq:def_P_cov} can be expressed as
\begin{align}
\mathbb{P}&\Big[ \text{SINR} > \Gamma,\, n^* \in \Phi_{i} \Big] = \mathbf{E}_{r_i^{n^*}}\Big[ \mathbb{P}[ \text{SINR}_i > \Gamma \mid r_i^{n^*}]  \Big] \notag  \\
&\stackrel{(a)}{=} \mathbf{E}_{r_i^{n^*}}\left\lbrace \mathbb{P}\left[ \frac{ |h_1|^2\Delta_1C_ir_i^{-\alpha_i}}{(I_L + I_N) + \sigma^2}  > \Gamma \mathrel{\Big|} r_i^{n^*}\right] \right\rbrace\notag \\
&=  \int_W^\infty  \mathbb{P}\left[ \frac{ |h_1|^2\Delta_1C_ir^{-\alpha_i}}{(I_L + I_N) + \sigma^2}  > \Gamma \mathrel{\Big|} r \right] \bar{f}^{(s)}_i(r) \mathrm{d}r \notag \\
&=  \int_W^\infty  \mathbb{P}\left[|h_1|^2 > \frac{ \Big[(I_L + I_N) + \sigma^2\Big]\Gamma r^{\alpha_i}}{\Delta_1 C_i}  \mathrel{\bigg|} r \right] \bar{f}^{(s)}_i(r) \mathrm{d}r,
\label{eq:appendix_P_cov_1}
\end{align}

where $(a)$ has been obtained by using the definition of SINR$_i$ in Eq. \eqref{eq:SINR}. 
Now, since $|h_1|^2$ is exponentially distributed with mean $\mu$, the probability term inside Eq.~\eqref{eq:appendix_P_cov_1} can be expressed as

\medmuskip=0mu
\thickmuskip=0mu

\Laplace

\medmuskip=6mu
\thickmuskip=6mu

where $(b)$ derives from the definition of Laplace functional $\mathcal{L}_{\mathcal{X}}(t) \triangleq \mathbf{E}[e^{-t\mathcal{X}}]$.
By substituting Eq.~\eqref{eq:appendix_P_cov_laplace} into Eq.~\eqref{eq:appendix_P_cov_1}, the coverage probability becomes

\medmuskip=0mu
\thickmuskip=0mu

\sinrfinal

Given that the test VN is associated to BS $n^*\in \Phi_{i}$, the Laplace functional of the interference from BSs $\in \Phi_j$ to the test VN, is obtained as follows:

\inside 

\medmuskip=6mu
\thickmuskip=6mu

where $(a)$ follows from the i.i.d. distribution of the interferes' channel parameters $|h_k|^2$ and from the further independence from the point process $\Phi_j$, $(b)$ derives from the symmetry of the scenario, and by applying the \textit{probability generating functional} of the PPP \cite{chiu2013stochastic}.
Moreover, using the \emph{moment generating function} of exponentially distributed $|h_k|^2$'s, that is
\beq
\mathbf{E}_{h_k}\left[ \exp\Big( -t |h_k|^2 v^{-\alpha_j}C_j\Delta_{I_k} \Big) \right] = 
\frac{\sfrac{1}{\mu}}{\sfrac{1}{\mu}+tv^{-\alpha_j}C_j \Delta_{I_k}},
\label{eq:MGF}
\eeq
and from the consideration that $\Delta_{I_k}$ are discrete random variables with $\mathbb{P}[\Delta_{I} = G_b G_{\rm VN}]=\theta_b/\pi$ and $\mathbb{P}[\Delta_{I} = g_b g_{\rm VN}]=1-(\theta_b/\pi)$, the expression for $\mathcal{L}^{(s)}_{I_i^j}(t)$ in \eqref{eq:L_first} can be further written as:
\medmuskip=0mu
\thickmuskip=0mu
 \squeeze_eq
By combining Eq. \eqref{eq:sinrfinal} and \eqref{eq:appendix_P_cov_2}, the proof is concluded.

\medmuskip=6mu
\thickmuskip=6mu

\section{Proof of Theorem \ref{th:P_C}}
\label{appendix:P_C}

\begin{figure}[t!]
     \centering
     \vspace*{-0.5cm}
     \includegraphics[trim= 0cm 0cm 0cm 0cm , clip=true, width= 0.95\textwidth]{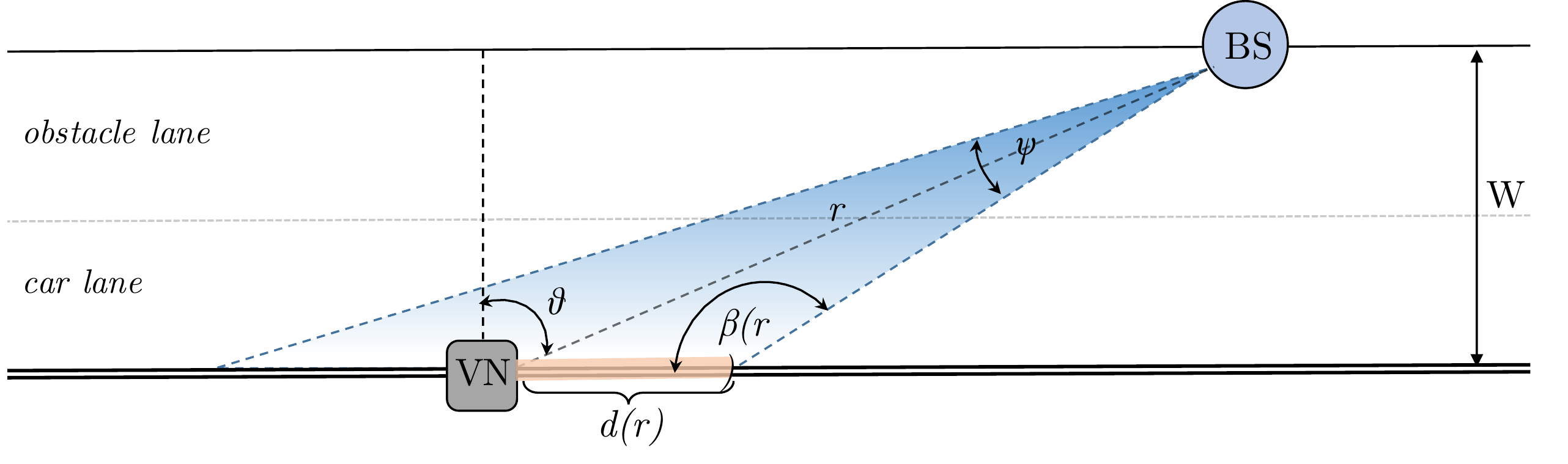}
\caption{Illustration of an half section of highway of width $W$, as a support to the proof of Theorem \ref{th:P_C}.}
\label{fig:scenario_d}
\end{figure}

Suppose that, at the beginning of a slot of duration $T_S$, the target VN is connected with a BS at distance $r$  (w.p. $P^{(s)}_{cov}(\Gamma)$). According to the considerations we made in Sec.~\ref{sec:system_model}, the main lobe center of the BS's transmit beam points at its associated~VN, as illustrated in Fig. \ref{fig:scenario_d}.
According to the \emph{law of sines}, the 	quantity  $d(r)$ that represents the maximum distance that the node can cover before leaving the communication range  of its serving BS is defined as
\beq
d(r) \triangleq \frac{r\sin(\psi/2)}{\sin(\beta(r))},
\label{eq:d_r}
\eeq
where $\psi$ is the beamwidth of the BS's main lobe and, from the trigonometric identities, $\beta(r)=\eta+\theta(r)$, with  $\eta = \pi/2-\psi/2$ and $\theta(r)=\arccos(W/r)$. Notice that $d(r)$ increases with $r$ due to the resulting wider geometric projection of the BS's beam onto the road surface.
That being said, the probability that the VN does not disconnect from its serving infrastructure can be viewed as the probability that the VN does not cover a distance greater than $d(r)$ within the slot. The distance covered by the VN, moving at speed $V$, within the slot of duration $T_S$ is $V T_S $,~therefore 
\begin{align}
P_{NL} &= \mathbb{P}[T_L > T_S] = \mathbb{P}\left[V  T_S  < d(r)\right]\notag\\
&{=} \mathbb{P}\left[V  T_S  <\frac{r\sin(\psi/2)}{\sin(\beta(r))}\right]\notag\\
& = \mathbb{P}\left[r  > \frac{VT_S\sin(\beta(r))}{\sin(\psi/2)}{}\right]\notag\\
&= \mathbb{P}\left[r  > \frac{VT_S\sin\Big(\overbrace{\pi/2-\psi/2}^{\eta}+\overbrace{\arccos(W/r)}^{\theta(r)}\Big)}{\sin(\psi/2)}{}\right]\notag\\
&\stackrel{(a)}{=}  \mathbb{P}\left[r  > \frac{VT_S}{\sin(\psi/2)}\Big(\sin(\eta)\cos(\arccos(W/r)) + \cos(\eta) \sin(\arccos(W/r)) \Big)\right]\notag\\
&\stackrel{(b)}{=} \mathbb{P}\left[ r > \frac{VT_S}{\sin(\psi/2)} \left( \frac{W}{r}\sin(\eta) + \sqrt{1-\left(\frac{W}{r}\right)^2}\cos(\eta)\right) \right], 
\label{eq:appendix_P_C}
\end{align}
where $(a)$ derives from the trigonometric addition formula $\sin(\eta + \theta(r)) = \sin(\eta)\cos(\theta(r)) + \cos(\eta)\sin(\theta(r))$, and $(b)$ from the trigonometric identities $\cos(\arccos(x))=x$ and $\sin(\arccos(x))=\sqrt{1-x^2}$. 
The expressions in \eqref{eq:appendix_P_C} and \eqref{eq:P_T_L} coincide, concluding the~proof.


\bibliographystyle{IEEEtran}
\bibliography{bibliography}

\begin{thebibliography}{10}
\providecommand{\url}[1]{#1}
\csname url@samestyle\endcsname
\providecommand{\newblock}{\relax}
\providecommand{\bibinfo}[2]{#2}
\providecommand{\BIBentrySTDinterwordspacing}{\spaceskip=0pt\relax}
\providecommand{\BIBentryALTinterwordstretchfactor}{4}
\providecommand{\BIBentryALTinterwordspacing}{\spaceskip=\fontdimen2\font plus
\BIBentryALTinterwordstretchfactor\fontdimen3\font minus
  \fontdimen4\font\relax}
\providecommand{\BIBforeignlanguage}[2]{{%
\expandafter\ifx\csname l@#1\endcsname\relax
\typeout{** WARNING: IEEEtran.bst: No hyphenation pattern has been}%
\typeout{** loaded for the language `#1'. Using the pattern for}%
\typeout{** the default language instead.}%
\else
\language=\csname l@#1\endcsname
\fi
#2}}
\providecommand{\BIBdecl}{\relax}
\BIBdecl

\bibitem{MOCAST_2017}
M.~Giordani, A.~Zanella, and M.~Zorzi, ``{Millimeter wave communication in
  vehicular networks: Challenges and opportunities},'' in \emph{2017 6th
  International Conference on Modern Circuits and Systems Technologies
  (MOCAST)}, May 2017.

\bibitem{bookV2X}
H.~Hartenstein and K.~Laberteaux, \emph{{VANET vehicular applications and
  inter-networking technologies}}.\hskip 1em plus 0.5em minus 0.4em\relax John
  Wiley \& Sons, 2009.

\bibitem{DSRC}
J.~B. Kenney, ``{Dedicated Short-Range Communications (DSRC) Standards in the
  United States},'' \emph{Proceedings of the IEEE}, vol.~99, no.~7, pp.
  1162--1182, Jul 2011.

\bibitem{3GPP_LTE}
3GPP, ``{Requirements for Evolved {U}{T}{R}{A} and Evolved {U}{T}{R}{A}{N}
  ({R}elease 7) },'' \emph{TR 25.913}, 2015.

\bibitem{Heath_surveyV2X}
\BIBentryALTinterwordspacing
V.~Va, T.~Shimizu, G.~Bansal, and R.~W. Heath, ``Millimeter wave vehicular
  communications: A survey,'' \emph{Foundations and Trends® in Networking},
  vol.~10, no.~1, pp. 1--113, 2016. [Online]. Available:
  \url{http://dx.doi.org/10.1561/1300000054}
\BIBentrySTDinterwordspacing

\bibitem{rappaportmillimeter}
T.~S. Rappaport, S.~Sun, R.~Mayzus, H.~Zhao, Y.~Azar, K.~Wang, G.~N. Wong,
  J.~K. Schulz, M.~Samimi, and F.~Gutierrez, ``{Millimeter Wave Mobile
  Communications for 5G Cellular: It Will Work!}'' \emph{IEEE Access}, vol.~1,
  pp. 335--349, May 2013.

\bibitem{magazine_IA}
M.~Giordani, M.~Mezzavilla, and M.~Zorzi, ``{Initial Access in 5{G} mm{W}ave
  Cellular Networks},'' \emph{IEEE Communications Magazine}, vol.~54, no.~11,
  pp. 40--47, Nov 2016.

\bibitem{keyelement_mmWave}
L.~Wei, R.~Q. Hu, Y.~Qian, and G.~Wu, ``{Key elements to enable millimeter wave
  communications for 5G wireless systems},'' \emph{IEEE Wireless
  Communications}, vol.~21, no.~6, pp. 136--143, Dec 2014.

\bibitem{Yamamoto_08}
A.~Yamamoto, K.~Ogawa, T.~Horimatsu, A.~Kato, and M.~Fujise, ``{Path-Loss
  Prediction Models for Intervehicle Communication at 60 GHz},'' \emph{IEEE
  Transactions on Vehicular Technology}, vol.~57, no.~1, pp. 65--78, Jan 2008.

\bibitem{JSAC_2017}
M.~Polese, M.~Giordani, M.~Mezzavilla, S.~Rangan, and M.~Zorzi, ``{Improved
  Handover through Dual Connectivity in 5G mmWave mobile networks},''
  \emph{IEEE Journal on Selected Areas in Communications}, vol.~35, no.~9, pp.
  2069--2084, Sept 2017.

\bibitem{TCP_DSRC}
B.~Khorashadi, A.~Chen, D.~Ghosal, C.~N. Chuah, and M.~Zhang, ``{Impact of
  Transmission Power on the Performance of UDP in Vehicular Ad Hoc Networks},''
  in \emph{IEEE International Conference on Communications}, Jun 2007, pp.
  3698--3703.

\bibitem{Highway_DSRC}
N.~Akhtar, S.~C. Ergen, and O.~Ozkasap, ``{Vehicle Mobility and Communication
  Channel Models for Realistic and Efficient Highway VANET Simulation},''
  \emph{IEEE Transactions on Vehicular Technology}, vol.~64, no.~1, pp.
  248--262, Jan 2015.

\bibitem{throughput_infrastructure_chen}
J.~Chen, G.~Mao, C.~Li, A.~Zafar, and A.~Y. Zomaya, ``{Throughput of
  Infrastructure-Based Cooperative Vehicular Networks},'' \emph{IEEE
  Transactions on Intelligent Transportation Systems}, Mar 2017.

\bibitem{Heath_V2X_magazine}
J.~Choi, V.~Va, N.~Gonzalez-Prelcic, R.~Daniels, C.~R. Bhat, and R.~W. Heath,
  ``{Millimeter-Wave Vehicular Communication to Support Massive Automotive
  Sensing},'' \emph{IEEE Communications Magazine}, vol.~54, no.~12, pp.
  160--167, Dec 2016.

\bibitem{Heath_automotive_radar}
P.~Kumari, N.~Gonzalez-Prelcic, and R.~W. Heath, ``{Investigating the IEEE
  802.11ad Standard for Millimeter Wave Automotive Radar},'' in \emph{2015 IEEE
  82nd Vehicular Technology Conference (VTC2015-Fall)}, Sept 2015.

\bibitem{Heath_high_speed_train}
V.~Va, X.~Zhang, and R.~W. Heath, ``{Beam Switching for Millimeter Wave
  Communication to Support High Speed Trains},'' in \emph{2015 IEEE 82nd
  Vehicular Technology Conference (VTC2015-Fall)}, Sept 2015.

\bibitem{Bacelli_book}
F.~Baccelli and B.~Błaszczyszyn, ``{Stochastic Geometry and Wireless Networks:
  Volume {I} Theory},'' \emph{{Foundations and Trends® in Networking}},
  vol.~3, no. 3–4, pp. 249--449, 2010.

\bibitem{Andrews_tractable_approach_stochastic}
J.~G. Andrews, F.~Baccelli, and R.~K. Ganti, ``{A Tractable Approach to
  Coverage and Rate in Cellular Networks},'' \emph{IEEE Transactions on
  Communications}, vol.~59, no.~11, pp. 3122--3134, Nov 2011.

\bibitem{stochastic_highway_DSRC}
M.~J. Farooq, H.~ElSawy, and M.~S. Alouini, ``{A Stochastic Geometry Model for
  Multi-Hop Highway Vehicular Communication},'' \emph{IEEE Transactions on
  Wireless Communications}, vol.~15, no.~3, pp. 2276--2291, Mar 2016.

\bibitem{stochastic_highway_DSRC_tong}
Z.~Tong, H.~Lu, M.~Haenggi, and C.~Poellabauer, ``{A Stochastic Geometry
  Approach to the Modeling of DSRC for Vehicular Safety Communication},''
  \emph{IEEE Transactions on Intelligent Transportation Systems}, vol.~17,
  no.~5, pp. 1448--1458, May 2016.

\bibitem{stochastic_highway_DSRC_Farooq}
M.~J. Farooq, H.~ElSawy, and M.~S. Alouini, ``{Modeling Inter-Vehicle
  Communication in Multi-Lane Highways: A Stochastic Geometry Approach},'' in
  \emph{2015 IEEE 82nd Vehicular Technology Conference (VTC2015-Fall)}, Sept
  2015.

\bibitem{Coverage_rate_analysis}
T.~Bai and R.~W. Heath, ``{Coverage and Rate Analysis for Millimeter-Wave
  Cellular Networks},'' \emph{IEEE Transactions on Wireless Communications},
  vol.~14, no.~2, pp. 1100--1114, Feb 2015.

\bibitem{on_the_feasibility}
A.~K. Gupta, J.~G. Andrews, and R.~W. Heath, ``{On the Feasibility of Sharing
  Spectrum Licenses in mmWave Cellular Systems},'' \emph{IEEE Transactions on
  Communications}, vol.~64, no.~9, pp. 3981--3995, Sept 2016.

\bibitem{rebato17}
M.~Rebato, J.~Park, P.~Popovski, E.~D. Carvalho, and M.~Zorzi, ``{Stochastic
  Geometric Coverage Analysis in mmWave Cellular Networks with a Realistic
  Channel Model},'' in \emph{IEEE Global Communications Conference (GLOBECOM)},
  Dec 2017.

\bibitem{Tassi17_Highway}
A.~Tassi, M.~Egan, R.~J. Piechocki, and A.~Nix, ``{Modeling and Design of
  Millimeter-Wave Networks for Highway Vehicular Communication},'' \emph{IEEE
  Transactions on Vehicular Technology}, vol.~66, no.~12, pp. 10\,676--10\,691,
  Dec 2017.

\bibitem{Mustafa}
M.~R. Akdeniz, Y.~Liu, M.~K. Samimi, S.~Sun, S.~Rangan, T.~S. Rappaport, and
  E.~Erkip, ``{Millimeter Wave Channel Modeling and Cellular Capacity
  Evaluation},'' \emph{IEEE Journal on Selected Areas in Communications},
  vol.~32, no.~6, pp. 1164--1179, June 2014.

\bibitem{coverage_het}
E.~Turgut and M.~C. Gursoy, ``{Coverage in Heterogeneous Downlink Millimeter
  Wave Cellular Networks},'' \emph{IEEE Transactions on Communications},
  vol.~65, no.~10, pp. 4463--4477, Oct 2017.

\bibitem{MedHoc2016}
M.~Giordani, M.~Mezzavilla, S.~Rangan, and M.~Zorzi, ``{{Multi-Connectivity} in
  {5G} mmWave Cellular Networks},'' in \emph{15th Annual Mediterranean Ad Hoc
  Networking Workshop (Med-Hoc-Net'16)}, Vilanova i la Geltru, Barcelona,
  Spain, Jun. 2016.

\bibitem{TWC2017}
\BIBentryALTinterwordspacing
------, ``{An Efficient Uplink Multi-Connectivity Scheme for 5{G} mm{W}ave
  Control Plane Applications},'' \emph{submitted to the IEEE Transactions of
  Wireless Communications, CoRR}, vol. abs/1610.04836, 2017. [Online].
  Available: \url{http://arxiv.org/abs/1610.04836}
\BIBentrySTDinterwordspacing

\bibitem{Park_channelModel}
J.~Park, S.~L. Kim, and J.~Zander, ``{Tractable Resource Management With Uplink
  Decoupled Millimeter-Wave Overlay in Ultra-Dense Cellular Networks},''
  \emph{IEEE Transactions on Wireless Communications}, vol.~15, no.~6, pp.
  4362--4379, Jun 2016.

\bibitem{akoum2012coverage}
S.~Akoum, O.~E. Ayach, and R.~W. Heath, ``{Coverage and capacity in mmWave
  cellular systems},'' in \emph{{ Conference Record of the Forty Sixth Asilomar
  Conference on Signals, Systems and Computers (ASILOMAR)}}, Nov 2012, pp.
  688--692.

\bibitem{tse_book}
D.~Tse and P.~Viswanath, \emph{Fundamentals of wireless communication}.\hskip
  1em plus 0.5em minus 0.4em\relax Cambridge University Press, 2005.

\bibitem{chiu2013stochastic}
S.~N. Chiu, D.~Stoyan, W.~S. Kendall, and J.~Mecke, \emph{Stochastic geometry
  and its applications}.\hskip 1em plus 0.5em minus 0.4em\relax John Wiley \&
  Sons, 2013.

\end{thebibliography}

\begin{IEEEbiography}
    [{\includegraphics[width=0.9in,clip,keepaspectratio]{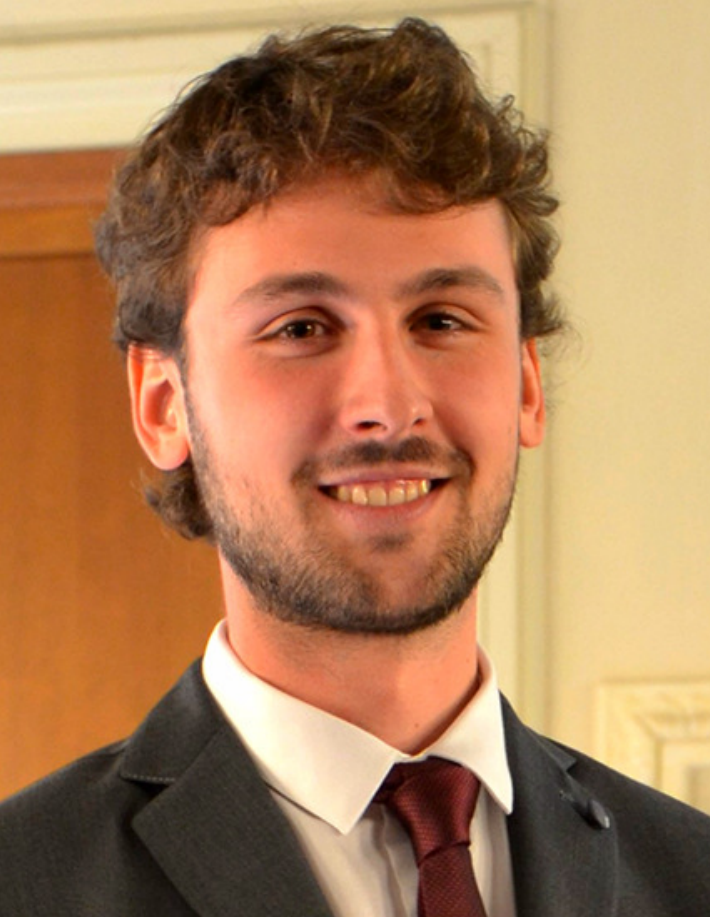}}]{Marco Giordani}
received his B.Sc. in Information Engineering in 2013 and the M.Sc. in
Telecommunication Engineering in 2015, both from the University of Padova
(Italy). Since October 2015 he has been a postgraduate researcher at the Department
of Information Engineering of the University of Padova (Italy), under the
supervision of prof. Michele Zorzi. From January to April 2016 he has been a visiting research
scholar at the New York University (NYU), USA.
  From October 2016, he is a Ph.D. student  at the Department of 
Information Engineering of the University of Padova (Italy). 
His research interests include design and validation of protocols for mobility and control-plane management in next-generation cellular networks (5G) and vehicular systems operating at millimeter waves.
\end{IEEEbiography}
\begin{IEEEbiography}
    [{\includegraphics[width=0.9in,clip,keepaspectratio]{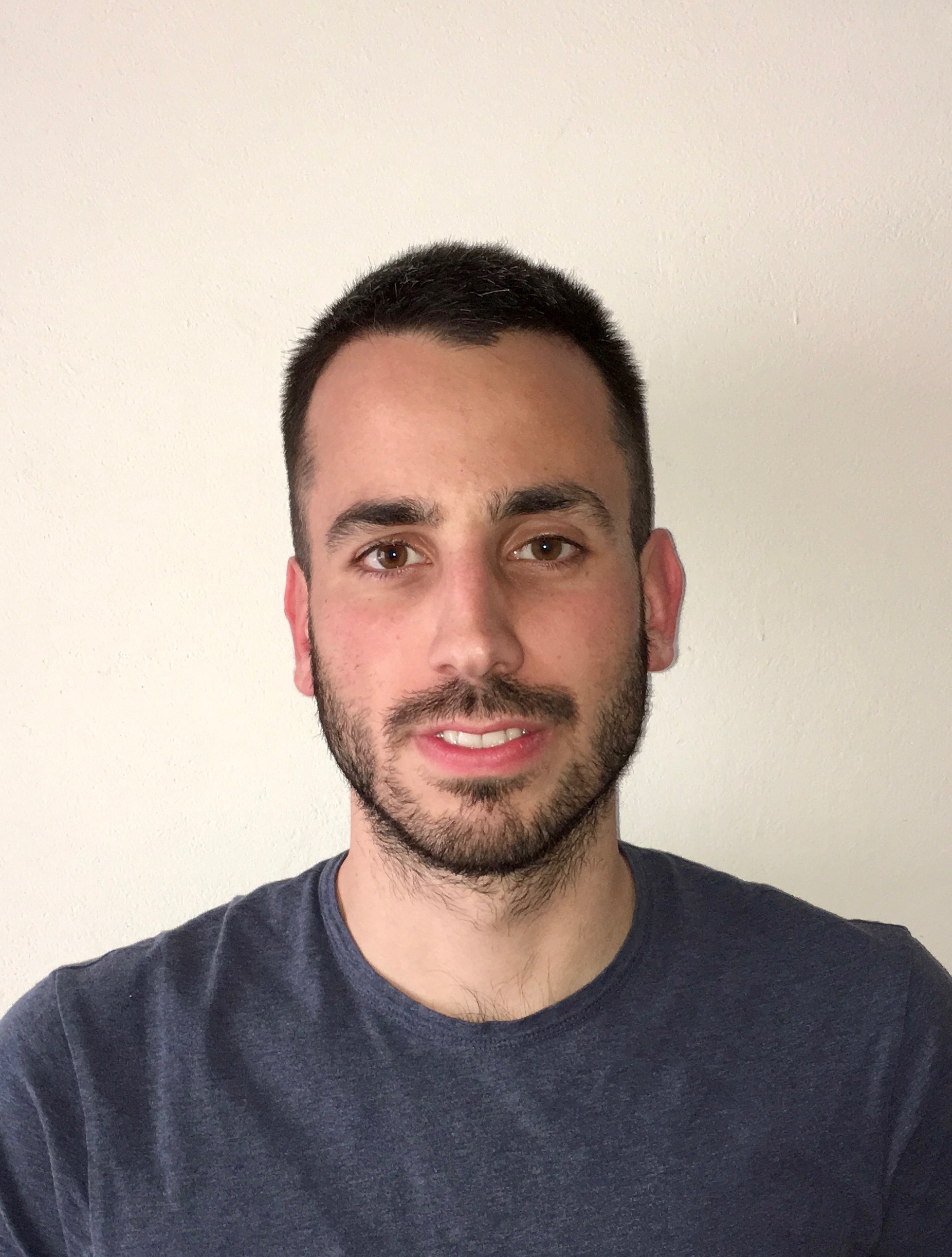}}]{Mattia Rebato}
received the B.Sc. degree in information engineering and the M.Sc. degree in telecommunication engineering from the University of Padova, Italy, in 2013 and 2015, respectively, where he is currently pursuing the Ph.D. degree with the Department of Information Engineering. Since 2015, he has been a Postgraduate Researcher with the Department of Information Engineering, University of Padova, under the supervision of Prof. M. Zorzi. In 2016, for three months, he was with New York University, NY, USA, as a Visiting Research Scholar. His research interests include design, analysis, and validation of protocols and applications for the next generation of cellular networks (5G), in particular for millimeter wave communication, resource sharing, and interference evaluation.
\end{IEEEbiography}
\begin{IEEEbiography}
    [{\includegraphics[width=0.9in,clip,keepaspectratio]{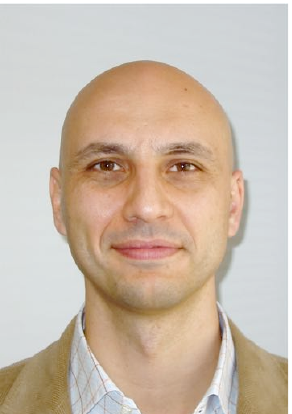}}]{Andrea Zanella}
[S'98-M'01-SM'13] is Associate Professor at the University of Padova, in Italy. He received the master degree in Computer Engineering and the Ph.D. degree in Electronic and Telecommunications Engineering from the same University, in 1998 and 2000, respectively. He has (co)authored more than 130 papers, five books chapters and three international patents in multiple subjects related to wireless networking and Internet of Things, Vehicular Networks, cognitive networks, and microfluidic networking. He serves as Technical Area Editor for the IEEE Internet of Things Journal, and as Associate Editor for the IEEE Communications Surveys \& Tutorials, and the IEEE Transactions on Cognitive Communications and Networking.
\end{IEEEbiography}
\begin{IEEEbiography}
    [{\includegraphics[width=0.9in,height=1.125in,clip,keepaspectratio]{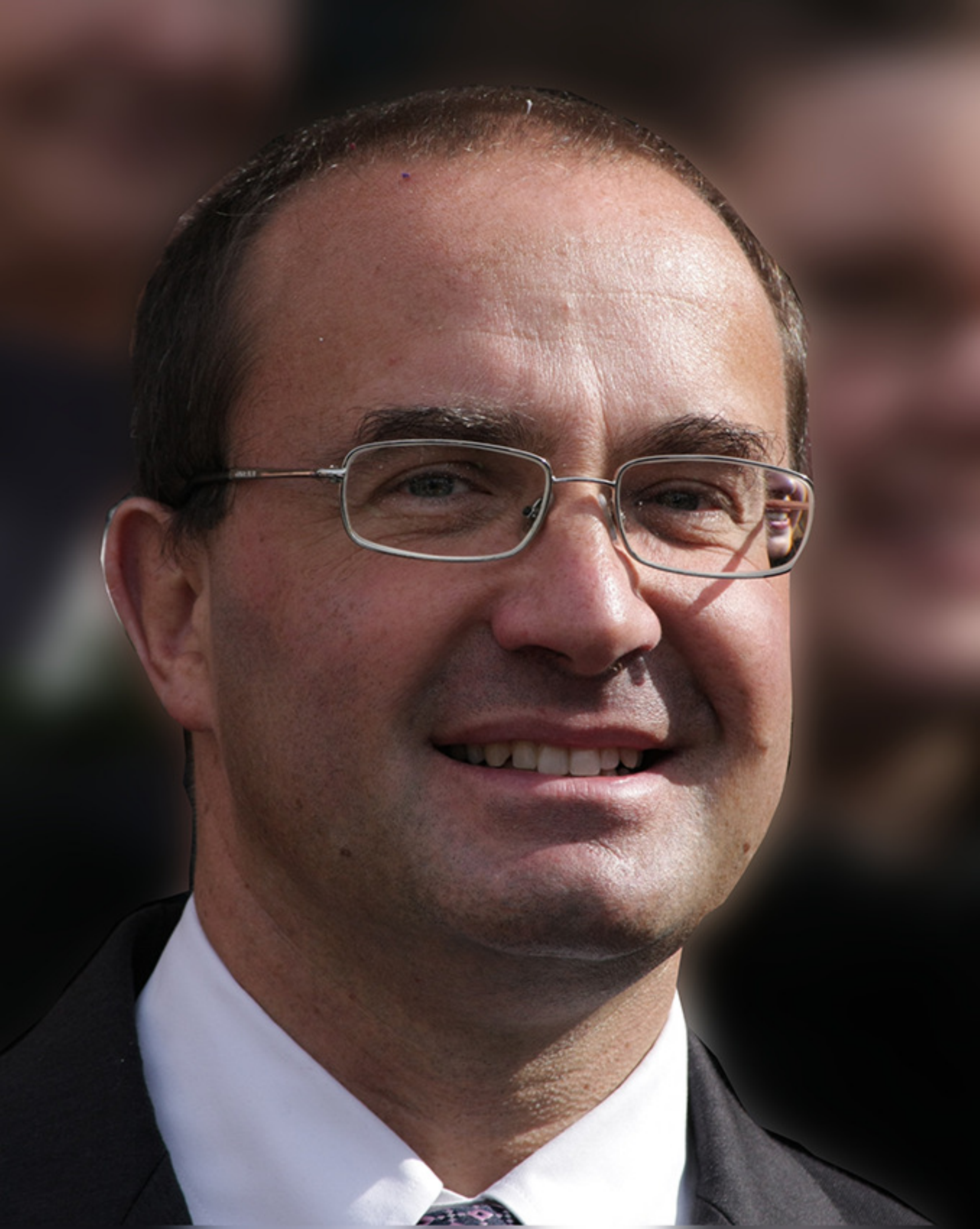}}]{Michele Zorzi}
   received his Laurea and PhD degrees in electrical engineering from the University of Padova in 1990 and 1994, respectively. During academic year 1992-1993 he was on leave at the University of California at San Diego (UCSD). In 1993 he joined the faculty of the Dipartimento di Elettronica e Informazione, Politecnico di Milano, Italy. After spending three years with the Center for Wireless Communications at UCSD, in 1998 he joined the School of Engineering of the University of Ferrara, Italy, where he became a professor in 2000. Since November 2003 he has been on the faculty of the Information Engineering Department at the University of Padova. His present research interests include performance evaluation in mobile communications systems, WSN and Internet-of-Things, cognitive communications and networking, 5G mmWave cellular systems, and underwater communications and networks. He is currently Editor in Chief of the IEEE Transactions on Cognitive Communications and Networking, and was Editor-In-Chief of IEEE Wireless Communications from 2003 to 2005 and of IEEE Transactions on Communications from 2008 to 2011. He served as a Member-at-Large of the Board of Governors of the IEEE Communications Society from 2009 to 2011, and as its Director of Education in 2014-15. He is a Fellow of the IEEE.
\end{IEEEbiography}

\end{document}